\documentclass[10pt, twocolumn, aps, pra,
longbibliography, superscriptaddress, nofootinbib]{revtex4-1}

\pdfoutput=1
\usepackage[utf8x]{inputenc}
\usepackage{microtype}
\usepackage{amsmath}
\usepackage{amsfonts}
\usepackage{amssymb}
\usepackage{makeidx}
\usepackage{cellspace,booktabs,array}
\usepackage{lipsum}
\usepackage{bm}
\usepackage{hyperref}
\usepackage{graphicx}
\usepackage[dvipsnames]{xcolor}
\usepackage{mathtools}
\usepackage{subcaption}

\renewcommand{\figurename}{\textbf{Fig.}}

\allowdisplaybreaks % allow pagebreak in equations

\usepackage{siunitx}
\usepackage{tabularx}

\definecolor{myorange}{RGB}{255,188,0}
\definecolor{mygreen}{RGB}{0,130,0}
\definecolor{PlotRed}{RGB}{236,37,0}
\definecolor{PlotBlue}{RGB}{0,92.5,175}

\begin{document}

\title{Hybrid Quantum Error Correction in Qubit Architectures}

\begin{abstract}
Noise and errors are inevitable parts of any practical implementation of a quantum computer. As a result, large-scale quantum computation will require ways to detect and correct errors on quantum information. Here, we present such a quantum error correcting scheme for correcting the dominant error sources, phase decoherence and energy relaxation, in qubit architectures, using a hybrid approach combining autonomous correction based on engineered dissipation with traditional measurement-based quantum error correction. Using numerical simulations with realistic device parameters for superconducting circuits, we show that this scheme can achieve a 5- to 10-fold increase in storage-time while using only six qubits for the encoding and two ancillary qubits for the operation of the autonomous part of the scheme, providing a potentially large
reduction of qubit overhead compared to typical measurement-based error correction schemes. Furthermore, the scheme relies on standard interactions and qubit driving available in most major quantum computing platforms, making it implementable in a wide range of architectures.
\end{abstract}
%\pacs{03.67.Lx, 42.50.Dv, 85.25.-j, 42.50.Pq}

\date{\today}
\author{Lasse Bjørn Kristensen}
\affiliation{Department of Physics and Astronomy, Aarhus University, DK-8000 Aarhus C, Denmark}
\author{Morten Kjaergaard}
\affiliation{Research Laboratory of Electronics, Massachusetts Institute of Technology, Cambridge, MA 02139, USA}
\author{Christian Kraglund Andersen}
\affiliation{Department of Physics, ETH Zürich, CH-8093 Zürich, Switzerland}
\author{Nikolaj Thomas Zinner}
\affiliation{Department of Physics and Astronomy, Aarhus University, DK-8000 Aarhus C, Denmark}
\affiliation{Aarhus Institute of Advanced Study, Aarhus University, DK-8000 Aarhus C, Denmark}

\maketitle
\section{Introduction}
Quantum mechanics as a tool for computation has gained significant traction due to the promising progress in quantum algorithms that could outperform classical computing~\cite{Shor1999, Aaronson2011, Devoret2013,Peruzzo2014, Schuld2018,Bravyi2018}. Nevertheless, quantum computers still face the challenge of inevitable interaction with the surrounding environment. 
Great progress has been made on the hardware side in creating well isolated and highly controllable systems with long coherence times, achieving improvements to coherence times of several orders of magnitude~\cite{Devoret2013,Kjaergaard2019}. Nevertheless, for many of the most promising architectures, 
achieving error rates low enough to perform 
quantum computational tasks to arbitrary precision still poses a significant challenge~\cite{Babbush2018,Kjaergaard2019}. 
To solve this problem, it will therefore be necessary to engineer systems where the physical errors can be detected and corrected. Many protocols have been developed for such quantum error correcting (QEC) schemes~\cite{Chesi2010,Gottesman2010,Fowler2012,Reed2012,Devitt2013,Bombin2015,Michael2016,Ofek2016,Johnson2017}, and a number of experiments have investigated their usefulness and implementability~\cite{Cory1998,Knill2001,Chiaverini2004,Schindler2011,Zhang2012,Nigg2014,Corcoles2015,Kelly2015,Takita2017,Andersen2019}. Common to many of these schemes is the concept of performing measurements of specific stabilizer operators in order to detect errors. However, measurements tend to take a significant amount of time, they can introduce additional errors, and they often require both additional quantum resources and additional classical signal analysis hardware~\cite{Riste2013,Negnevitsky2018,Andersen2019}. As a result, recent work have been done on investigating how engineered interactions and dissipation can be used to replace these measurements, leading to systems capable of autonomously correcting errors without the need for
potentially error-introducing measurements and corresponding real-time classical logic~
\cite{Pastawski2009,Kapit2015,Kapit2016,Cohen2017,Reiter2017,Lihm2018}. In the case of platforms that require low temperatures, 
a major advantage of these autonomous schemes is a reduced 
information transfer between the cryogenic environment and a room temperature classical electronics system for feedback, 
leading to a decrease in generated entropy and increased energy-efficiency.

Here we present a novel hybrid error correction scheme that utilizes simple interactions and engineered dissipation to implement partially self-correcting QEC 
codes. This scheme provides a means to reduce the overhead required in typical measurement-based QEC schemes. To demonstrate the principles of this hybrid QEC scheme, we consider a system with 8 physical two-level (qubit) systems, two of which are coupled strongly to the environment, and show that it can store one logical bit of quantum information, i.e. a single logical qubit. Even with such a simple setup, the scheme achieves a 5- to 10-fold increase of storage lifetimes. While the scheme
is general, we consider superconducting circuits as a case in point. Using parameters corresponding to state-of-the-art superconducting transmon qubits, we demonstrate that storage lifetimes of around 1 millisecond are achievable using hybrid QEC.

\begin{figure*}[hbtp]
  \centering
   \includegraphics[width=1.91\columnwidth]{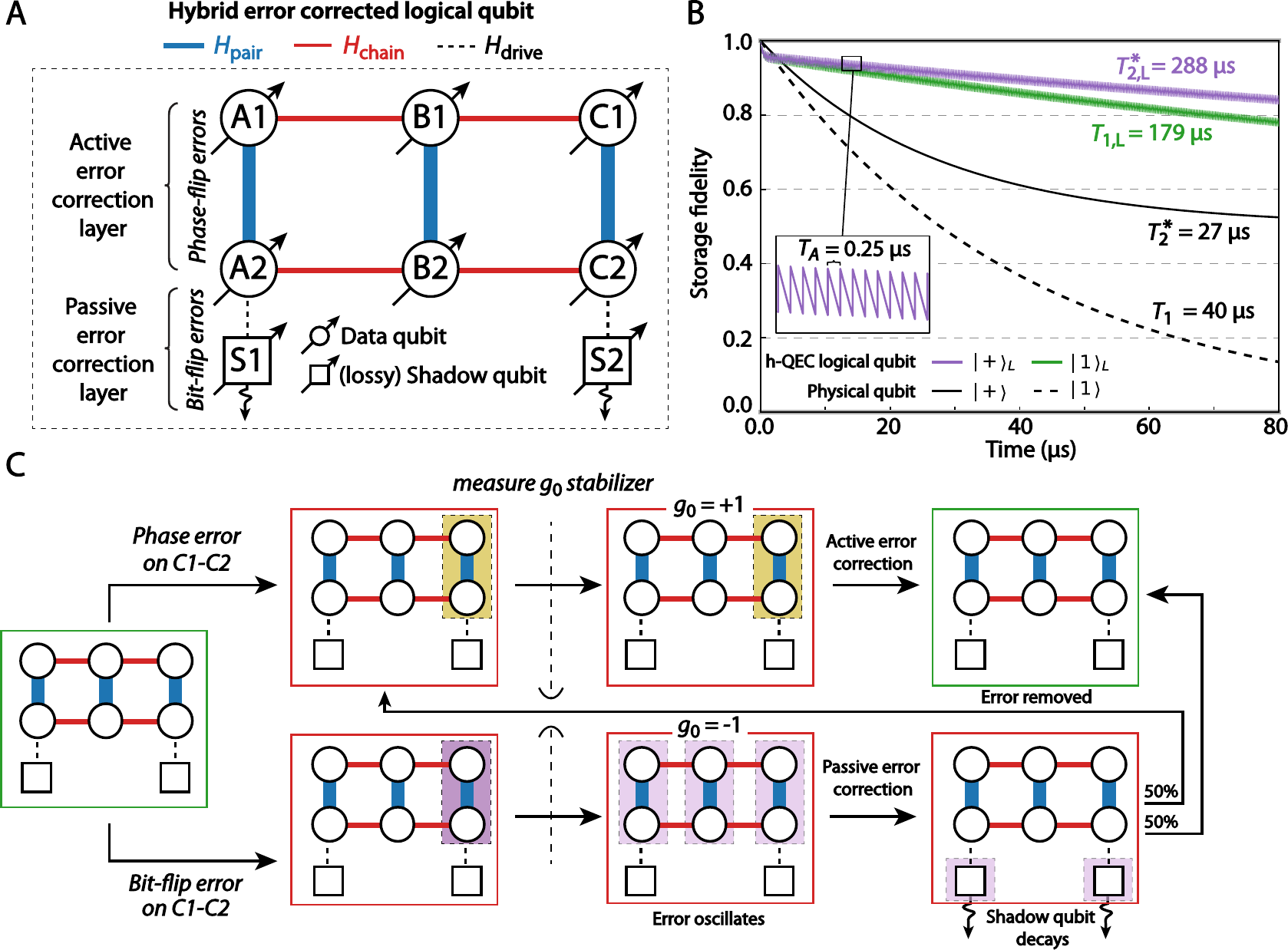}
   \caption{\textbf{Geometry of the hQEC system, schematic depiction of the corresponding error correction scheme, and a plot of the resulting error correction performance.} (\textbf{A}): Schematic depiction of the six data and two 'shadow' qubits setup. The vertical interactions in 
	each block (A,B, and C) are of Heisenberg $XXZ$ type, while the horizontal interactions between blocks are 
	exchange $XX$ interactions.
	(\textbf{B}): Simulation results for the storage fidelity achieved for the states $\left| \, 1 \, \right>_L$ and $  \frac{1}{\sqrt{2}} \left( \left| 0 \right> + \left| 1 \right> \right)_L$ in the 6-qubit hybrid scheme from (a), with results for an uncorrected single qubit included for comparison. The inset depicts behaviour on short time-scales resulting from the interplay of continuous decoherence and discontinuous active correction steps. 
	(\textbf{C}): A schematic depiction of the error correction steps. Starting from a valid state on the left, the system can either undergo phase-errors or loss-errors. If a phase-error occurs, $g_0=+1$ and the active correction is performed to correct the error (Top). If a loss-error occurs, the error oscillates until it is either corrected or converted to a phase error by shadow-qubit decay (Bottom). During the oscillations, $g_0=-1$, thus active correction is disabled.}
  \label{fig:Mega}
\end{figure*}

\section{Results}

The minimal setup of this hybrid quantum error correcting scheme (hQEC) requires a set
of qubits that interact with standard 
spin-spin exchange of strength $J_x$ and Ising interactions of strength $J_z$, in combination
a Heisenberg $XXZ$ interaction.
These are realizable in most practical platforms 
of relevance for quantum computing and quantum information processing. A subset of our qubits are 
assumed to have energy relaxation rates, i.e. to decay fast to their ground state. This can be achieved 
by coupling the qubit to a controllable environment that absorbs energy on a fast timescale. Additionally, 
we apply driving to the qubits, and we use ancilla qubits to measure 'syndrome' operators 
on a subset of the qubits in our architecture. All of these
requirements are standard tools in quantum technology platforms. 
As a model for the general noise on the system, we assume two channels; 
excitation loss (energy relaxation of the qubit) and phase noise. 
Our hybrid QEC scheme can overcome both of these noise channels by a combination 
of passive and active error correction steps. 

\subsection{Example of Implementation}
As a concrete example to illustrate the physics of the corrective 
procedure and its performance, we consider an implementation with six 
'data' qubits (circles) shown 
in Fig.~\ref{fig:Mega}. The example uses an additional two high-loss qubits, so-called 
'shadow' qubits (squares), to passively protect one logical qubit against spontaneous decay 
(this can be reduced to a single shadow---see App. \ref{sec:Single_Shadow}).
To protect against phase noise, 
we implement an active measurement protocol using a relatively 
standard QEC syndrome protocol that we describe below.
The geometry of the implementation consists of a set of three two-qubit pairs that are 
coupled via an XXZ term, while the pairs are coupled via 
exchange interactions, see Fig. \ref{fig:Mega}A. The full 
Hamiltonian of the system is given by $H= H_0 +  H_{\text{pair}} + H_{\text{chain}}$
where
\begin{align}
H_0 =& \hspace{0.5cm} \sum_{\mathclap{\substack{\phantom{n}\\i \in \{A,B,C\} \\ j \in \{1,2\}}}} \; \frac{1}{2} \Omega \, \sigma^z_{i,j} + \frac{1}{2}\Omega_S \, \sigma_{S_1}^z + \frac{1}{2}\Omega_S \, \sigma_{S_2}^z \;, \\
H_{\text{pair}} =& \hspace{0.5cm} \sum_{\mathclap{i \in \{A,B,C\}}} \left[ J_x \left( \sigma_{i,1}^x  \sigma_{i,2}^x + \sigma_{i,1}^y \, \sigma_{i,2}^y \right) + J_z \, \sigma_{i,1}^z  \sigma_{i,2}^z \right]  \; , \\
H_{\text{chain}} =& \, \delta \sum_{j=1}^2  \left( \sigma_{A,j}^x \sigma_{B,j}^x + \sigma_{A,j}^y \sigma_{B,j}^y \right) \nonumber\\
&+ \delta \sum_{j=1}^2  \left( \sigma_{B,j}^x \sigma_{C,j}^x + \sigma_{B,j}^y \sigma_{C,j}^y \right) \; . 
\end{align}
On these couplings, we impose the requirements 
\begin{align}
\delta \ll J_x, J_z \ll \Omega, \Omega_S,
\label{eq:Size_Relations}
\end{align}
which implies that $H_{\text{pair}}$ dominates the dynamics and 
constrains the terms in $ H_{\text{chain}}$ to predominantly 
operate in subspaces that are degenerate under $H_{\text{pair}}$.
The essential role of $H_{chain}$ is to couple the pairs in a chain such that errors can propagate from one pair to the others and, eventually, to the shadow qubits, where the autonomous error correction takes place.
The shadow qubits are coupled
to the data qubits through a driving term which has the 
interaction-picture form~\cite{Poletto2012}
\begin{align}
\label{eq:Hdriv}
H_d =& A \left( \sigma_{C,2}^+ \sigma_{S_2}^+ \, \exp \left(\frac{i \left( 2 J_z-J_x \right) t}{\hbar}\right) + \text{h.c.} \right) \nonumber\\
&+ A \left( \sigma_{A,2}^+ \sigma_{S_1}^+ \, \exp \left(\frac{i \left( 2 J_z+J_x \right) t}{\hbar}\right) + \text{h.c.} \right),
\end{align}
where the driving strength $A$ is comparable to $\delta$, i.e. weak compared to 
$H_{\text{pair}}$. Along with the large dissipative loss in 
the shadow, this driving term gives a state-dependent effect
that is the key to the autonomous part of the QEC process. 
The full model used in the 
simulations below is given in App. \ref{system}. Note that while autonomous correction does not require the parameters of the model to change, tunability of the interactions is likely to be an advantage when performing state preparation, gates, and readout.

The autonomous 
part of the correction is schematically illustrated in Fig.~\ref{fig:Mega}C; in the event of a photon loss in 
one of the pairs, the state of the system is projected to an orthogonal subspace, and 
the resulting error state can travel along the chain. Once the error reaches the shadow at 
one of the ends, the coupling of shadow and data qubit is 
engineered to detect the presence of the error, prompting the driving to either correct the error completely or transform
it to a phase error that is handled by the active QEC.

The few-qubit model presented in Fig.~\ref{fig:Mega} consist 
of three blocks (A,B, and C), each with two 
qubits. We first define two-qubit subspaces 
for each pair of the form
\begin{align}
\left| \pm \right> &= \frac{1}{\sqrt{2}} \left( \left| \uparrow \uparrow \right> \pm \left| \downarrow \downarrow \right> \right).
\end{align}
These are degenerate eigenstates of $H_{\text{pair}}$ with 
eigenvalues $E_\pm=J_z$. If a photon loss error occurs in a single 
block, the resulting state will be a linear combination of 
\begin{align}
\left| T \right> &= \frac{1}{\sqrt{2}} \left( \left| \uparrow \downarrow \right> + \left| \downarrow \uparrow \right> \right) 
\\
\left| S \right> &= \frac{1}{\sqrt{2}} \left( \left| \uparrow \downarrow \right> - \left| \downarrow \uparrow \right> \right),
\end{align}
with eigenvalues $E_T=J_x-J_z$ and $E_S=-J_x-J_z$, 
respectively. Hence, the error projects our state into an
orthogonal subspace of error states, and this error is then free to move through the system.
The full encoding of the logical qubit uses a 
repetition code across the blocks as follows
\begin{align}
\alpha \left| \downarrow \right> &+ \beta \left| \uparrow \right> \nonumber \\
&  \mapsto \;  \alpha \left| \downarrow \right>_L + \beta \left| \uparrow \right>_L \nonumber\\
&= \alpha \left| \downarrow \right>\left| + \; + \; + \right>\left| \downarrow \right> + \beta \left| \downarrow \right>\left| - \; - \; - \right>\left| \downarrow \right>,
\end{align}
where the spin-down on either side represent the two 'shadow' qubits, 
$S_1$ and $S_2$ in Fig.~\ref{fig:Mega}. When there are no errors, the 
shadows will be in their ground (spin-down) state.

\subsection{Correction Processes}
To see how errors affect our 
system, assume that a decay error occurs in one of the two qubits
in block A, resulting in the state
\begin{align}
\alpha & \left| \downarrow \right> \left| \uparrow  \downarrow \; + \; + \right> \left| \downarrow \right> + \beta \left| \downarrow \right> \left| \uparrow \downarrow \; - \; - \right> \left| \downarrow \right> \nonumber \\
 =& \alpha \left( \left| \downarrow \right> \left| T \; + \; + \right>\left| \downarrow \right> + \left| \downarrow \right>\left| S \; + \; + \right>\left| \downarrow \right> \right) \\
&+ \beta \left( \left| \downarrow \right>\left| T \; - \; - \right>\left| \downarrow \right> + \left| \downarrow \right>\left| S \; - \; - \right> \left| \downarrow \right>\right) \nonumber.
\end{align}
The dynamics of the four different states that result from the 
error are similar, and the correction mechanism can 
be explained using
the state $\left| \downarrow \right> \left| T \; + \; + \right>\left| \downarrow \right>$ as an example. 
We first observe that this state has the same energy w.r.t. 
$H_{\text{pair}}$ as 
$\left| \downarrow \right> \left| + \; T \; + \right>\left| \downarrow \right>$ and $\left| \downarrow \right> \left| + \; + \; T \right>\left| \downarrow \right>$. This implies that 
the $XX$ exchange interactions (horizontal lines in Fig.~\ref{fig:Mega}A) will couple these 
states, and hence the $T$ error will propagate along the chain. 
We now engineer a parametric drive (i.e. the first term of Eq. \eqref{eq:Hdriv})  on the right-hand shadow qubit, $S_2$ in Fig.~\ref{fig:Mega}, such that a component of the form
\begin{align*}
\left| \downarrow \right> \left| + \; + \; T \right> \left| \downarrow \right>,
\end{align*} 
will be driven into the state
\begin{align}
\left| \downarrow \right> \left| + \; + \; \uparrow \uparrow \right> \left| \uparrow \right> =& \nonumber\\
 \left| \downarrow \right> \left| + \; + \; +  \right> &\left| \uparrow \right> + \left| \downarrow \right> \left| + \; + \; -  \right> \left| \uparrow \right> \; .
\end{align} 
Due to its large loss rate, the shadow will quickly decay and leave 
the system in the state
\begin{align}
\label{eq:final_eq}
\left| \downarrow \right> \left| + \; + \; +  \right> \left| \downarrow \right> + \left| \downarrow \right> \left| + \; + \; - \right> \left| \downarrow \right> \; .
\end{align}
The essential observation is that our scheme has converted a decay error into either a corrected state or a
phase error, {\it without} performing any measurements on the system.
Similarly, the $S$ error states are corrected on the left-hand side by the 
shadow $S_1$. An important advantage of our implementation is the symmetry of 
correction on the left- and right-hand shadows, which implies a symmetry of the 
autonomous operation on two components of the logical qubit, $\alpha$ and $\beta$, 
hence avoiding any unwanted and random phase among these components.
The full technical details of this mechanism are discussed in App. \ref{sec:Autonomous_EC} and \ref{sec:Alternative_Schemes}, 
including a discussion of how one can eliminate one of the shadows at the cost of 
a minor reduction in performance.

It is particularly illuminating to consider the 
autonomous correction in the framework of stabilizer codes~\cite{Gottesman1997}. It is straightforward
to see that the three stabilizers for each of the blocks are given by 
\begin{align*}
g_1 &= \left( Z \, Z \, I \, I \, I \, I \right)\\
g_2 &= \left( I \, I \, Z \, Z \, I \, I \right)\\
g_3 &= \left( I \, I \, I \, I \, Z \, Z \right),
\end{align*}
where the notation refers to operations on the six qubits (A1,A2,B1,B2,C1,C2).
One way to understand the action of the autonomous part of the code is that it automatically checks $g_1$, $g_2$ and $g_3$. If an error is detected, it is either corrected or turned into a phase error, as in \eqref{eq:final_eq}.

The final step to complete our hybrid quantum memory is the correction 
of phase errors through an active measurement strategy. Here we can use
standard weight-4 parity checks~\cite{Takita2017,royer2018}. As 
shown in App. \ref{sec:Active_EC}, one needs to measure the two stabilizers 
\begin{align*}
g_4 &= \left( X \, X \, X \, X \, I \, I  \right)\\
g_5 &= \left( I \, I \, X \, X \, X \, X  \right), 
\end{align*}
to detect and correct any of the phase errors, i.e. to perform corrections of the form
\begin{align}
\left. \begin{tabular}{l}
$\left| + \; + \; - \right> \,$\\
$\left| + \; - \; + \right> \,$\\
$\left| - \; + \; + \right> \,$
\end{tabular} \right\} \; \longrightarrow \; \left| + \; + \; + \right> \nonumber \\
\left. \begin{tabular}{l}
$\left| - \; - \; + \right> \,$\\
$\left| - \; + \; - \right> \,$\\
$\left| + \; - \; - \right> \,$
\end{tabular} \right\} \; \longrightarrow \; \left| - \; - \; - \right>.
\end{align}
There is one important caveat to applying active phase correction 
in this manner, namely that it may interfere with the 
passive error correction of decay errors (see App. \ref{sec:Active_EC} for 
details). To prevent this error channel, we introduce the stabilizer 
\begin{align*}
g_0 = \left( Z \right) \left( Z \, Z \, Z \, Z \, Z \, Z \right) \left( Z \right).
\end{align*}
A measurement of $g_0=-1$ indicates that a decay loss 
is being processed and implies that we should refrain from doing the 
active phase correction measurements. In Fig.~\ref{fig:Mega}C, 
we show a schematic of how the full scheme works.

\subsection{Performance of the Scheme}
In Fig.~\ref{fig:Mega}B we show the memory preserving quality of the hQEC code for two different logical states, assuming physical qubit decay on a timescale of $T_1=\SI{40}{\micro \second}$ and pure dephasing 
noise on a timescale of $T_{\phi}=\SI{80}{\micro \second}$, 
and assuming that we measure $g_0$ every $T_A=\SI{0.25}{\micro \second}$ (see below for discussions of these timescales).
As seen in the plot, the code significantly outperforms
the single-qubit lifetime, giving a ten-fold enhancement
of the lifetime for the superposition states 
$(|\uparrow\rangle_L+|\downarrow\rangle_L)/\sqrt{2}$, while 
giving almost a factor of 5 enhancement on the state 
$|\uparrow\rangle_L$.
App. \ref{sec:Performance} and \ref{sec:Main_Results} gives the details
of the calculation and the parameters, including a discussion 
of the initial reduction in the storage fidelity that is a 
generic feature of many quantum error correction schemes.
We note that these numbers do not 
require any fine-tuning of parameters beyond a 
few basic criteria on the Heisenberg interactions in 
the chain (see App. \ref{sec:Details_on_Main_Scheme}). The lifetime 
grows with increasing strength of the autonomous 
correction and with the single-qubit lifetime. 
For single-qubit lifetimes of $\SI{80}-\SI{100}{\micro \second}$~\cite{Wang2019,Burnett2019},
the hybrid scheme can potentially achieve lifetimes of 
around 1 millisecond. Additionally, the scheme is able to outperform single qubits 
even when the time between measurements is increased significantly. For instance, lifetimes of the state $|\uparrow\rangle_L$ of over $\SI{100}{\micro \second}$ are still achievable when $T_A$ is increased to  $\SI{750}{\micro \second}$, a timescale currently achievable using off-the-shelf hardware~\cite{Andersen2019,Riste2013}. Interestingly, this reduction in correction frequency actually yields improved
storage of the superposition state due to an interplay between the symmetries 
inherent in the scheme and the reduction of quantum-Zeno effects related to the measurements (See App. \ref{sec:Scalings_with_Ts} for details on the scaling of performance with $T_A$ and physical-qubit lifetimes).

\section{Discussion}
In conclusion, we have shown how active and passive 
error correction elements can be leveraged in 
hybrid QEC schemes for efficient quantum memories. 
As a concrete example, we studied a qubit model 
with six data qubits, but the working principle 
of combining passive correction of particular 
stabilizers with active correction of others 
is a generic one. As an illustration of this point, the appendices contains a 
small selection of different schemes adapted from the one presented in 
the main text. These show how hybrid schemes afford flexibility for 
trading complexity and performance---Simpler implementations with fewer 
driving tones or with only a single shadow qubit is possible, at the 
cost of slightly reduced performance. Conversely, improved performance 
can be achieved by suppressing second order processes, at the cost of a 
slight increase in the complexity of the interactions of the scheme.

An interesting question is how to expand on 
these quantum memory models in order to 
also perform a set of fault-tolerant gates on 
the systems (see App. \ref{sec:General_Properties_of_Code} for more 
details). We expect that the type of hybrid codes
presented above may be well-suited as a 
first layer of a concatenated code to provide 
enhanced performance over raw physical qubits. 
For instance, the residual errors in the concrete examples
presented above are dominated by logical bit-flips, 
thus these schemes would be well-suited for concatenation
with a simple bit-flip repetition-code~\cite{Kelly2015}.

While the physical 
parameters were adopted to superconducting 
circuits, our schemes requires tunable 
Heisenberg interactions and single-qubit
driving, both of which are standard tools of major 
quantum technology platforms. Additionally, great 
flexibility with respect to parameters exist, with 
little to no fine-tuning required. While large 
interaction strengths and relatively rapid 
active correction is preferable, the hQEC
scheme improve coherence for a wide 
range of parameters, and display very 
favourable scaling with respect to physical-qubit 
lifetimes.

It is our hope that the generic and flexible nature 
of the hybrid error-correction ideas presented here
will afford them great applicability, leading to new
and exciting avenues for incorporating autonomous 
correction principles into a wide range of qubit 
architectures

\section*{Acknowledgements} 
The authors would like to thank David Petrosyan for fruitful discussions. Additionally, L.B.K. would like to thank the Aspuru-Guzik group for their hospitality during his research visit, during which parts of the manuscript was prepared. This research was funded in part by the U.S. Army Research Office Grant No.  W911NF-17-S-0008 and performed in part at the Aspen Center for Physics, which is supported by National Science Foundation grant PHY-1607611. L.B.K. and N.T.Z acknowledge support from the Carlsberg Foundation and The Danish National Research Council under the Sapere Aude program.  M.K. acknowledges
support from the Carlsberg Foundation.

\appendix

\renewcommand{\figurename}{\textbf{Fig.}}
\renewcommand{\tablename}{\textbf{Table}}
\renewcommand{\thesubfigure}{\Alph{subfigure}}

\setcounter{equation}{0}
\setcounter{figure}{0}
\setcounter{table}{0}

\section{Detailed Analysis of the hQEC Scheme}
\label{sec:Details_on_Main_Scheme}
This section contains a more detailed description of the hQEC-scheme from the main text, along with a few mentions of (and references to) some of the results and alternative schemes presented in other parts of the supplementary material.\\

\subsection{The System}
\label{system}
The system we will be considering is depicted in Fig. \ref{fig:Mega2}. It consists of a total of 8 spin-1/2 systems, which we will refer to as qubits in keeping with quantum information tradition. 6 of these qubits constitute our data qubits. It is this part of the system that will store the quantum information that we wish to protect from the effects of errors. This protection will be partially supplied by the remaining two qubits, referred to as shadow qubits. It is these shadow qubits that we will later subject to engineered dissipation. We will denote the two states of each qubit as $\left| \downarrow \right>$ and $\left| \uparrow \right>$ respectively. Inspired by superconducting qubits, we will think of the $\left| \downarrow \right>$ as the ground state of each qubit and $\left| \uparrow \right>$ as an excited state with significantly higher energy. Additionally, we will imagine that our system is weakly coupled to a surrounding environment with low thermal energy compared to the energy difference between the two levels. As a result, an excited qubit is likely to decay by emitting energy to the surroundings, for example in the form of a photon, but the low temperature of the surroundings will make the inverse process of photon capture unlikely. Thus we will consider our system as subjected to only two types of noise: Photon losses flipping the spin from "up" to "down", and dephasing errors disturbing the relative sign of these two wave function components. It is these errors we will now aim to detect and correct. To more easily facilitate the discussion of how this works, let us represent the state of a given qubit as a vector using the association
\begin{align*}
\left| \uparrow \right> &= \begin{pmatrix}
1 \\
0
\end{pmatrix} & 
\left| \uparrow \right> &= \begin{pmatrix}
0 \\
1
\end{pmatrix} \; . 
\end{align*}
This allows us to represent any linear operator on the state space of the qubit as a 2 by 2 matrix, which in turn can be written in terms of the 3 Pauli matrices $\sigma^x, \sigma^y, \sigma^z$ and the identity $I$. In order to specify what qubit a given operator acts on we will use subscripts, with the data qubits described by both a column index $i \in \{ A, B, C\}$ and row index $j \in \{ 1,2\}$, and the shadow qubits referred to simply by the subscripts $S_1$ and $S_2$ (see Fig. \ref{fig:Mega2}). With these conventions sorted, we can now approach the problem of how to use engineered dynamics and dissipation to protect information stored in the data qubits. Consider therefore a Hamiltonian of the form
\begin{align*}
H= H_0 +  H_{\text{pair}} + H_{\text{chain}} \; ,
\end{align*}
where
\begin{align}
H_0 =& \hspace{0.5cm} \sum_{\mathclap{\substack{i \in \{A,B,C\}\\ j \in \{1,2\}}}} \;  \frac{1}{2} \Omega \, \sigma^z_{i,j} + \frac{1}{2}\Omega_S \, \sigma_{S_1}^z + \frac{1}{2}\Omega_S \, \sigma_{S_2}^z \;, \nonumber\\
H_{\text{pair}} =& \hspace{0.5cm} \sum_{\mathclap{i \in \{A,B,C\}}} \left[ J_x \left( \sigma_{i,1}^x  \sigma_{i,2}^x + \sigma_{i,1}^y \, \sigma_{i,2}^y \right) + J_z \, \sigma_{i,1}^z  \sigma_{i,2}^z \right]  \; , \nonumber\\
H_{\text{chain}} =& \, \delta \sum_{j=1}^2  \left( \sigma_{A,j}^x \sigma_{B,j}^x + \sigma_{A,j}^y \sigma_{B,j}^y \right) \label{eq:Hamiltonian}\\
&+ \delta \sum_{j=1}^2  \left( \sigma_{B,j}^x \sigma_{C,j}^x + \sigma_{B,j}^y \sigma_{C,j}^y \right) \; . \nonumber
\end{align}
The first of these terms, $H_0$, is simply the energy difference between the two states mentioned above. It is included here mostly for conceptual consistency, since it commutes with the other contributions and plays little role in most of the subsequent arguments, although it will come in handy once driving is introduced to the system. The second term constitutes a Heisenberg XXZ-coupling between the two qubits in a given column. We will assume that the strengths of this coupling is smaller than the single qubit energies $\Omega$, $\Omega_S$, but that it is much larger than the interaction strength of the third term, i.e. that
\begin{align}
\delta \ll J_x, J_z \ll \Omega, \Omega_S \; .
\label{eq:Order_of_Magnitude}
\end{align}
This will allow the Heisenberg interaction to act as a backbone for the dynamics of the system, constraining other, weaker terms like $H_{\text{chain}}$ to predominantly induce dynamics within the subspaces degenerate with respect to the large $H_{\text{pair}}$ (as can be seen from rotating wave arguments). Finally, $H_{\text{chain}}$ constitutes an interaction between neighbouring qubits within a row. It is this interaction that will allow errors to move along the rows of the system, and which will therefore allow us to extract and correct errors in the central $B$-column of data qubits without having to couple a separate shadow qubit to this pair.\\
At this point, it is worth noting that this Hamiltonian is relatively simple in form. It contains only 2-spin interactions among nearest neighbours, all expressible using products of two Pauli matrices, and with each interaction conserving the total $z$-component of the spin--or equivalently, the total single qubit energy as defined by $H_0$. Hopefully, this relative simplicity of the model increases its potential for experimental implementation, e.g. in the superconducting circuits that inspired its construction.\\
As a first step towards understanding how the dynamics of this simple model helps mitigate errors, let us start by getting rid of  $H_0$ by changing to the interaction picture related to this Hamiltonian. This yields an interaction-picture Hamiltonian consisting of only the two interactions
\begin{align*}
H_I = e^{i \frac{H_0 t}{\hbar}} (H-H_0) e^{-i \frac{H_0 t}{\hbar}} =  H_{\text{pair}} + H_{\text{chain}} \; .
\end{align*}
Since the first of these is assumed significantly stronger than the second, the natural states for understanding the dynamics must be the eigenstates of $H_{\text{pair}}$. We therefore define the following notation for the state of two qubits within a column:
\begin{align}
\left| \pm \right> &= \frac{1}{\sqrt{2}} \left( \left| \uparrow \uparrow \right> \pm \left| \downarrow \downarrow \right> \right) \nonumber\\
\left| T \right> &= \frac{1}{\sqrt{2}} \left( \left| \uparrow \downarrow \right> + \left| \downarrow \uparrow \right> \right) \label{eq:State_Defs}\\
\left| S \right> &= \frac{1}{\sqrt{2}} \left( \left| \uparrow \downarrow \right> - \left| \downarrow \uparrow \right> \right) \nonumber\; .
\end{align}
Each of these is an eigenstate of the Heisenberg XXZ-interaction, with eigenvalues $E_{\pm}=J_z$, $E_T= J_x-J_z$ and $E_S=-J_x-J_z$ respectively. We will specify the state of the system as a whole by combining three of these symbols to specify the state of the data qubits and two arrows to specify the states of the shadow qubits, i.e. a state with the first qubit pair in the '$+$' state, the second in the '$S$' state, the third in the '$-$' state and both the left and right shadow qubits in the '$\downarrow$' state will be written as $\left| \downarrow \right> \left| + \; S \; - \right> \left| \downarrow \right>$.\\
The final step in setting up the system is adding the engineered dissipation through the shadow qubits at the edge of the system. Consider therefore these qubits to be constructed in such a way that they are very dissipative, i.e. in such a way that they are very prone to the photon-loss error introduced above. Additionally, assume that we add to the interaction-picture Hamiltonian a driven interaction between shadow qubits and data qubits of the form 
\begin{align}
H_d =& A \left( \sigma_{C,2}^+ \sigma_{S_2}^+ \, \exp \left(\frac{i \left( 2 J_z-J_x \right) t}{\hbar}\right) + \text{h.c.} \right) \nonumber\\
&+ A \left( \sigma_{A,2}^+ \sigma_{S_1}^+ \, \exp \left(\frac{i \left( 2 J_z+J_x \right) t}{\hbar}\right) + \text{h.c.} \right) \; ,
\label{eq:Driving}
\end{align}
where "h.c." is the hermitian conjugate and $\sigma^{\pm} = \frac{1}{2} \left( \sigma^x \pm i \sigma^y \right)$ are the ladder-operators that increase ($+$) or decrease ($-$) the z-component of the corresponding spins. The strength $A$ of this interaction is chosen so that it is on the same order as $\delta$, i.e. weak compared to our backbone Hamiltonian. As a result, a rotating wave argument shows that the driving is only capable of coupling states where the the difference in energy related to the Heisenberg interaction matches the frequency of the driving. As a result, the first driving term is only able to induce oscillations of the form
\begin{align*}
& \dots \, \left| T \right> \left| \downarrow \right> \\
& \hspace{1cm} \updownarrow \\
& \dots \, \left| \uparrow  \; \uparrow \right> \left| \uparrow \right> \; ,
\end{align*}
manipulating the state of the last qubit column and the right shadow qubit. Similarly, the second term only operates on the first data qubit column, inducing transitions of the form
\begin{align*}
&\left| \downarrow \right> \left| S \right> \dots \\
& \hspace{0.8cm} \updownarrow \\
&\left| \downarrow \right> \left| \uparrow  \; \uparrow \right> \dots \; .
\end{align*}
It is worth noting that an effective driving of the form \eqref{eq:Driving} can be achieved using relatively simple sinusoidal driving. The first term can be achieved using the Schrödinger-picture driving terms of the form
\begin{align}
2 A \cos\left( \frac{ \Omega +  \Omega_S + 2J_z-J_x }{\hbar} \, t \right) \sigma_{C,2}^x \sigma_{S_2}^x \; ,
\label{eq:Driving1}
\end{align}
since the large energy scale of $\Omega$ and $\Omega_S$ means that rotating wave-like arguments will allow all other contributions than the ones in Eq. \eqref{eq:Driving} to be neglected once the shift to the rotating frame has been performed. Similarly, sinusoidal driving of the form
\begin{align}
2 A \cos\left( \frac{ \Omega +  \Omega_S + 2J_z+J_x}{\hbar} \, t \right) \sigma_{A,2}^x \sigma_{S_1}^x
\label{eq:Driving2}
\end{align}
achieves the second driving term.

\subsection{Error Correction of Photon Loss}
\label{sec:Autonomous_EC}
Having introduced the system and the notation used to describe it, we can now illustrate the error correcting properties of this system. We will encode one qubit worth of information (called a logical qubits) into the state of the data qubits  as follows:
\begin{align}
\alpha \left| \downarrow \right> &+ \beta \left| \uparrow \right> \nonumber \\
&  \mapsto \;  \alpha \left| \downarrow \right>_L + \beta \left| \uparrow \right>_L \nonumber\\
&= \alpha \left| \downarrow \right>\left| + \; + \; + \right>\left| \downarrow \right> + \beta \left| \downarrow \right>\left| - \; - \; - \right>\left| \downarrow \right> \; .
\label{eq:Encoding}
\end{align}
Both of these two components are eigenstates of the backbone Heisenberg-XXZ part of the Hamiltonian, and so the encoded information will be left alone by this interaction. However, they are strictly speaking not eigenstates of $H_{\text{chain}}$ and $H_d$. Let us due to the symmetry of the situation focus on the state $ \left| + \; + \; + \right>$, omitting the state of the shadow qubits for brevity. This state is coupled by $H_{\text{chain}}$ to the states
\begin{align}
\left| S \; S \; + \right> & & &\left| T \; T \; + \right> \nonumber \\
\left| + \; S \; S \right> & & &\left| + \; T \; T \right> \; .
\label{eq:Mediator_States} 
\end{align}
Luckily, the backbone Hamiltonian ensures that these states will be detuned by an energy $\Delta E = 4 J_z \pm 2 J_x$. As long as both of these energies are much larger than the scale $\delta$ of $H_{\text{chain}}$, transitions to these states due to this contribution to the Hamiltonian will be heavily suppressed. In other words, our backbone Hamiltonian is performing its duty and constraining $H_{\text{chain}}$ from interfering with the storage of information in the absence of errors. A small complication to this reasoning is that the states of \eqref{eq:Mediator_States} in turn couple to states such as $\left| + \; - \; - \right>$ that have the same energy as the original state. As a result, transitions to these states will sometimes occur due to second order processes occurring through a virtual excitation of the states in \eqref{eq:Mediator_States}. However, the rate of such transitions scale with $\left( \delta / \Delta E \right)^2$, so by keeping $\delta$ sufficiently small compared to the scales of the backbone Hamiltonian these transitions can be made to only occur on timescales of about $\SI{100}{\micro \second}$, which allows them to be partially suppressed via. quantum-Zeno effects related to the active error correction that will be introduced in App. \ref{sec:Active_EC} below. Additionally, it is possible to further suppress these second order effects through further engineering of the interactions of the model---see App. \ref{sec:Second_Order_Effects}. We may therefore conclude that $H_{\text{chain}}$ will leave our stored information alone. A similar conclusion can easily be drawn about $H_d$---Simply put, the driving frequencies of this term do not match any of the possible transitions that $H_d$ could induce starting from the states in Eq. \eqref{eq:Encoding}. We thus conclude that information can be safely stored in our system, at least as long as it is not disturbed by noise or decays.

Assume, however, that an error does occur by way of a decay in the second qubit of the first data qubit column. This results in our system picking up a component of the form:
\begin{align}
\label{eq:Error_States}
\alpha & \left| \downarrow \right> \left| \uparrow  \downarrow \; + \; + \right> \left| \downarrow \right> + \beta \left| \downarrow \right> \left| \uparrow \downarrow \; - \; - \right> \left| \downarrow \right> \nonumber \\
 =& \alpha \left( \left| \downarrow \right> \left| T \; + \; + \right>\left| \downarrow \right> + \left| \downarrow \right>\left| S \; + \; + \right>\left| \downarrow \right> \right) \\
&+ \beta \left( \left| \downarrow \right>\left| T \; - \; - \right>\left| \downarrow \right> + \left| \downarrow \right>\left| S \; - \; - \right> \left| \downarrow \right>\right) \nonumber 
\end{align}
due to the loss of one of the excitations from the $\left| \uparrow \uparrow \right>$-component previously stored in the first column of qubits. The dynamics of the system will process each of the four states above in a very symmetrical manner, so let us for definiteness consider what happens to the state $\left| \downarrow \right> \left| T \; + \; + \right>\left| \downarrow \right>$. The first thing to note is that this state has the same energy with respect to the Heisenberg backbone as the states $\left| \downarrow \right> \left| + \; T \; + \right>\left| \downarrow \right>$ and $\left| \downarrow \right> \left| + \; + \; T \right>\left| \downarrow \right>$. As a result, nothing prevents $H_{\text{chain}}$ from coupling these states. This essentially allows the error, represented by the "$T$" in the notation of the state, to move along the chain of qubit pairs from left to right. At some point, this motion will result in a component of the form
\begin{align*}
\left| \downarrow \right> \left| + \; + \; T \right> \left| \downarrow \right> ; ,
\end{align*} 
at which point our driven interaction will take over, inducing a transition to
\begin{align}
\left| \downarrow \right> \left| + \; + \; \uparrow \uparrow \right> \left| \uparrow \right> =& \nonumber\\
 \left| \downarrow \right> \left| + \; + \; +  \right> &\left| \uparrow \right> + \left| \downarrow \right> \left| + \; + \; -  \right> \left| \uparrow \right> \; .
\label{eq:EC_State}
\end{align} 
In reality, the oscillations among this state and the three states containing a "$T$" can be rather more complicated than indicated by this explanation, but the central point remains: as time passes, components of the form \eqref{eq:EC_State} is picked up due to oscillations induced by $H_{\text{chain}}$ and $H_d$. Although this new state looks a lot like the components present before the error, these oscillations would be of little help to us without the engineered dissipation, since we would in time simply oscillate back into the original error states again. However, by making the shadow qubit lossy, the decay of this object can be used as a valve in the Hilbert space. Every time some probability accumulates for finding the system in the state \eqref{eq:EC_State}, there will be a chance of a decay, and hence within a few oscillations we can be almost certain that a decay has occurred and that we are therefore now in the state
\begin{align*}
\left| \downarrow \right> \left| + \; + \; +  \right> \left| \downarrow \right> + \left| \downarrow \right> \left| + \; + \; - \right> \left| \downarrow \right> \; .
\end{align*}
Both of these states are left alone by the Hamiltonian as explained above, and hence our journey of autonomous error correction ends here. The process is depicted schematically in Fig. \ref{fig:Simple_Corection} and \ref{fig:Simple_CorectionE}. We see that the process has terminated in a superposition of two states. The first of these is exactly the $\left| + \; + \; + \right>$-state that gave rise to the $\left| T \; + \; + \right>$ when the error occurred. In other words, this component corresponds to a corrected error. As for the second component, we note that it is not the corrected $\left| + \; + \; + \right>$-state that we want, but rather this state with a $\sigma^z$ (i.e. a phase) applied to one of the qubits in the first pair. The second term therefore still contains an error, but the type of error has been converted from a photon loss to a phase error. We will deal with such phase errors separately using active error correction below. Thus, as long as such an active error correction step can be applied before any further errors occur, this phase error will also be extracted from the second component. The end result is that the error has been corrected in both components, and thus the erroneous $\left| T \; + \; + \right>$ has been removed and the entire population of it transferred back to the correct $\left| + \; + \; + \right>$-state.\\
The error correction of the other 3 components of Eq. \eqref{eq:Error_States} proceed in a similar manner, with the only difference being that it is the left-hand shadow qubit $S_1$ that corrects the states containing an "$S$", rather than the right-hand shadow qubit $S_2$. It is important to note that although this gives asymmetry between the correction of $S$- and $T$-errors, there is still rigid symmetry in the way in which the component related to the amplitude $\alpha$ and the component related to the amplitude $\beta$  corrected. This symmetry is essential, since the dynamics of the error states will tend to result in the accumulation of phases that will depend on the exact nature of the oscillations and how long they are allowed to run before a decay in the shadow qubit occurs. Had the error components related to $\alpha$ and $\beta$ components been treated asymmetrically, for instance because they were corrected by different shadow qubits, they would therefore tend to pick up non-identical random phases, resulting in an effective dephasing of the logical qubit:
\begin{align*}
\alpha \left| 0 \right>_L + \beta \left| 1 \right>_L \; \longrightarrow \; \alpha e^{i \phi_1} \left| 0 \right>_L + \beta e^{i \phi_2}  \left| 1 \right>_L \; .
\end{align*}
Only because of the degree of symmetry in the correction can we be sure that $\phi_1=\phi_2$, and thus that the phases become a meaningless global phase.

The correction of errors in other qubit pairs proceeds through the same type of oscillations. Indeed, the error states that would appear in these decays are exactly the ones already participating in the oscillations of Fig. \ref{fig:Simple_Corection} and \ref{fig:Simple_CorectionE}. One thing to note is that it is important to engineer the way that the states couple in such a way that there is nowhere within the error-state subspace (i.e. the subspace spanned by the states in Eq. \eqref{eq:Error_States}) that does not experience oscillations to the state with an excitation in the shadow qubit. If such a subspace were to exist, it would not be corrected, resulting in the slow accumulation of these erroneous components, resulting in a breakdown of the error correction. In this case, such a subspace cannot exist simply due to the topology of how the states couple. It is for this reason we require $J_x$ non-zero, since this makes the energies of $S$ and $T$-states become different, thereby allowing us to address these two states individually. One can actually circumvent this requirement by by adjusting the geometry of the setup. In this way, only a single driving frequency $ \omega= \left(  \Omega +  \Omega_S + 2J_z-J_x \right) / \hbar$ is required, at little to no expense of code performance. Additional details on this scheme can be found in App. \ref{sec:Alternative_Driving}.

An obvious and somewhat related related question is whether we really need two separate shadow qubits to perform correction similar to the one above. It turns out that this is not strictly speaking necessary, but forcing a single shadow qubit to pull double duty tends to result in poorer overall performance of the autonomous correction. Additional details can be found in App. \ref{sec:Single_Shadow}.

\subsection{Error Correction of Phase Errors}
\label{sec:Active_EC}
As mentioned above, we need to be able to handle phase errors, both for their own sake and to handle the phase errors occurring as a result of the photon-loss correction. We envision this to be done using standard active error correction methods wherein syndrome measurements are performed and the results of the measurements are used to determine what error correction steps are needed. The desired operation is the correction
\begin{align}
\left. \begin{tabular}{l}
$\left| + \; + \; - \right> \,$\\
$\left| + \; - \; + \right> \,$\\
$\left| - \; + \; + \right> \,$
\end{tabular} \right\} \; \longrightarrow \; \left| + \; + \; + \right> \nonumber \\
\left. \begin{tabular}{l}
$\left| - \; - \; + \right> \,$\\
$\left| - \; + \; - \right> \,$\\
$\left| + \; - \; - \right> \,$
\end{tabular} \right\} \; \longrightarrow \; \left| - \; - \; - \right> 
\label{eq:Active}
\end{align}
that is, a simple majority rule among the signs is to be performed. This can be done by measuring the so-called stabilizers
\begin{align*}
g_4 &= \sigma_{A,1}^x \sigma_{A,2}^x \sigma_{B,1}^x \sigma_{B,2}^x \\
g_5 &= \sigma_{B,1}^x \sigma_{B,2}^x \sigma_{C,1}^x \sigma_{C,2}^x \; .
\end{align*}
These can be written more succinctly by numbering the data qubits using their $(i,j)$-indices as $n=2i+j$ and switching to a shorter but equally conventional notation:
\begin{align*}
g_4 &= \left( X \, X \, X \, X \, I \, I  \right)\\
g_5 &= \left( I \, I \, X \, X \, X \, X  \right) \; ,
\end{align*}
wherein the $n$'th entry in the vector represents the operator acting on the $n$'th qubit, and the sigmas have been removed for legibility. Measuring these to properties and performing the operation
\begin{align}
\label{eq:Active_EC_Alg}
g_4 = +1, \, g_5 = +1 \; &\longrightarrow  \; \left( I \, I \, I \, I \, I \, I  \right) \nonumber \\
g_4 = -1, \, g_5 = +1 \; &\longrightarrow  \; \left( Z \, I \, I \, I \, I \, I  \right)  \\
g_4 = +1, \, g_5 = -1 \; &\longrightarrow  \; \left( I \, I \, I \, I \, Z \, I  \right) \nonumber \\
g_4 = -1, \, g_5 = -1 \; &\longrightarrow  \; \left( I \, I \, Z \, I \, I \, I  \right) \nonumber
\end{align}
performs the desired correcting operation.

While this simple procedure works as advertised on the states of \eqref{eq:Active}, it does not play well with the autonomous part of the error correction. To see why this is the case, consider again the situation where an error has put us into the state from Eq. \eqref{eq:Error_States}. In fact, let us now consider a more general case where the error could have occurred on either of the two qubits in the first column. In this case, the state would read
\begin{align*}
 \left| \psi_{\pm} \right> =& \, \alpha \left(  \left| T \; + \; + \right> \pm \left| S \; + \; + \right> \right)\\
 &+ \beta \left( \left| T \; - \; - \right> \pm \left| S \; - \; - \right> \right)
\end{align*}
with the sign determined by which qubit experienced the decay. Since the active error correction would not be able to distinguish a $\left| T \right>$ from a $\left| + \right>$ and an $\left| S \right>$ from a $\left| - \right>$, running an active error correction step on this state would result in the state
\begin{align*}
\left| \psi_{\pm} \right> \; &\overset{\text{measure}}{\longrightarrow} \; \left\{ \begin{tabular}{ll}
$\alpha \left| T \; + \; + \right> \pm \beta \left| S \; - \; - \right>$ & $\; g_1=+1$\\
$\alpha \left| S \; + \; + \right> \pm \beta \left| T \; - \; - \right>$ & $\; g_1=-1$
\end{tabular} \right.\\
&\; \overset{\text{correct}}{\longrightarrow} \; \alpha  \left| T \; + \; + \right> \pm \beta \left| S \; - \; - \right> \; .
\end{align*}
This final state explicitly breaks the symmetry of the passive error correction of the $\alpha$ and $\beta$ components--one now proceeds through the correction of $S$ by the left shadow qubit and the other proceeds through the correction of the $T$ error by the right shadow qubit. As a result, different phases will be picked up, and logical dephasing will occur. Even more damningly, the state may actually already have experienced dephasing in the form of the sign that may have appeared between the two components---a distortion of the encoded information that in fact occurred as soon as $g_1$ was measured. In other words, active correction on a system that has experienced photon loss will by itself cause dephasing, and will set up the autonomous dynamics to cause even more.

Since the problem arise already at the measurement phase, solving this problem must require that we change the detection scheme. By adding a measurement of both data and shadow qubits 
\begin{align*}
g_0 = \left( Z \right) \left( Z \, Z \, Z \, Z \, Z \, Z \right) \left( Z \right)
\end{align*}
as a first step to our active error correction procedure, we are able to detect if the passive error correction is currently in effect ($-1$) or not ($+1$). Aborting the active error correction step if $-1$ is measured therefore means we leave the passive correction alone, thus avoiding any unwanted interplay between the active and passive error correction. As an added benefit, avoiding measurements on the states when the passive error correction is running also allows us to avoid quantum-Zeno effects that would result from repeatedly observing the state of our system, and which could freeze the oscillatory dynamics that are so essential to the passive error correction.

\subsection{Code Performance}
\label{sec:Performance}
\begin{figure*}[hbtp]
  \centering
\begin{subfigure}[b]{1.0\textwidth}
   \includegraphics[scale=0.95]{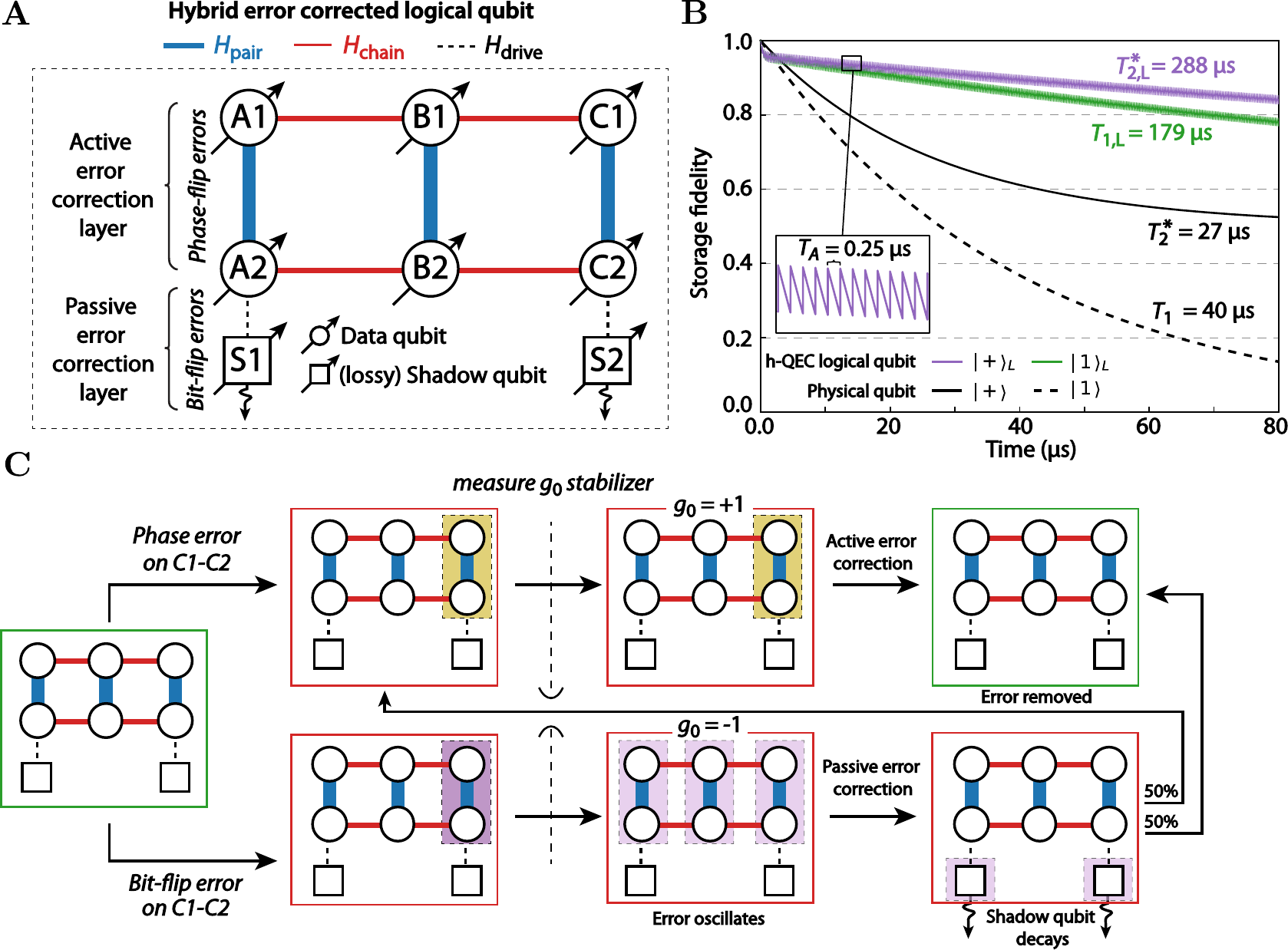}
  \captionlistentry{} \label{fig:Mega2} 
  \captionlistentry{} \label{fig:Mega2B}
  \captionlistentry{} \label{fig:Mega2C} 
\end{subfigure}\\ \vspace{0.4cm}
\begin{subfigure}[b]{1.0\textwidth}
\includegraphics[scale=0.23]{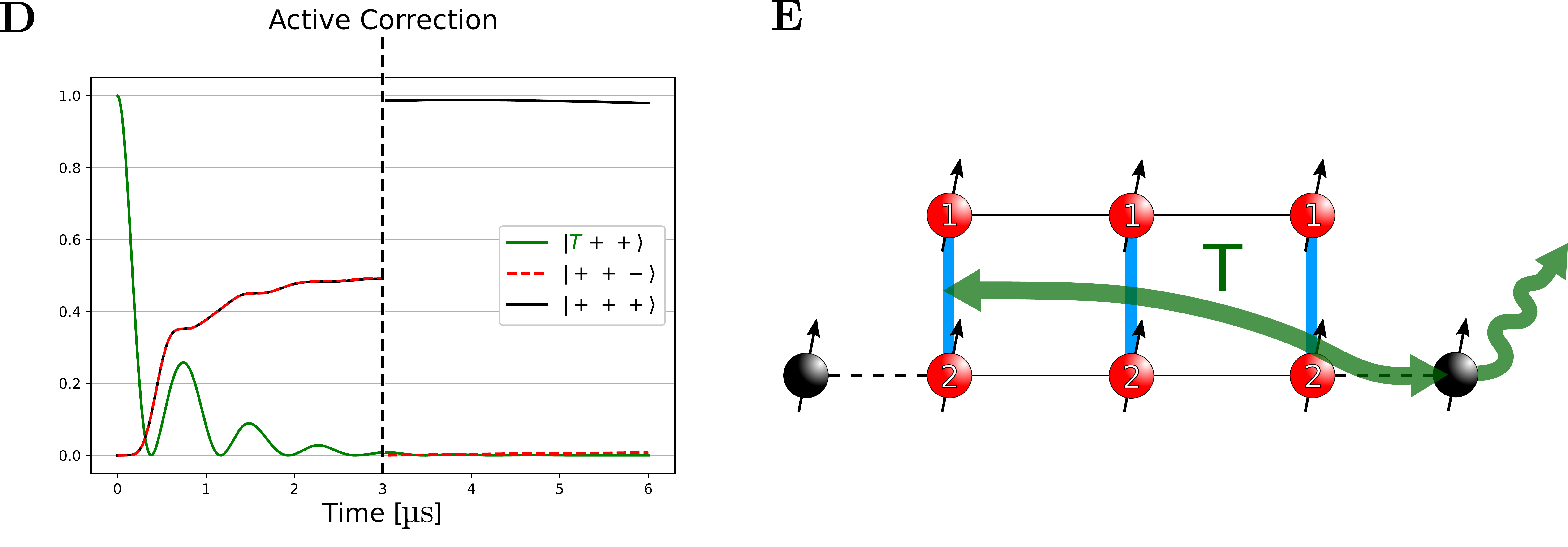}
\captionlistentry{}\label{fig:Simple_Corection}
\captionlistentry{}\label{fig:Simple_CorectionE}
\end{subfigure}\\ \vspace{0.4cm}
\end{figure*}

Having looked at both the passive and active correction, as well as the interplay between them, we now have all of the components needed for the proposed error correction scheme. The result of these processes in operation can be seen on Fig. \ref{fig:Mega2B}. In this figure, an initial state $\left| \psi_0 \right>$ was encoded into both a single qubit and two copies of our 6-qubit code, one with and one without the error correction in operation. Using the python package Qutip~\cite{Johansson2012}, system dynamics were then simulated, including the presence of photon-losses on a timescale of $T_1=\SI{40}{\micro \second}$ and dephasing noise on a timescale of $T_{\phi}=\SI{80}{\micro \second}$. In the figure is depicted how well each system retained the encoded information, quantified as the storage fidelity
\begin{align}
F_{\text{storage}}\left( t \right) = \text{Tr} \left( \rho \left( t \right) \left| \psi_0 \right> \left< \psi_0 \right| \right) \; ,
\label{eq:Fstorage}
\end{align}
where $\rho(t)$ is the density matrix representing the state of the system at time $t$. In other words, storage performance is quantified simply as the likelihood of observing the system to be in the original state $\left| \psi_0 \right>$ after time has passed and errors have attempted to corrupt the encoded information. To make sure that our code is operating as a fully fledged quantum memory, the initial state was chosen as the superposition state 
\begin{align}
\left| T \right> &= \frac{1}{\sqrt{2}} \left( \left| \uparrow \right> + \left| \downarrow \right> \right) \; .
\label{eq:Benchmark_1}
\end{align}
However, the symmetric way the dynamics treat the two states $\left| \uparrow \right>_L$ and $\left| \downarrow \right>_L$ used for encoding indicates that the noise experienced by the encoded logical qubit is likely to be similarly symmetric, for instance by manifesting as logical $\sigma_L^x$-like noise. This noise does not affect \eqref{eq:Benchmark_1}, thus to detect it we run the simulations also for the state $\left| \uparrow \right>$.

As can be seen from Fig \ref{fig:Mega2B}, the code  manages to outperform the single qubit despite having 6 times as many qubits and hence 6 times as many errors to deal with. Indeed, with the parameters used in this figure (see Fig. \ref{tbl:Parameters}), the code manages to increase the lifetime related to the exponential decay of the storage fidelity for the state $\frac{1}{\sqrt{2}} \left( \left| 0 \right> + \left| 1 \right> \right)$ from $T_2^*=\SI{27}{\micro \second}$ for the single qubit to $T_2^*\simeq \SI{288}{\micro \second}$ for the six-qubit code. As expected, the $\left| \uparrow \right>$-state is not protected quite as well, but the code still manages an increase in coherence time from $T_1=\SI{40}{\micro \second}$ to $T_1=\SI{179}{\micro \second}$. Note that for the determination of these values, we have neglected the initial decay of storage fidelity during the first few microseconds. This decay is very common in error correcting schemes, and is the result of the system having to have a small but non-zero probability of containing an error before the error correction can start operating. In other words, the early dynamics will consist of the accumulation of population in the error states, and only once this has happened for a little while will the error correction start operating, leading to the pseudo-equilibrium between of errors and correction that characterizes the later behaviour. The initial decrease in fidelity, measured as the intersection of the exponential fit with the $t=0$ axis, is about 4.2\%.

The performance of the code is of course dependent on the choice of parameters used for the simulation, with increases in the size of the constants of the Hamiltonian and the loss-rate of the shadow qubits leading to better passive correction, and active correction with a higher frequency generally leading to stronger active protection (see \ref{sec:TA_Dependency} for details). Note, however, that no fine-tuning of parameters beyond the criteria of Eq. \eqref{eq:Order_of_Magnitude} and an ability to accurately adjust the driving frequency is required for the scheme to operate. The parameters used in the simulation in Fig. \ref{fig:Mega2B} is reproduced in the table on Fig. \ref{tbl:Parameters}, and the performance of the code with other choices is depicted in Fig. \ref{fig:Parameter_Scaling_0} and \ref{fig:Parameter_Scaling}. Addtional information on the parameters used in this figure can be found in App. \ref{sec:Params}.

\subsection{General Properties of Proposed Code}
\label{sec:General_Properties_of_Code}
As suggested by the stabilizer formalism appearing in section \ref{sec:Active_EC}, the error correcting code in Eq. \eqref{eq:Encoding} is in fact at its heart a simple stabilizer code. The full set of stabilizers of the code are
\begin{align*}
g_1 &= \left( Z \, Z \, I \, I \, I \, I \right)\\
g_2 &= \left( I \, I \, Z \, Z \, I \, I \right)\\
g_3 &= \left( I \, I \, I \, I \, Z \, Z \right)\\
g_4 &= \left( X \, X \, X \, X \, I \, I  \right)\\
g_5 &= \left( I \, I \, X \, X \, X \, X  \right) \; ,
\end{align*}
with the three stabilizers $g_1$, $g_2$ and $g_3$ enforced by the autonomous error correction and only the enforcement of the two final stabilizers (and $g_0$) requiring measurements and external pulses. A central question when dealing with such an encoding is how easy or difficult it is to perform operations on the stored data. One particularly interesting class of operations are transverse operations, defined as the operations that can be performed by independently applying operations to each of the six qubits in the code, or in the case of two-qubit gates: by independently applying operations between each pair of corresponding qubits of two 6-qubit blocks. It can be shown that any stabilizer code allows for a transverse implementation of logical $X$ and $Z$ operators if the right basis is chosen for the encoding~\cite{Gottesman2010}, and this is indeed the case here:
\begin{align*}
Z_L &= \left( X \, X \, I \, I \, I \, I \right)\\
X_L &= \left( Z \, I \, Z \, I \, Z \, I \right)\; .
\end{align*}
Additionally, the two-qubit CNOT~\cite{Nielsen2002} gate may be transversally performed between two logical qubits of our encoding by performing CNOTs between each pair of corresponding qubits in the two data-qubit blocks, although the target and control roles will be reversed for the logical qubits compared to the operations on the data qubits and an additional $Z_{L}$ is needed on the control qubit: 
\begin{align*}
\text{CNOT}_L &(\text{control}, \text{target}) \\
&= Z_{L,\text{control}} \prod_{i=1}^6 \text{CNOT}(\text{target}_i, \text{control}_i) \; .
\end{align*}
One thing to keep in mind is that the transversality of the gates above does not imply that they are fault tolerant, i.e. that they operate correctly in the presence of errors\cite{Gottesman2010}. The reason that this rather common implication does not hold in this case is that it only applies to codes capable of correcting arbitrary single qubit errors, and the code presented here does not have this property. Indeed, for the code to be able to correct arbitrary single-qubit errors, it would need to be able to correct $\sigma^x$-errors on both the first and second data qubits. However, the effect of such an error on the first qubit is identical to performing a $Z_L$ and then performing the error on the second qubit. Since the code has no way of knowing whether a $Z_L$ was done by the experimenter or an error---indeed it is not even allowed to detect the state of the logical qubit at all lest it destroy superposition states---it cannot possibly hope to correct both of these errors, because correctly countering one of them would necessarily lead to the other one experiencing a spurious $Z_L$ operation. The result of this is that only the $X_L$-operator is actually fault tolerant, and only with respect to photon-loss and dephasing errors. Thus more involved methods would likely be necessary in order to develop a fault tolerant gate set. Due to the fact that fault tolerance is a property related to concatenation of codes~\cite{Gottesman2010} and the fact that concatenation of autonomous code would require couplings of rapidly increasing complexity in order to couple logical qubits in a manner similar to Eq. \eqref{eq:Hamiltonian}, one interpretation of this is that the scheme presented here would be most useful as the first layer of a series of concatenated codes. In this way, the layers of code build on top of our scheme could ensure fault tolerance while benefiting from the improved coherence times of our scheme compared to raw physical qubits.

\setcounter{figure}{0}
\begin{figure*}[ht!]
  \centering
  \begin{subfigure}[b]{0.45\textwidth}
    \setcounter{subfigure}{5}
   \includegraphics[scale=0.17]{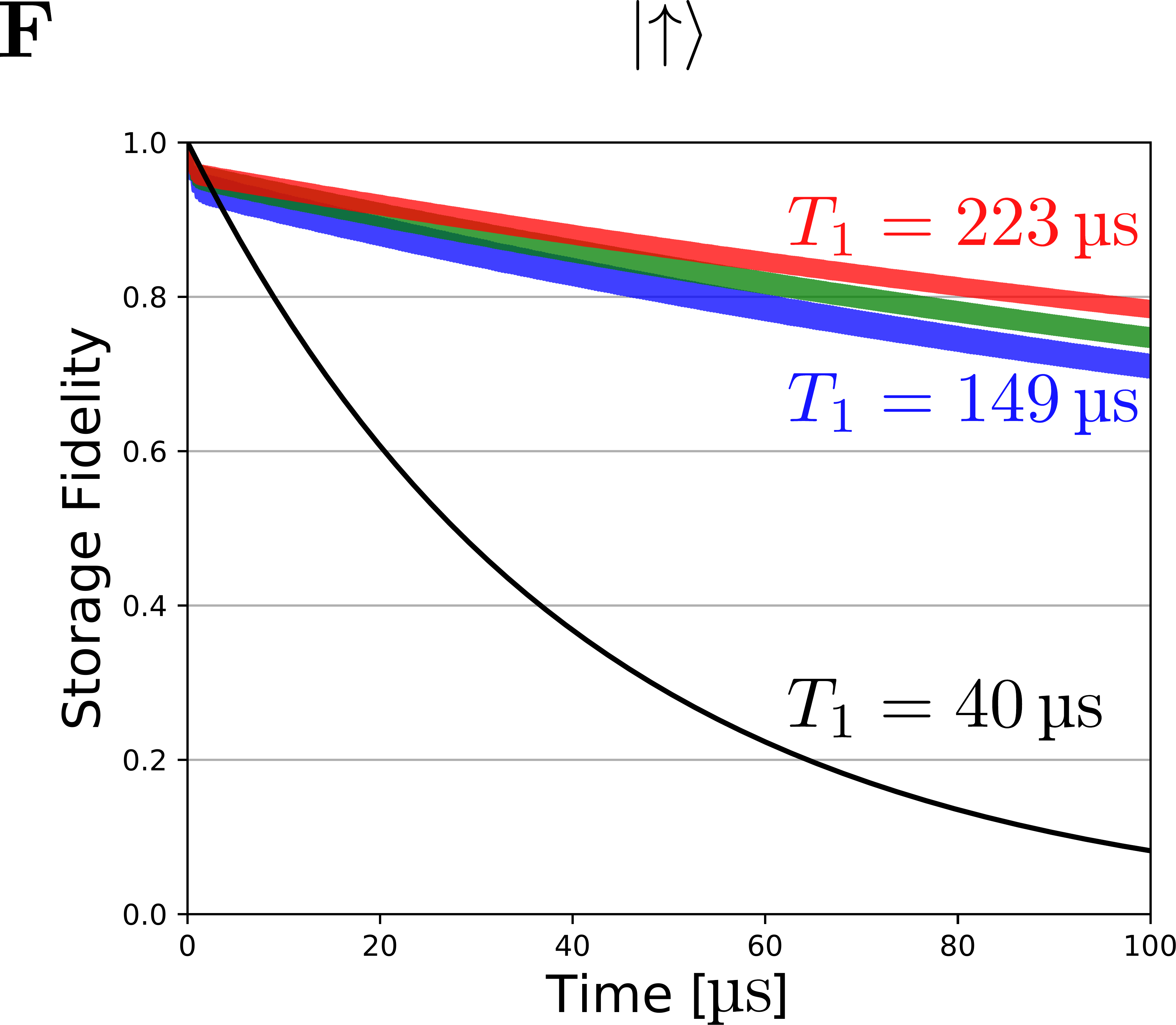}
      \captionlistentry{} \label{fig:Parameter_Scaling_0}
   \end{subfigure}
   ~
    \begin{subfigure}[b]{0.45\textwidth}
   \includegraphics[scale=0.17]{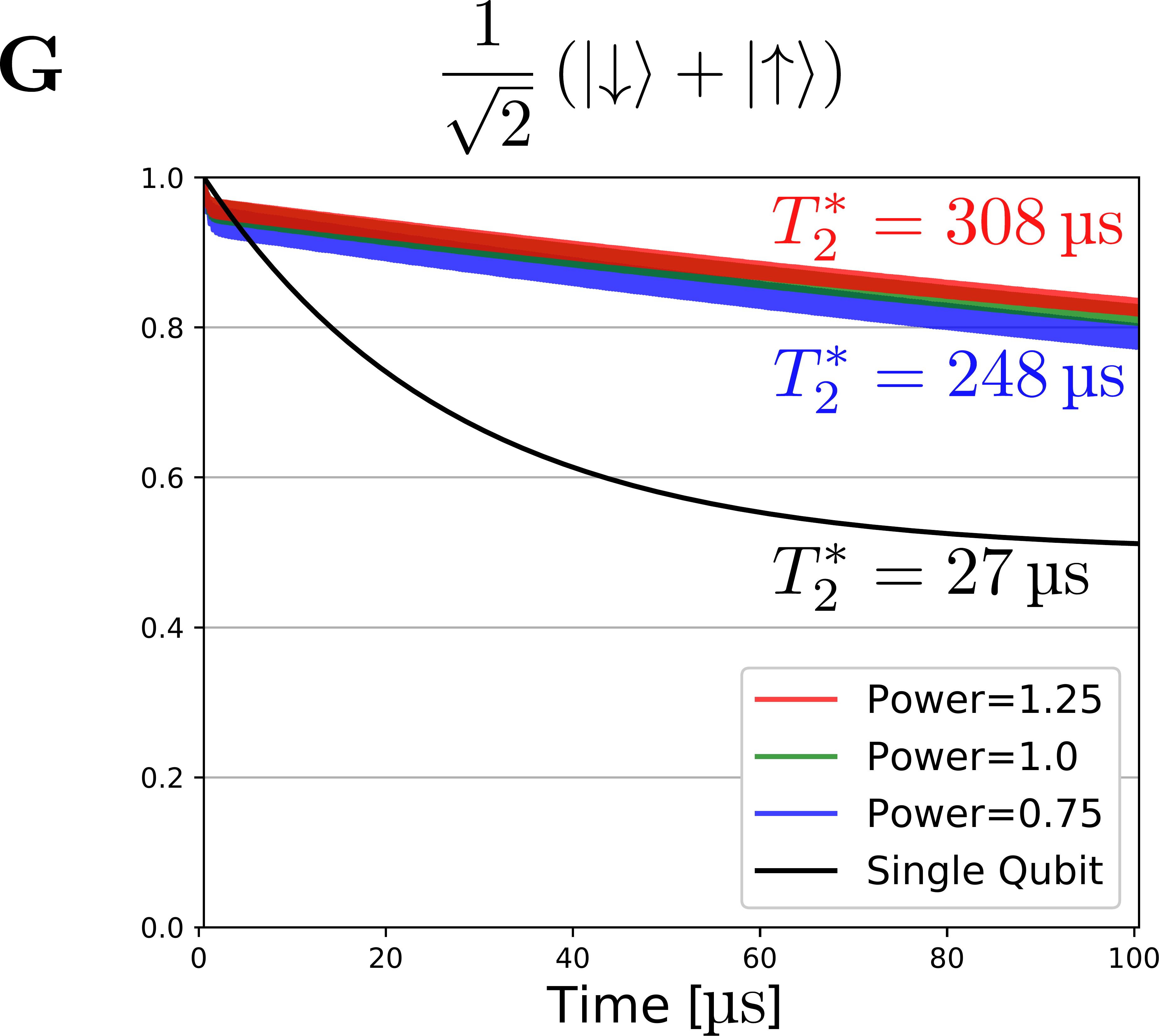}
   \captionlistentry{} \label{fig:Parameter_Scaling}
   \end{subfigure}\\ \vspace{0.4cm}
   \begin{subfigure}[b]{1.0\textwidth}
\includegraphics[scale=1.0]{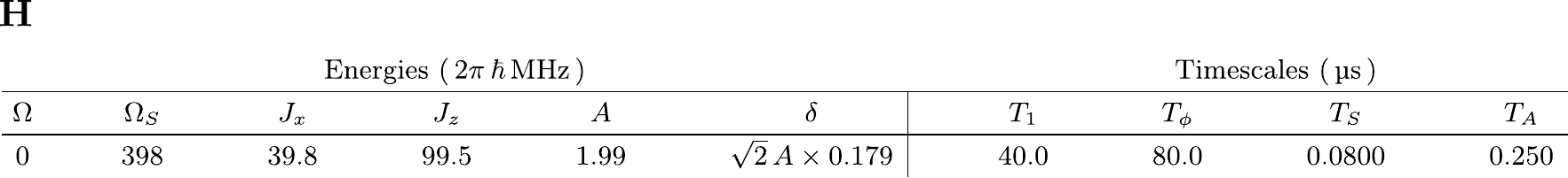}
\captionlistentry{} \label{tbl:Parameters}
\end{subfigure}
\caption{\textbf{Structure, properties and performance of the main hQEC scheme.} (\textbf{A}) Schematic depiction of the six data and two 'shadow' qubits setup of the main hQEC scheme. The vertical interactions in each block (A,B, and C) are of Heisenberg $XXZ$ type, while the horizontal interactions between blocks are exchange $XX$ interactions. (\textbf{B}) Simulation results for the storage fidelity achieved for the states $\left| \, 1 \, \right>_L$ and $\frac{1}{\sqrt{2}} \left( \left| 0 \right> + \left| 1 \right> \right)_L$ in the 6-qubit hybrid scheme from (A), with results for an uncorrected single qubit included for comparison. The inset depicts behaviour on short time-scales resulting from the interplay of continuous decoherence and discontinuous active correction steps. (\textbf{C})  A schematic depiction of the error correction steps. Starting from a valid state on the left, the system can either undergo phase-errors or loss-errors. If a phase-error occurs, $g_0=+1$ and the active correction is performed to correct the error (Top). If a loss-error occurs, the error oscillates until it is either corrected or converted to a phase error by shadow-qubit decay (Bottom). During the oscillations, $g_0=-1$, and thus active correction is disabled. For further details, see App. \ref{sec:Autonomous_EC} and \ref{sec:Active_EC}. (\textbf{D}) Simulation of the oscillations along the chain occurring when an eror has ocurred and the hQEC-system has been put into the state $\left| T \; + \; + \right>$. The plotted quantities are the probability of observing the data qubits in the three states $\left| T \; + \; + \right>$, $\left| + \; + \; - \right>$ and $\left| + \; + \; + \right>$ as a function of time after the initial decay. Note the fact that the state is initially corrected to a superposition of $\left| + + + \right>$ and $\left| + + - \right>$. Only once an active error correction step is applied in the middle of the simulation is the erroneous of these components eliminated and the correction to the state $\left| + + + \right>$ completed. (\textbf{E}) Schematic depiction of the oscillations of the $T$-error and subsequent error correction by decay of the shadow qubit. (\textbf{F}-\textbf{G}) Plots depicting the scaling of the performance of the code presented in the main text for storing the two states $\left| \, 1 \, \right>_L$ and $\frac{1}{\sqrt{2}} \left( \left| 0 \right> + \left| 1 \right> \right)_L$ when the interaction-strengths, shadow loss-rates and the frequency of active correction have all been scaled by either a factor of 0.75, 1.0 or 1.25 (see App. \ref{sec:Params} for details). (\textbf{H}) Parameters used for the simulations in Fig. \ref{fig:Mega2}. Note that the simulations were performed in the rotating frame related to $H_0$, except a small amount of $\Omega_S$ was kept since the sum $\Omega+\Omega_S$ plays a role in limiting what transitions are driven by $H_{\text{driv}}$. The specified timescales are the timescales related to photon loss ($T_1$) and phase errors ($T_{\phi}$) in the data quits, photon loss in the shadow qubits ($T_S$), and the time between the active error correction steps ($T_A$). For more details on the choice parameters see App. \ref{sec:Params}.}
\end{figure*}

\clearpage

\section{Alternative Schemes}
\label{sec:Alternative_Schemes}
In this section we provide further details on the structure required to make an autonomous error correction scheme work. From these considerations, we then introduce a few alternative schemes to the one from the main text, including a scheme using simpler interactions and only a single driving-tone, and a scheme using only a single shadow-qubit.

\subsection{Alternative Driving Schemes}
\label{sec:Alternative_Driving}
\begin{figure*}[hbtp]
  \centering
\begin{subfigure}[b]{1.0\textwidth}
   \includegraphics[scale=0.35]{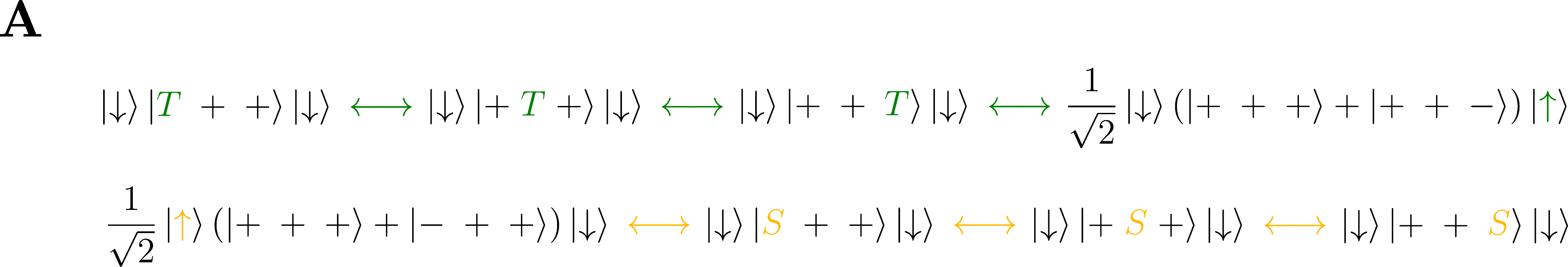}
  \captionlistentry{} \label{fig:Standard_Driving}
\end{subfigure}\\ \vspace{0.8cm}
\begin{subfigure}[b]{1.0\textwidth}
   \includegraphics[scale=0.35]{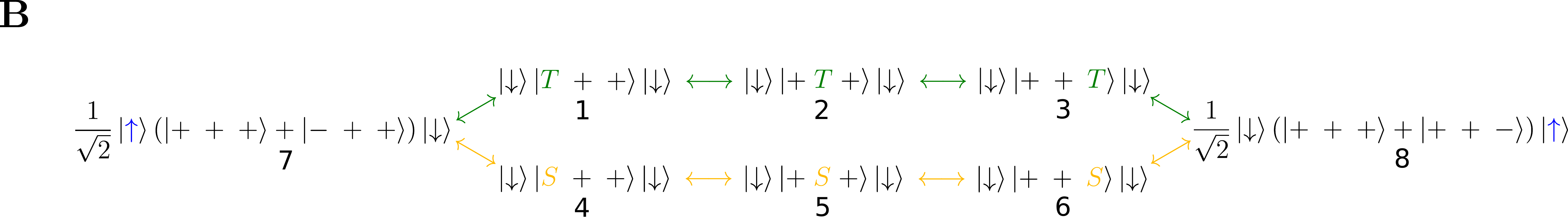}
  \captionlistentry{} \label{fig:Bad_Driving}
\end{subfigure}\\ \vspace{1.3cm}
\begin{subfigure}[b]{1.0\textwidth}
   \includegraphics[scale=0.14]{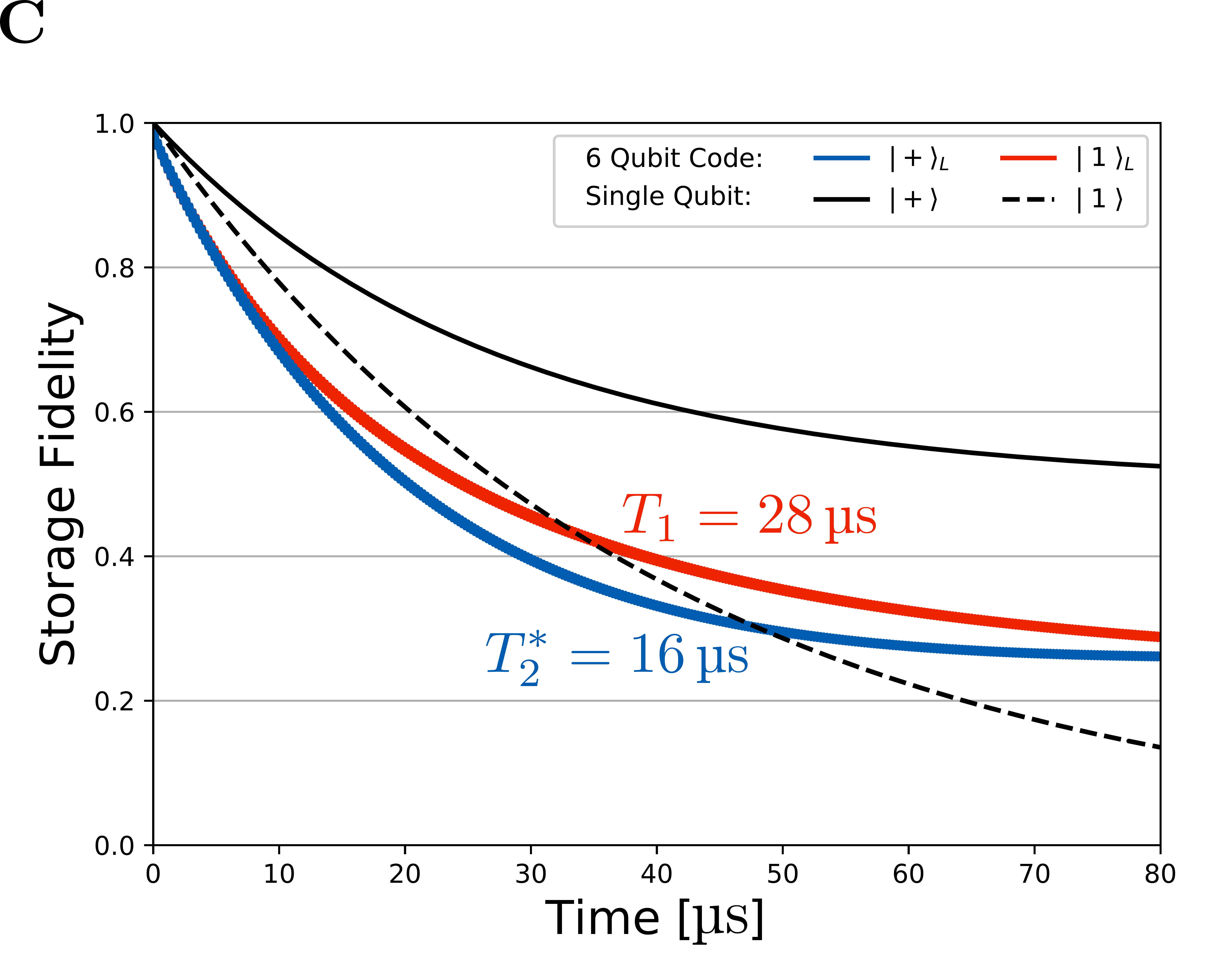}
  \captionlistentry{}  \label{fig:Failure_plot}
\end{subfigure}\\ \vspace{0.2cm}
\caption{\textbf{Breakdown of autonomous error correction for $J_x=0$.} (\textbf{A}) Schematic representation of the way error states and error correcting states are coupled by $H_{\text{chain}}$ and $H_d$ for the scheme presented in the main text. (\textbf{B}) Corresponding couplings when $J_x$ is set  to zero. (\textbf{C}) Plot depicting the storage fidelity as a function of time for the coupling toplogy depicted in (B), i.e for the scheme presented in the main text in the case where $J_x$ is set to zero with all other parameters unchanged. Note the significantly reduced performance compared to Fig. \ref{fig:Mega2B}, indicating the presence of non-corrected subspaces within the space of error-states.}
\label{fig:JX0Megafig}
\end{figure*}

%\begin{figure*}[hbtp]
%  \centering
%   \includegraphics[width=1.8\columnwidth]{Figures/Standard_Driving_Geometry.pdf}
%   \caption{Schematic representation of the way error states and error correcting states are coupled by $H_{\text{chain}}$ and $H_d$ for the scheme presented in the main text.}
%  \label{fig:Standard_Driving}
%\end{figure*}
A central challenge in making autonomous error correction is to engineer the system dynamics in such a way that the states resulting from an error always oscillate into a corrected state with an excitation in a shadow qubit. As we saw above, one way to achieve this is to couple the relevant states in the way schematically depicted in Fig. \ref{fig:Standard_Driving}. Specifically, in the rotating frame of $H_0 + H_{\text{pair}}$, the dynamics of the four states $\{ \left| T \, + \, + \right> \left| \downarrow \downarrow \right>, \left| + \, T \, + \right> \left| \downarrow \downarrow \right>, \left| + \, + \, T \right> \left| \downarrow \downarrow \right>, \left| + \, + \, \uparrow \uparrow \right> \left| \uparrow \downarrow \right> \}$ can be described using rotating wave approximation arguments by the Hamiltonian
\begin{align}
\tilde{H} &= \begin{pmatrix}
0 & 2\delta & 0 & 0 \\
2\delta & 0 & 2\delta & 0 \\
0 & 2\delta & 0 &  \frac{A}{\sqrt{2}} \\
0 & 0 & \frac{A}{\sqrt{2}} & 0
\end{pmatrix} \; , \label{eq:Driving_Matrix}
\end{align}
which is of the general form
\begin{align*}
\tilde{H} = \alpha \begin{pmatrix}
0 & 1 & 0 & 0 \\
1 & 0 & 1 & 0 \\
0 & 1 & 0 &  \beta \\
0 & 0 & \beta & 0
\end{pmatrix} \; .
\end{align*}
One can easily show that for any non-zero $\beta$, this matrix has no eigenvectors without a non-zero component of the last, error correcting state. It turns out that this is equivalent to saying that no matter what state in the subspace we pick as the initial state, it will at some point pick up a non-zero error correcting component, and hence be successfully corrected. To see why this is the case, imagine that we did have an initial state $\vec{v}$ that never picked up a correcting component. We could then write this state in terms of the eigenstates $\vec{v}_i$ of the Hamiltonian as follows:
\begin{align*}
\vec{v} = \sum_{i=1}^4 a_i \vec{v}_i \; ,
\end{align*} 
with the $a_i$'s a suitable set of complex coefficients. The time-evolved state would then be given by
\begin{align*}
\vec{v}(t) = \sum_{i=1}^4 a_i \vec{v}_i e^{-i \frac{E_i}{\hbar} t} \; ,
\end{align*}
where $E_i$ is the energy of the eigenstate $\vec{v}_i$. The fact that the last component is always zero would translate to the criterion
\begin{align*}
\left(\vec{v}(t)\right)_4 = \sum_{i=1}^4 a_i \left(\vec{v}_i\right)_4 e^{-i \frac{E_i}{\hbar} t} = 0 & &\forall t \; ,
%%\label{eq:Math_Sum_to_zero}
\end{align*}
where $\left(\vec{v}\right)_k$ is the $k$'th component of the vector $\vec{v}$. But if this expression is to be zero at all times, the independence of complex exponentials tells us that the coefficients related to each oscillation frequency must sum to zero separately. In other words, for each degenerate subset $D_j \subseteq \{ 1,2,3,4\}$ of the eigenstates of the Hamiltonian, we must have that
\begin{align}
\sum_{i \in D_j} a_i \left(\vec{v}_i\right)_4 = 0 & & \forall j \; .
\label{eq:Math_Sum_to_zero_Deg}
\end{align}
Assuming we started with a properly normalized state, we also know that $\sum_i \left| a_i \right|^2=1$, meaning at least one of the $a_i$-coefficients is non-zero. Letting $D$ be the degenerate subspace related to such a non-zero coefficient, the object
\begin{align*}
\tilde{v} = \frac{1}{\sqrt{\sum_{i \in D} \left|a_i \right|^2}} \sum_{i \in D} a_i \vec{v}_i
\end{align*}  
must then a well-defined normalized state. Since it is constructed as a superposition of eigenstates from the same degenerate subspace, it is even an eigenstate of the Hamiltonian. Additionally, examining Eq. \eqref{eq:Math_Sum_to_zero_Deg} reveals that the fourth component of this new state must necessarily be zero. Thus we see that having a state that has no correcting component at any point in time  will necessarily allow us to construct an eigenvector of the Hamiltonian with the same property. We therefore conclude that if no eigenstates without correcting component exist, no states that never pick up correcting components can exist, and hence no error state exist that escapes the grasp of our autonomous correction. The above considerations easily generalize, allowing us to easily check if any driving scheme will work or not simply by checking if it has eigenstates without components in the states responsible for the autonomous correction. Note that our problem is reminiscent of an inverse dark-state problem--rather than engineering the system to operate without population in the unstable states of the system, we explicitly try to make sure that such a population will always arise.

\begin{figure*}[hbtp]
  \centering
\begin{subfigure}[b]{1.0\textwidth}
   \includegraphics[scale=0.35]{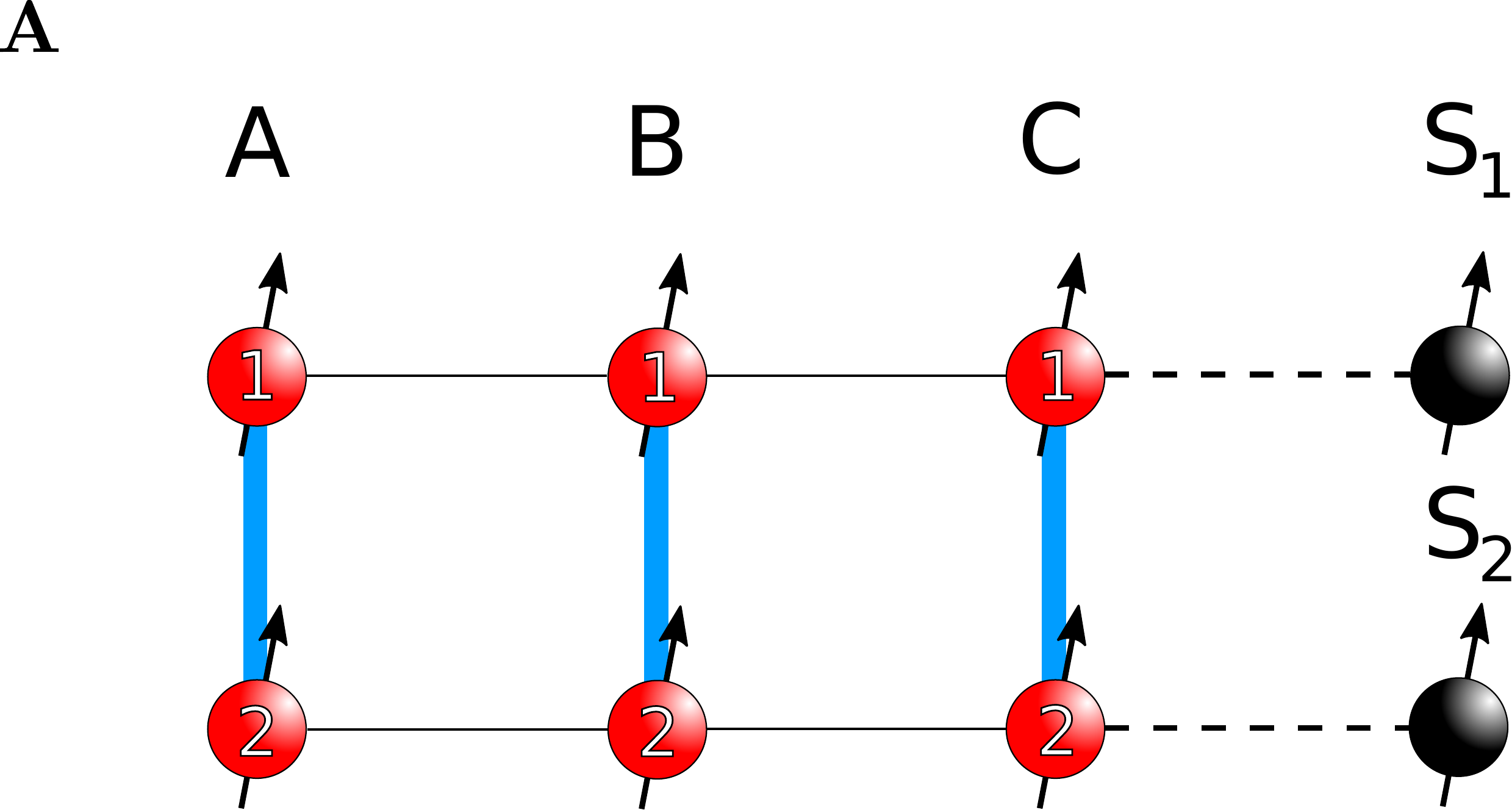}
  \captionlistentry{} \label{fig:New_Geometry_Realspace}
\end{subfigure}\\ \vspace{1.4cm}
\begin{subfigure}[b]{1.0\textwidth}
   \includegraphics[scale=0.46]{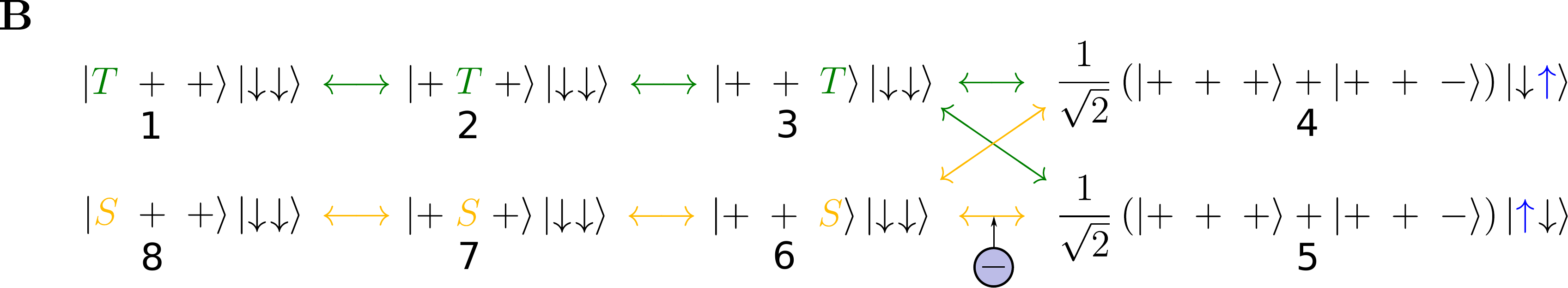}
  \captionlistentry{}  \label{fig:New_Geometry_Statespace}
\end{subfigure}\\ \vspace{1.3cm}
\begin{subfigure}[b]{1.0\textwidth}
   \includegraphics[scale=0.6]{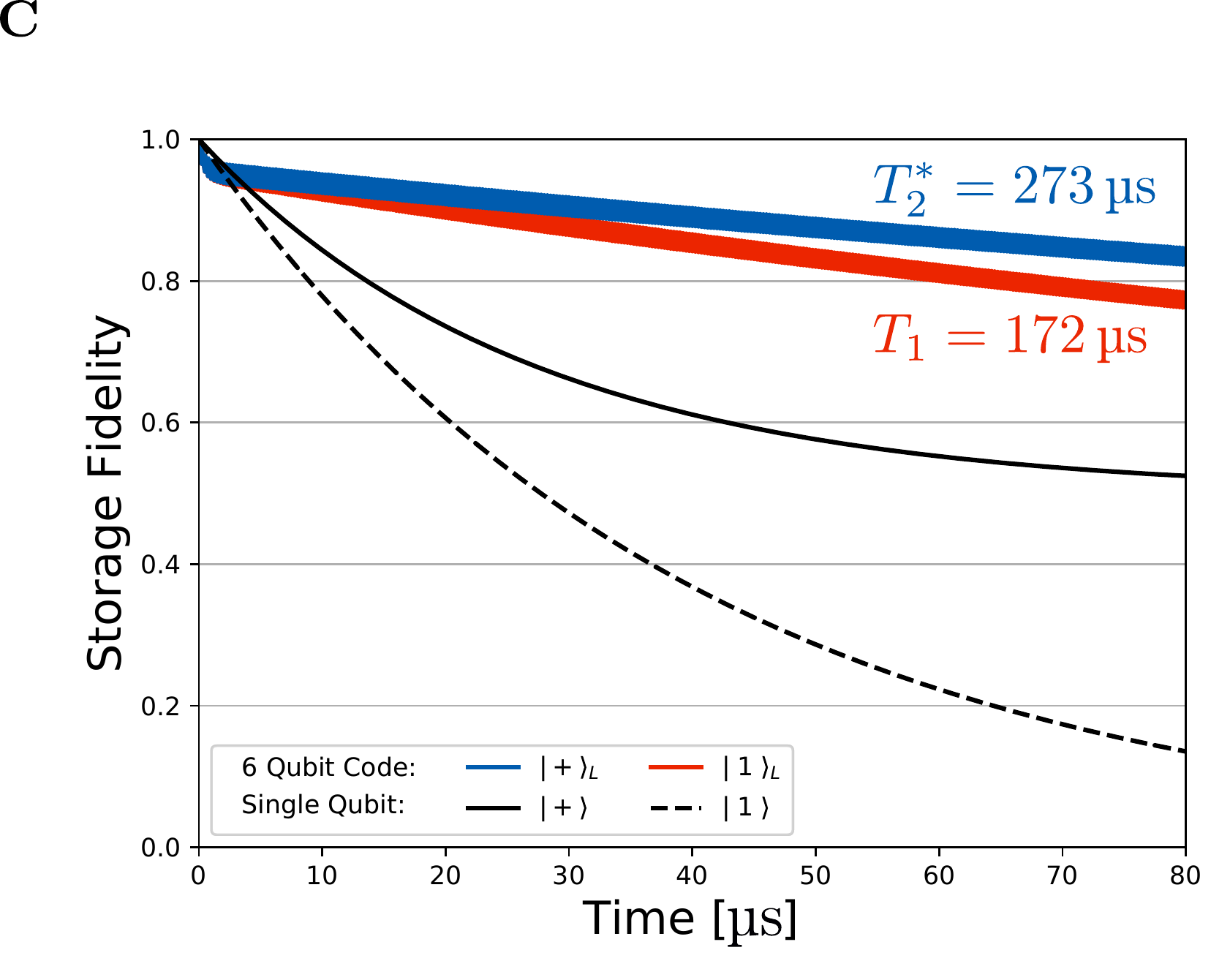}
  \captionlistentry{}  \label{fig:New_Geom_Performance}
\end{subfigure}\\ \vspace{0.2cm}
\caption{\textbf{hQEC with an alternative coupling geometry.} (\textbf{A}) Schematic depiction of an alternative coupling geometry that allows for autonomous correction even in the case where $J_x$ vanishes, in contrast to the scheme from the main text (See Fig. {fig:JX0Megafig}). (\textbf{B}) Schematic depiction of the coupling topology related to this geometry in the case $J_x=0$. (\textbf{C}) Plot depicting the storage-performance of the alternative coupling geometry. The parameters used are identical to those used to generate Fig. \ref{fig:Mega2B}, except $J_x$ has been set to zero.}
\end{figure*}

For an example of driving that does not work, consider the case where $J_x=0$. In this case, the driving frequencies related to transitions from $S$ and $T$ to $\left| \uparrow \uparrow \right>$ become identical, and as a result both of the shadow qubits will interact with both the $T$- and $S$-error subspaces. The resulting coupling geometry will therefore be the one depicted in Fig.\ref{fig:Bad_Driving}, with an effective coupling Hamiltonian that takes the form
\begin{align*}
\tilde{H} &=  \alpha \begin{pmatrix}
0 & 1 & 0 & 0 & 0 & 0 & \beta_1 & 0 \\
1 & 0 & 1 & 0 & 0 & 0 & 0 & 0 \\
0 & 1 & 0 & 0 & 0 & 0 & 0 & \beta_2 \\
0 & 0 & 0 & 0 & 1 & 0 & \beta_1 & 0 \\
0 & 0 & 0 & 1 & 0 & 1 & 0 & 0 \\
0 & 0 & 0 & 0 & 1 & 0 & 0 & \beta_2\\
\beta_1^* & 0 & 0 & \beta_1^* & 0 & 0 & 0 & 0 \\
0 & 0 & \beta_2^* & 0 & 0 & \beta_2^* & 0 & 0 \\ 
\end{pmatrix}
\end{align*}
when the states are ordered as indicated by the numbers on Fig. \ref{fig:Bad_Driving} and a star is used to denote complex conjugates. It is then clear that a state of the form 
\begin{align*}
\frac{1}{2}\begin{pmatrix}
1 & 0 & -1 & -1 & 0 & 1 & 0 & 0
\end{pmatrix}^T
\end{align*}
will be an eigenstate of the system with zero energy and no component in the error correcting subspace related to the two final entries. As a result, this state will tend to accumulate uncorrectable population, leading to a breakdown of the error correction (see Fig. \ref{fig:Failure_plot}). Note that this persists even if the couplings along the chain (i.e. the two $\delta$'s appearing in $H_{\text{chain}}$) are not identical, thus resulting in a Hamiltonian of the more general form
\begin{align*}
\tilde{H} &=\begin{pmatrix}
0 & \alpha_1 & 0 & 0 & 0 & 0 & \beta_1 & 0 \\
\alpha_1^* & 0 & \alpha_2 & 0 & 0 & 0 & 0 & 0 \\
0 & \alpha_2^* & 0 & 0 & 0 & 0 & 0 & \beta_2 \\
0 & 0 & 0 & 0 & \alpha_1 & 0 & \beta_1 & 0 \\
0 & 0 & 0 & \alpha_1^* & 0 & \alpha_2 & 0 & 0 \\
0 & 0 & 0 & 0 & \alpha_2^* & 0 & 0 & \beta_2\\
\beta_1^* & 0 & 0 & \beta_1^* & 0 & 0 & 0 & 0 \\
0 & 0 & \beta_2^* & 0 & 0 & \beta_2^* & 0 & 0 \\ 
\end{pmatrix} \; .
\end{align*}
In this case, the troublesome eigenstate is simply
\begin{align*}
\frac{1}{\sqrt{2 \left| \alpha_1 \right|^2 + 2 \left| \alpha_2 \right|^2 }} \begin{pmatrix}
\alpha_2 & 0 & -\alpha_1^* & -\alpha_2 & 0 & \alpha_1^* & 0 & 0
\end{pmatrix}^T \; ,
\end{align*}
supporting the claim made in App. \ref{sec:Details_on_Main_Scheme} that the success or failure of a driving scheme tends to depend more strongly on the topology of how states are coupled than on specific parameter choices.
To further underline this point, imagine that we instead used the geometry depicted in Fig. \ref{fig:New_Geometry_Realspace} while still letting $J_x$ vanish. In this case, the couplings take the form from Fig. \ref{fig:New_Geometry_Statespace}. Note that because the two shadow qubits now couple to different data qubit rows, a relative sign appears the matrix elements related to the coupling of the shadow qubits to the $S$-states. This sign, essentially inherited from the sign in the definition of $\left| S \right>$ in Eq. \eqref{eq:State_Defs}, means the Hamiltonian now takes the form
\begin{align*}
\tilde{H} &= \begin{pmatrix}
0 & \alpha_1 & 0 & 0 & 0 & 0 & 0 & 0 \\
\alpha_1^* & 0 & \alpha_2 & 0 & 0 & 0 & 0 & 0 \\
0 & \alpha_2^* & 0 & \beta_1 & \beta_2 & 0 & 0 & 0 \\
0 & 0 & \beta_1^* & 0 & 0 & \beta_1^* & 0 & 0 \\
0 & 0 & \beta_2^* & 0 & 0 & -\beta_2^* & 0 & 0 \\
0 & 0 & 0 & \beta_1 & -\beta_2 & 0 & \alpha_2^* & 0 \\
0 & 0 & 0 & 0 & 0 & \alpha_2 & 0 & \alpha_1^* \\
0 & 0 & 0 & 0 & 0 & 0 & \alpha_1 & 0 
\end{pmatrix} \; ,
\end{align*}
inheriting again the state ordering from the corresponding Fig. \ref{fig:New_Geometry_Statespace}. Looking for eigenvectors of the form
\begin{align*}
\begin{pmatrix}
a_T & b_T & c_T & 0 & 0 & c_S & b_S & a_S
\end{pmatrix}^T
\end{align*}
with some arbitrary eigenvalue $\lambda$ then yields the equations
\begin{align}
\beta_1^* \left( c_T + c_S \right) &= 0 \label{eq:eigenval}\\
\beta_2^* \left( c_T - c_S \right) &= 0 \nonumber \; ,
\end{align}
which, assuming non-zero parameters $\beta_1, \beta_2 \neq 0$, leads to the requirement that
\begin{align*}
c_T=c_S=0 \; .
\end{align*}
Looking at the $T$-part of the eigenvalue equation:
\begin{align*}
\alpha_1 b_T &= \lambda a_T\\
\alpha_1^* a_T + \alpha_2 c_T &= \lambda b_T\\
\alpha_2^* b_T &= \lambda c_T
\end{align*} 
and using $c_T=0$ now yields that $a_T$ and $b_T$ must also vanish as long as $\alpha_1$ and $\alpha_2$ do not. Similarly, $a_S$ and $b_S$ must vanish. We thus conclude that the only way to construct a vector that fulfils the eigenvalue-equation $\tilde{H} \vec{v} = \lambda \vec{v}$ and that has zeroes on the entries related to error correction is to have $\vec{v}=0$, confirming that no eigenvector without corrector-state components exist. Note that the sign appearing in Eq. \eqref{eq:eigenval} due to having shadows coupled to both rows of data qubits played a vital role in these arguments, further indicating the topological nature of the driving question.\\
What we may conclude is that the operation of the autonomous error correcting code depends in an essential way on the topology of how the error states and error correcting states are coupled. In contrast, the exact choice of parameters is of lesser importance, although it can still play a role in determining the speed of the correction and the degree of validity of the rotating wave arguments used to derive the Hamiltonians of this section. Additionally, we can conclude that by altering the real-space geometry of how the qubits are connected to one another, we are actually allowed to forego the rather complicated Heisenberg XXZ-couplings in favour of simpler couplings of the form $\sigma_{j,1}^z \sigma_{j,2}^z$. This has the added benefit that the two driving frequencies mentioned in the main text become identical, meaning only a single driving tone is needed in order to run the error correction. From an experimental stand point, it is  likely that this scheme is easier to implement, even though the dynamics of the system are not quite as intuitive as for the system investigated in the main text. One thing to keep in mind, however, is that setting $J_x$ to zero may make the scheme more fragile with respect to asymmetries in the couplings induced by $H_{\text{chain}}$ along the top and bottom row of data qubits. To see why, imagine that $H_{\text{chain}}$ takes the more general form
\begin{align*}
H_{\text{chain}} =&\, \left( \delta_1 + \Delta_1 \right) \left( \sum_{k \in \{x,y\}}  \sigma_{A,1}^k \sigma_{B,1}^k \right) \\
&+ \left( \delta_1 - \Delta_1 \right) \left( \sum_{k \in \{x,y\}}  \sigma_{A,2}^k \sigma_{B,2}^k \right) \\
&+ \left( \delta_2 + \Delta_2 \right)  \left(  \sum_{k \in \{x,y\}}  \sigma_{B,1}^k \sigma_{C,1}^k \right) \\
&+ \left( \delta_2 - \Delta_2 \right) \left(  \sum_{k \in \{x,y\}}  \sigma_{B,2}^k \sigma_{C,2}^k \right)\\
=&\, \delta_1 \left( \sum_{k \in \{x,y\}} \left[ \sigma_{A,1}^k \sigma_{B,1}^k + \sigma_{A,2}^k \sigma_{B,2}^k \right] \right)\\
&+ \delta_2 \left( \sum_{k \in \{x,y\}} \left[ \sigma_{B,1}^k \sigma_{C,1}^k + \sigma_{B,2}^k \sigma_{C,2}^k \right] \right)\\
&+  \Delta_1 \left( \sum_{k \in \{x,y\}} \left[ \sigma_{A,1}^k \sigma_{B,1}^k - \sigma_{A,2}^k \sigma_{B,2}^k \right] \right)\\
&+ \Delta_2 \left( \sum_{k \in \{x,y\}} \left[ \sigma_{B,1}^k \sigma_{C,1}^k - \sigma_{B,2}^k \sigma_{C,2}^k \right] \right)\; .
\end{align*}
In this case, only the two first terms appearing in this expression would actually participate in driving the transitions depicted in Fig. \ref{fig:Standard_Driving}. The two terms related to the asymmetries $\Delta_1$ and $\Delta_2$ would instead drive transitions such as 
\begin{align}
\left| T \; + \; + \right> \; \longleftrightarrow \; \left| - \; S \; + \right> \; .
\label{eq:Asym_Process}
\end{align}
In other words, they would still allow errors to move through the system, but would in the process allow the nature of both the error and the $\pm$-states to change. This leads to processes that corrupt the stored information, thus breaking the autonomous scheme. Luckily, such transitions are suppressed as long as $J_x$ is much larger than the $\Delta_i$'s, since in this case the $4 J_x$ energy difference between $T$- and $S$-states effectively suppresses transitions like the one in \eqref{eq:Asym_Process}. Even though the $J_x=0$-scheme lacks this protection, it is still able to achieve $T_2^*$ coherence times in the range $\SI{239}{\micro \second}$ to $\SI{325}{\micro \second}$ and $T_1$ coherence times in the range $\SI{144}{\micro \second}$ to $\SI{226}{\micro \second}$ as long as rigid symmetry is maintained (i.e. $\Delta_1=\Delta_2=0$), thus exhibiting similar performance to the scheme from the main text in this limit. A plot illustrating the performance of the $J_X=0$-scheme is depicted in Fig. \ref{fig:New_Geom_Performance}.

\subsection{Error Correction with a Single Shadow Qubit}
\label{sec:Single_Shadow}
\begin{figure*}[hbtp]
  \centering
\begin{subfigure}[b]{1.0\textwidth}
   \includegraphics[scale=0.83]{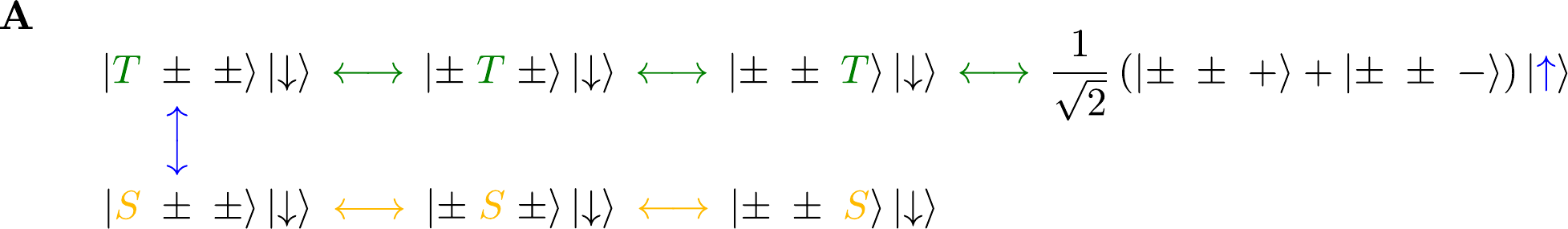}
  \captionlistentry{}  \label{fig:Single_Shadow_Top}
\end{subfigure}\\ \vspace{1.0cm}
\begin{subfigure}{.47\textwidth}
   \includegraphics[scale=0.57]{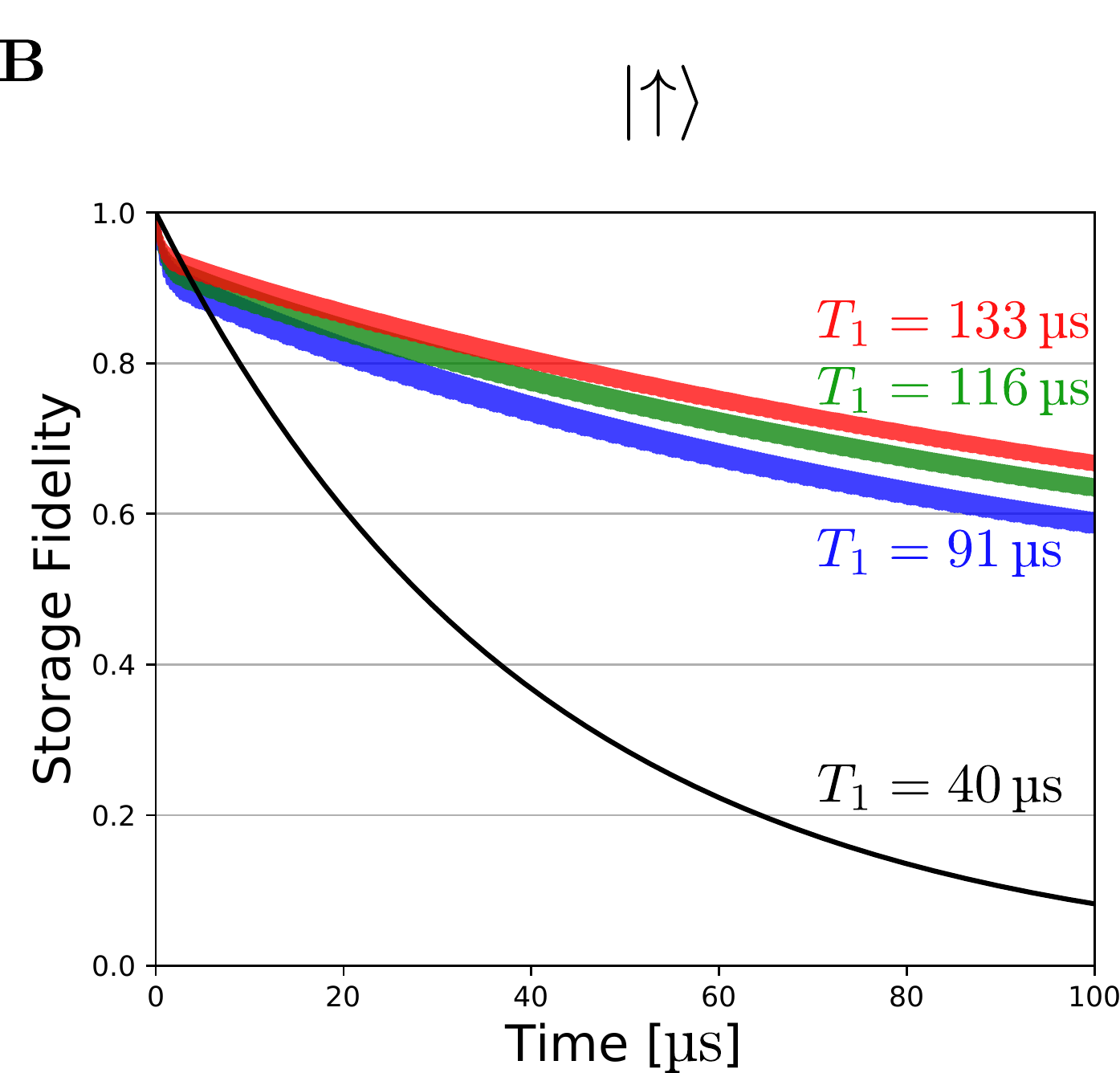}
     \captionlistentry{} \label{fig:Single_Shadow_Parameter_Scaling}
\end{subfigure}%
\begin{subfigure}{.47\textwidth}
   \includegraphics[scale=0.57]{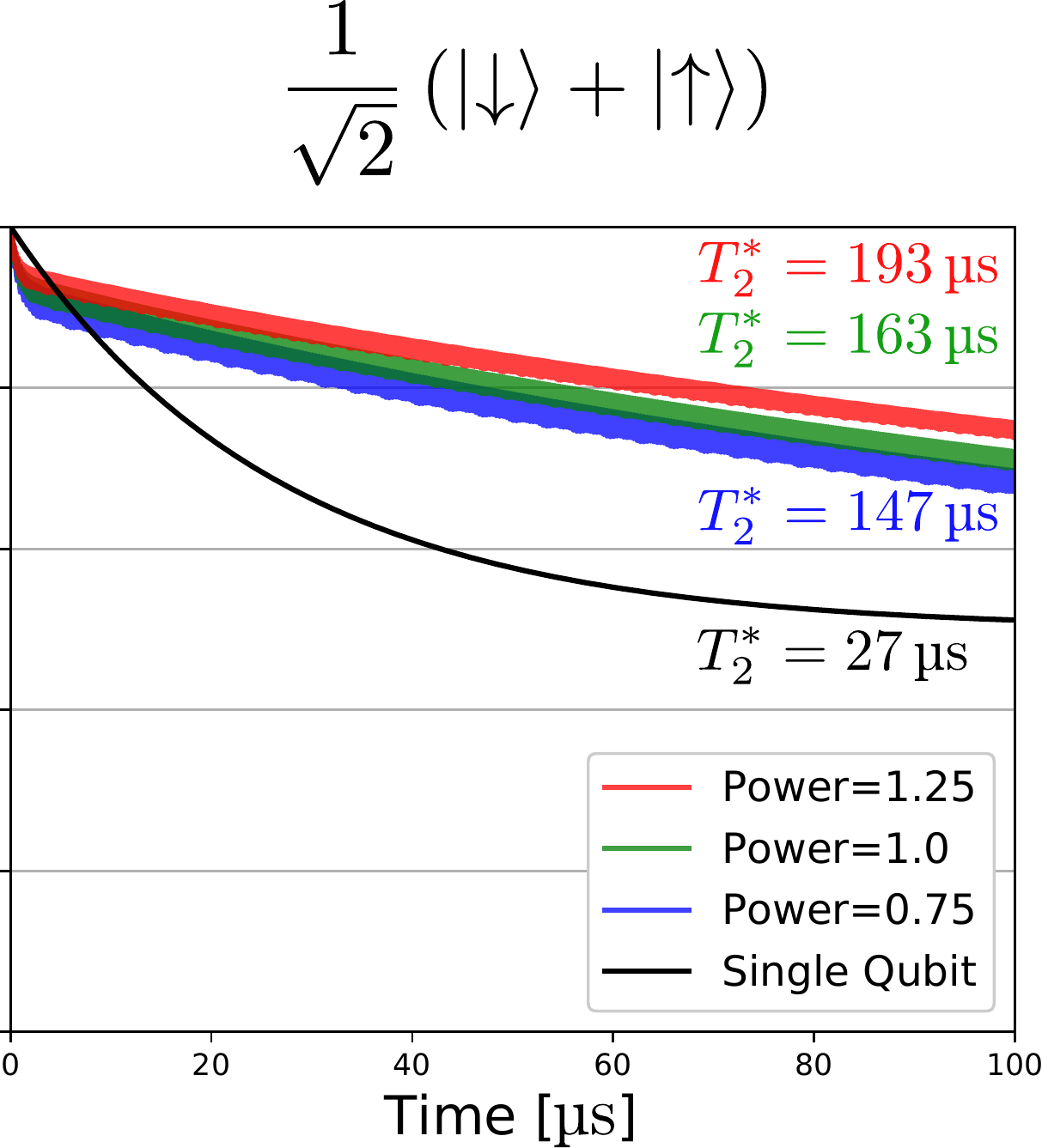}
\end{subfigure}\vspace{1.2cm}
\begin{subfigure}[b]{1.0\textwidth}
   \includegraphics[scale=0.9]{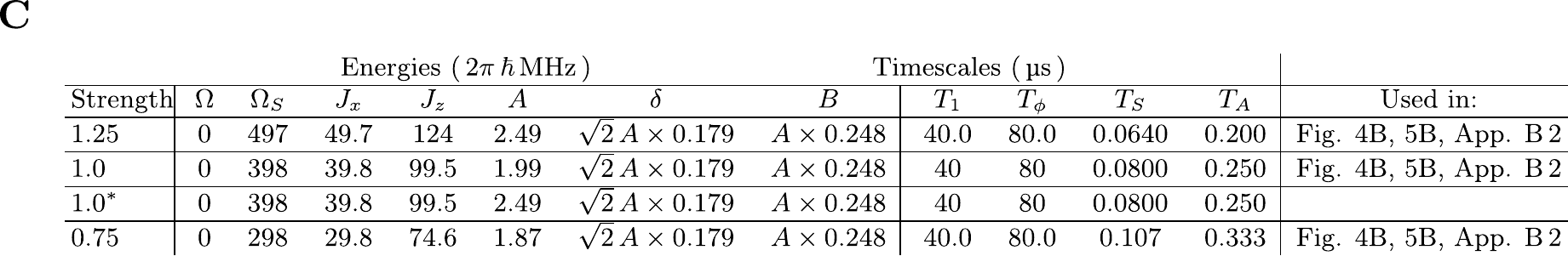}
  \captionlistentry{}  \label{tbl:Parameters_Single_Shadow}
\end{subfigure}\\ \vspace{0.8cm}
\caption{\textbf{Single-shadow hQEC-scheme.} (\textbf{A}) Schematic depiction of the coupling topology of the single-shadow hQEC-scheme presented in App. \ref{sec:Single_Shadow}. (\textbf{B}) Storage-fidelity of this scheme for the states $\left| \uparrow \right>$ (Left) and $\frac{1}{\sqrt{2}} \left( \left| \uparrow \right> + \left| \downarrow \right> \right)$ (Right) for different values of the parameters of the model. For reference the storage performance of a single unprotected qubit is also reproduced on both plots. (\textbf{C}) Parameters used to generate the plots in (B), as well as throughout App. \ref{sec:Single_Shadow}. }
\end{figure*}

In the scheme presented in the main text, the correction of $S$- and $T$-errors proceeded independently, with each error correction process having its own shadow qubit and driving term in the Hamiltonian. However, looking at the error state from Eq. \eqref{eq:Error_States}, it seems like this separation might be unnecessary. If we could instead find a way to have the full subspace of error states
\begin{align*}
\left| T \; \pm \; \pm \right>\left| \downarrow \right>, \left| \pm \; T \; \pm \right>\left| \downarrow \right>, \left| \pm \; \pm  \; T \right>\left| \downarrow \right>\\
\left| S \; \pm \; \pm \right>\left| \downarrow \right>, \left| \pm \; S \; \pm \right>\left| \downarrow \right>, \left| \pm \; \pm \; S \right>\left| \downarrow \right>
\end{align*}
oscillate to a state of the form $\left| \pm \; \pm \; \uparrow \, \uparrow \right> \left| \uparrow \right>$ in such a way that the two signs $\pm$ are treated symmetrically, then error correction should be able to proceed through a single shadow qubit in exactly the same '\textit{oscillate-decay-active correction}'-sequence as in the main scheme. As mentioned in App. \ref{sec:Alternative_Driving}, the coupling would have to be constructed somewhat carefully in order to avoid inert states that dodge the error correction protocol. Inspired by the driving implemented in the main text, one way to implement working error correction would be to couple the  states as depicted on Fig. \ref{fig:Single_Shadow_Top}. It is easy to show using the methods of App. \ref{sec:Alternative_Driving} that this coupling topology indeed implements working error correction. In order to achieve such a set of effective couplings, four ingredients are required. The first three are straightforward: We still need the Hamiltonian $H_{\text{chain}}$ in order to couple the data qubit pairs to each other along the rows, we still need a Hamiltonian of the form
\begin{align*}
H_{\text{driv},A} = 2 A \cos\left( \frac{ \Omega + \Omega_S + 2J_z-J_x }{\hbar} \, t \right) \sigma_{C,2}^x \sigma_{S}^x
\end{align*} 
to couple the single remaining shadow qubit to the chain of data qubits in such a way that oscillations of the form 
\begin{align*}
& \dots \, \left| T \right> \left| \downarrow \right> \\
& \hspace{1cm} \updownarrow \\
& \dots \, \left| \uparrow  \; \uparrow \right> \left| \uparrow \right> 
\end{align*}
are induced, and we still need our backbone Hamiltonian $H_{\text{pair}}$ to constrain all of these oscillations. In other words, we need to keep the Hamiltonian terms that allow for the correction of the $T$ states. The fourth and final ingredient we need is a way to couple the $S$-states to the $T$-states so that they can also be corrected. This can be achieved by adding the following term:
\begin{align}
H_{\text{driv},B} = 2 B \cos\left( \frac{2 J_x t }{\hbar} \right) \sigma_{A,1}^z \; .
\label{eq:HdrivB}
\end{align}
With this added term, all of the couplings of Fig. \ref{fig:Single_Shadow_Top} are present and accounted for. The only consideration left is therefore whether any couplings surplus to these couplings have appeared. It turns out that this is not the case as long as $J_x$ is non-zero. Indeed, letting $J_x$ be zero would allow both $S$- and $T$-states to couple to the shadow qubit, thus giving us a different coupling topology than the one in Fig. \ref{fig:Single_Shadow_Top} and ultimately resulting in a breakdown of the autonomous correction due to similar  arguments to those introduced in App. \ref{sec:Alternative_Driving}. In fact, setting $J_x$ to zero introduces even more problems than just a breakdown of the error correction---it also introduces additional errors. To see why, note that setting $J_x=0$ would mean that the two transitions
\begin{align}
\left| + \right> \; &\longleftrightarrow \; \left| - \right> \label{eq:PM_trans} \\
\left| T \right> \; &\longleftrightarrow \; \left| S \right> \label{eq:TS_trans}
\end{align}
will both be transitions between states that have the same energy as defined by $H_{\text{pair}}$. As a result, neither transition will require time-dependent driving, and thus both of these transitions will be induced equally well by the term \eqref{eq:HdrivB} when $J_x=0$. But we recognize the first of these transitions as exactly the transition that would result from single qubit phase noise. In other words, setting $J_x$ to zero effectively turns $H_{\text{driv},B}$ into an additional source of phase noise on the first pair of qubits! The only way to avoid this is to make sure that the transition  \eqref{eq:TS_trans} is detuned from the zero-energy \eqref{eq:PM_trans}-transition so that we can drive one without driving the other. Since adjusting the relative energies of $S$- and $T$ states was exactly what $J_x$ was useful for, this further underlines the importance of $J_x$ if correction using a single shadow qubit is to work.

The performance of the single-shadow scheme with non-zero $J_x$ is depicted on Fig. \ref{fig:Single_Shadow_Parameter_Scaling}. As can be seen from this figure, the performance of the single shadow scheme is not as good as the two-shadow scheme, likely because of the smaller ratio of correcting states to error states leading to smaller relative populations in the correcting states and thus slower corrective decay. Nevertheless, $T_1$ lifetimes in the range $\SI{90.7}{\micro \second}$ to $\SI{133}{\micro \second}$ and $T_2^*$ lifetimes in the range $\SI{146}{\micro \second}$ to $\SI{192}{\micro \second}$ can be achieved using the parameters of Fig. \ref{tbl:Parameters_Single_Shadow}.

\clearpage

\section{Details on Results and Methods}
\label{sec:Main_Results}
Below, we provide further details on the parameters used in the hQEC schemes presented, as well as a more detailed description of the methods used to investigate the error-correction performance of these schemes.

\subsection{Parameters and Results}
\label{sec:Params}
\begin{figure*}[hbtp]
  \centering
\begin{subfigure}[b]{1.0\textwidth}
   \includegraphics[scale=0.9]{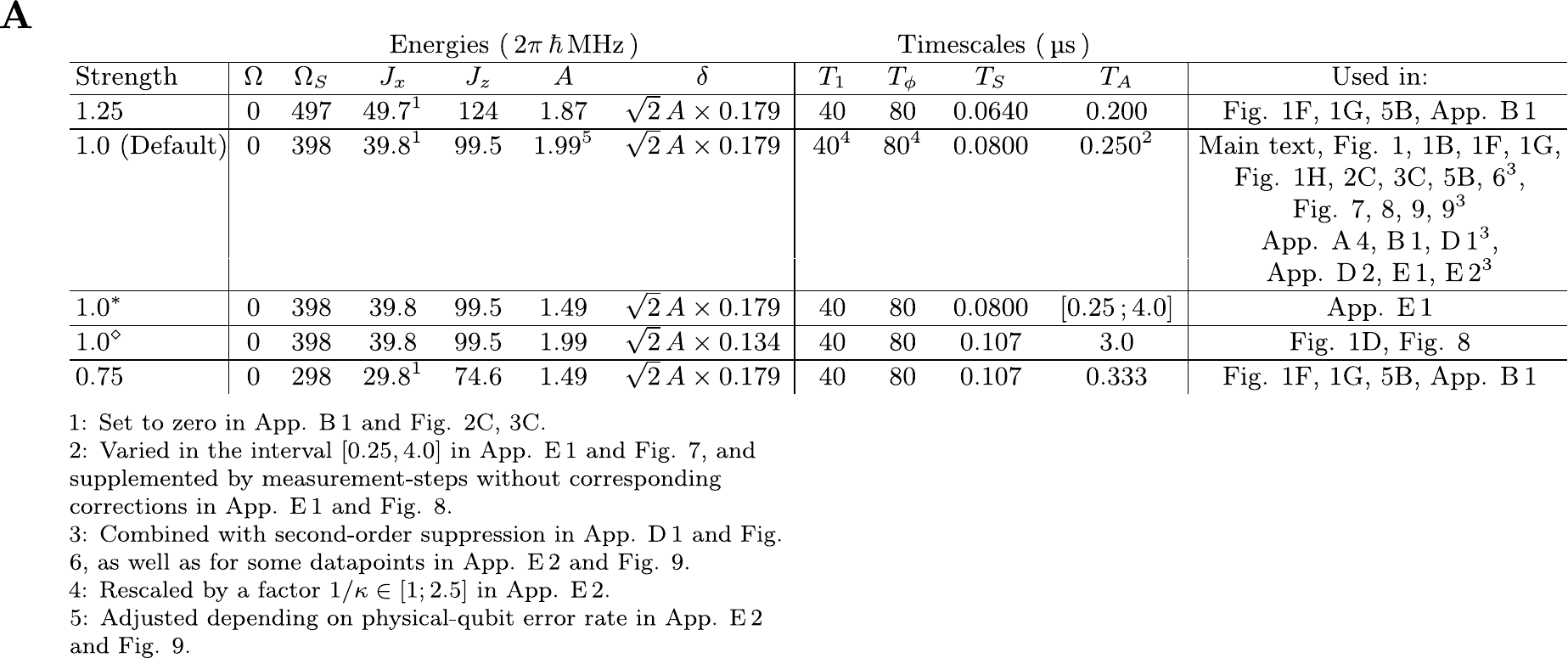} 
  \captionlistentry{}  \label{tbl:Parameters_large}
\end{subfigure}\\ \vspace{1.0cm}
\begin{subfigure}[b]{1.0\textwidth}
   \includegraphics[scale=0.9]{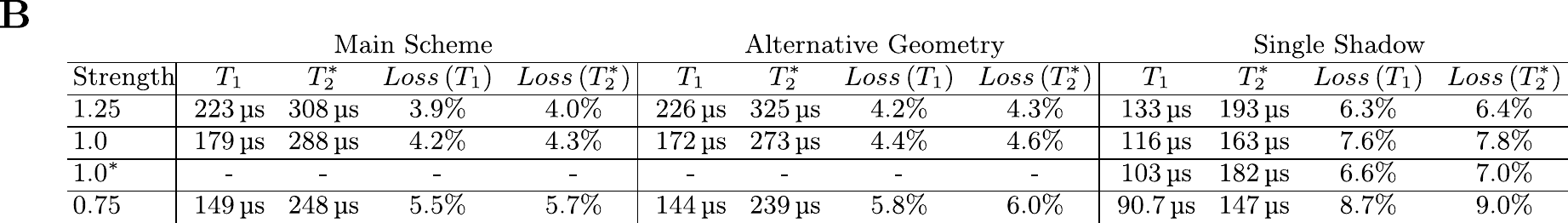}
  \captionlistentry{}  \label{tbl:Detailed_Results}
\end{subfigure}\\ \vspace{0.7cm}
\caption{\textbf{Parameters and main results.} (\textbf{A}) Sets of parameters used throughout the article in all of the schemes containing 2 shadow qubits. Whenever an interval of coherence times are given, the lowest and highest values correspond to the parameter sets '0.75' and '1.25', respectively. Note that a special set of parameters were used for Fig. \ref{fig:Simple_Corection} in order to better emphasize the different parts of the error correction process. (\textbf{B}) Overview of the simulation results presented throughout the text. $T_1$ and $T_2^*$ denote the characteristic timescales for the decay of the storage-fidelities of the state $\left| \; \uparrow \; \right>$ and $\frac{1}{\sqrt{2}} \left( \left| \downarrow \right> + \left| \uparrow \right> \right)$, respectively. Similarly, $Loss(T_1)$ and $Loss(T_2^*)$ denote the  initial loss of fidelity before error correction kicks in for this pair of benchmark-states. For details, see the App. \ref{sec:Methods}.}
\end{figure*}

Most of the simulations presented throughout this paper used the same set of parameters, namely the ones depicted in Fig. \ref{tbl:Parameters}. However, occasionally other parameters were used, either for pedagogical reasons or to indicate how the correction scales with the magnitude of the parameters.
Specifically, a separate set of parameters were used to generate the more schematic Fig. \ref{fig:Simple_Corection}, and scaled parameters were used to generate two of the three error-corrected graphs on Fig. \ref{fig:Parameter_Scaling}, as well as to provide the upper and lower bounds on the $T_1$ and $T_2^*$-intervals given for the alternative geometry in App. \ref{sec:Alternative_Driving}. The complete set of parameters used for 2-shadow schemes is depicted in Fig. \ref{tbl:Parameters_large}.
As mentioned in both the main text and App. \ref{sec:Alternative_Driving}, the precise values of these parameters do not play an essential role in the operation of the error correction scheme. As a result, only a limited number of considerations have gone into picking these parameters. The central guiding principle inherited from Eq. \eqref{eq:Order_of_Magnitude} was the requirement that
\begin{align*}
\Omega_S \ll J_x, J_z, 4 J_z \pm 2 J_x \ll A, \delta \; .
\end{align*}
Additionally, requiring the ability to combat realistic decoherence-rates sets a lower bound on the rate of error correction, and thus on the magnitude of $J_x$, $J_z$ and $A$. Combining these considerations with a bit of trial and error yielded the values of $\Omega_S$, $J_x$, $J_z$ and $A$  depicted in Fig. \ref{tbl:Parameters_large}. Since the dynamics within the subspace of error states is highly dependent on the ratio of $A$ and $\delta$, a small analytical investigation was performed in order to optimize this ratio. Specifically, a maximization of the magnitude of the correcting component picked up by the three states resulting from decay in the first, second or third qubit pair was performed. As a result of this investigation, the ratio of the two parameters was fixed at $0.179 \, \sqrt{2}$. Next, the value of $T_S$ was picked as the lowest value where further increases did not improve error correction performance any further when using the fixed interaction strengths already picked. Additionally, the rate of active correction was chosen sufficiently large that dephasing was kept in check. Thus we arrived at the parameters above. In fact, we specifically arrived at the 'Strength=0.5'-parameters, with the rest of the parameters being almost a direct rescaling of these parameters. The only deviation from a straight scaling was a small adjustment of the relative strength of the driving, i.e. of the size of $A$. The reason for this was that numerical investigations indicated that at low overall error correction strengths, the correction scheme needs as much help as it can get, and thus benefits from a stronger driving to make up for the weaker overall correction. In contrast, stronger overall error correction do not need a correspondingly stronger driving, and thus better performance can be achieved by making $A$ smaller so that the problematic second-order processes mentioned in App. \ref{sec:Active_EC} are better suppressed. Note a separate set of parameters ("1.0$\vphantom{1}^\diamond$") was used for Fig. \ref{fig:Simple_Corection} in order to better emphasize both the oscillatory dynamics and the role of the active correction.

The parameters used for the alternative driving scheme from App. \ref{sec:Alternative_Driving} are a direct copy of the ones used for the main scheme. Thus, it is possible that small improvements could be achieved for this scheme by adjusting the parameters, for instance by changing the value of $A$ or by altering the ratio of $A$ and $\delta$ fixed by analytical investigations of the standard scheme. Similarly, the parameters used for the single-shadow scheme introduced in App. \ref{sec:Single_Shadow} (Fig. \ref{tbl:Parameters_Single_Shadow}) were also adapted directly from the parameters of the 2-shadow schemes. In order to fix the new $B$-parameter, a small set of analytical and numerical investigations were performed and the strength of this interaction relative to $A$ was fixed. Additionally, the size of the parameter $A$ was adjusted slightly compared to the 2-shadow schemes since the overall weaker error correction of  the single shadow qubit meant a larger overall driving strength was needed in order to partially compensate this weakness. Note that the single-shadow scheme showed increased sensitivity to the size of $A$ compared to the 2-shadow schemes. To illustrate this effect and the tradeoffs that it can lead to, two sets of '1.0' parameters, identical except for the value of $A$, were tested (see Fig. \ref{tbl:Parameters_Single_Shadow}). The coherence times of both sets of parameters are depicted in Fig. \ref{tbl:Detailed_Results} along with a general overview of the rest of the results of this paper. 

\subsection{More on Methods}
\label{sec:Methods}
The method used for obtaining the data on coherence times and initial fidelity loss was as follows: First, the behaviour of the system was simulated using the Python package QuTiP~\cite{Johansson2012} for both of the two initial states $\left| \; \uparrow \; \right>$ and $\frac{1}{\sqrt{2}} \left( \left| \downarrow \right> + \left| \uparrow \right> \right)$. The Hamiltonian used in the simulation was either a sum of the terms in Eq. \eqref{eq:Hamiltonian},  \eqref{eq:Driving1} and \eqref{eq:Driving2} for the main scheme or the corresponding modified Hamiltonians for the alternative-geometry scheme and the single-shadow schemes. Photon-loss and phase-noise on the data-qubits and photon loss on the shadow qubits was modelled as white noise using the Lindblad collapse-operator methods built into the QuTiP framework. While this is a relatively good model of photon-loss noise in many types of superconducting qubits~\cite{Astafiev2004, Bylander2011, Kapit2016, Paladino2014, Yan2013}, it is not strictly speaking an accurate description of phase noise, which tends to be of the more correlated telegraph- or $1/f$-varieties~\cite{Astafiev2004,  Bylander2011, Kapit2016, Paladino2014, Anton2012, OMalley2015, Yan2016, Yoshihara2006}. The reasons why this noise model was chosen despite its shortcomings are two-fold. Firstly, we can note that most of the states connected by the phase-noise operator $\sigma^z$ (for instance $\left| \; + \; + \; + \right>$ and $\left| + \; + \; - \right>$) have the same energies. As a result, no Ramsey spin-echo-like effects are in play, and thus the difference between the results obtained from different types of noise spectra should be negligible. If anything, the suppressed nature of telegraph- and $1/f$-noise at higher frequencies should make these forms of noise easier to guard against than the white noise used for our simulations, and thus we can think of the results presented here as a worst case scenario for our error correcting schemes. Secondly, there is significant additional computational cost associated to simulating non-white noise spectra. As a result, doing $\SI{100}{\micro \second}$-scale simulations of the dynamics of the system under the influence of non-white noise turned out to be practically unfeasible.

To include active correction, the simulation was done in short intervals of length $T_A$, with an active error correction step at the end of each interval. The active correction was implemented by conjugating the density matrix output from the simulation by suitable operators. In other words, it was treated in an instantaneous idealized form rather than modelled as a realistic extended readout and correction process. This was mostly done for conceptual simplicity in order to better illustrate the performance of the code without also having to worry about details related to the readout procedure. After performing the active correction procedure, the simulation was restarted using the resulting density matrix as the initial state. During each simulation interval, a set of data points $\{ \left(t_i, F_{\text{storage}}(t_i) \right) \}$ describing the time evolution of the storage fidelity was recorded and stored for later use.

The simulation-process would continue in this start-stop fashion until the total time evolution reached $\SI{200}{\micro \second}$, at which point the it was terminated. Thus the output of the simulation was a large set of points chronicling the decay of the storage fidelity over a timespan of $\SI{200}{\micro \second}$ after a perfect initialization of the code--that is, sets of data points like the ones depicted on graphs throughout the paper. In order to extract the decoherence times and initial fidelity loss from these data, a fit of the form $f(t)=A \exp\left(- B t \right) + C$ was performed on the data in the time interval $\left[ \SI{10}{\micro \second} ; \SI{200}{\micro \second} \right]$. The reason for omitting the first \SI{10}{\micro \second} was to avoid the anomalous short-term behaviour that is not well approximated by the exponential decay model, and whose shape provides little information about the long-term performance of the code. The final step was to translate the fitting parameters $A$, $B$ and $C$ into corresponding decoherence times $T$ and initial fidelity losses $Loss\left(T\right)$ using the expressions: 
\begin{align*}
T &= \frac{1}{B}\\ 
Loss(T) &= 1-(A+C) = 1-f(0) \; .
\end{align*}
In this way, simulations of the storage of the state $\left| \; \uparrow \; \right>$ provided the values of $T_1$ and $Loss\left(T_1\right)$, and simulations of the storage of the state $\frac{1}{\sqrt{2}} \left( \left| \downarrow \right> + \left| \uparrow \right> \right)$ provided the values of $T_2^*$ and $Loss\left(T_2^*\right)$.

\clearpage

\section{Second Order Effects}
\label{sec:Second_Order_Effects}
While rotating-wave approximations allow for relatively simple and accurate explanations of most of the error correcting dynamics, the weaker higher-order effects will on long timescales begin to influence the dynamics as well. In this section, we look at how these higher-order effects influence the storage-performance of our scheme, and investigate how this influence can be suppressed using weak engineered interactions.

\subsection{Suppression of Second Order Transitions}
\label{sec:Second_Order_Suppression}
A central part of the analysis of the dynamics induced by the Hamiltonians of Eq. \eqref{eq:Hamiltonian} was the assumption that the term $H_{\text{pair}}$ was sufficiently strong that any dynamics that did not preserve the energy related to this term would be heavily suppressed due to rotating-wave-like effects, at least unless appropriate driving was supplied to compensate, as in the case of $H_{\text{d}}$. This energy-conservation constraint is of central importance in suppressing unwanted transitions that may otherwise be induced by the term $H_{\text{chain}}$. For instance, $H_{\text{chain}}$ is capable of inducing transitions of the form
\begin{align}
\left| \pm \, \pm \, \pm \right> \; \longleftrightarrow \; \begin{array}{c c}
\left| T \, T \, \pm \right>, & \left| \pm \; T \; T \right>\\
\left| S \, S \, \pm \right>, & \left| \pm \; S \; S \right> \; ,
\end{array}
\label{eq:FirstOrderTransitions}
\end{align}
which would lead to the corruption of the two logical states $\left| \pm \, \pm \, \pm \right>$ when combined with the error-correction mechanisms of our hQEC scheme. Only because the coupling of $2 \delta$ between these states is much smaller than the $4J_z \pm 2 J_x$ detuning resulting from $H_{\text{pair}}$ can we be certain that the transitions are suppressed, and thus certain that our information is not corrupted by the very same interactions responsible for part of the hQEC-scheme itself. However, while no significant population will occur in the states  on the right-hand side of \eqref{eq:FirstOrderTransitions} due to these energy-conservation arguments, population is still able to flow through them to other states. As a result, they are able to facilitate transitions that \emph{do} conserve the energy of $H_{\text{pair}}$. Specifically, a small investigation reveals that $H_{\text{chain}}$ also couple the states on the right hand side of \eqref{eq:FirstOrderTransitions} to another set of states, yielding a more complete coupling-diagram:
\begin{align}
\left| \pm \, \pm \, \pm \right> \; \longleftrightarrow \; \begin{array}{c c}
\left| T \, T \, \pm \right>, & \left| \pm \; T \; T \right>\\
\left| S \, S \, \pm \right>, & \left| \pm \; S \; S \right>
\end{array} \; \longleftrightarrow \;  \begin{array}{c c} 
\left| \pm \; \mp \; \mp \right> \\
 \left| \mp \; \mp \; \pm \right> 
\end{array} \; .
\label{eq:Second_Order_Full_Process}
\end{align}
The states on the left hand side and the states on the right hand side have the same $H_{\text{pair}}$-energy, and thus transitions between them are not forbidden by arguments related to conservation of energy. Of course, such a transition would have to happen through what is essentially a tunneling-event through the energy-barrier of the middle states, meaning the transition-rates from left to right will likely be heavily suppressed. Nevertheless, we will see below that these transitions are still able to influence the behaviour of our scheme on long timescales. This is problematic, because a transition from left to right followed by a measure-correct step would result in a logical bit-flip:
\begin{align*}
\left| \pm \, \pm \, \pm \right> \; \longrightarrow \; \left| \mp \, \mp \, \mp \right> \; ,
\end{align*}
and thus to a corruption of  the stored information, at least unless the system happens to be storing a state immune to this $\sigma^x_L$-like bit-flip noise.

\begin{figure*}[hbtp]
  \centering
\begin{subfigure}[b]{1.0\textwidth}
   \includegraphics[scale=0.55]{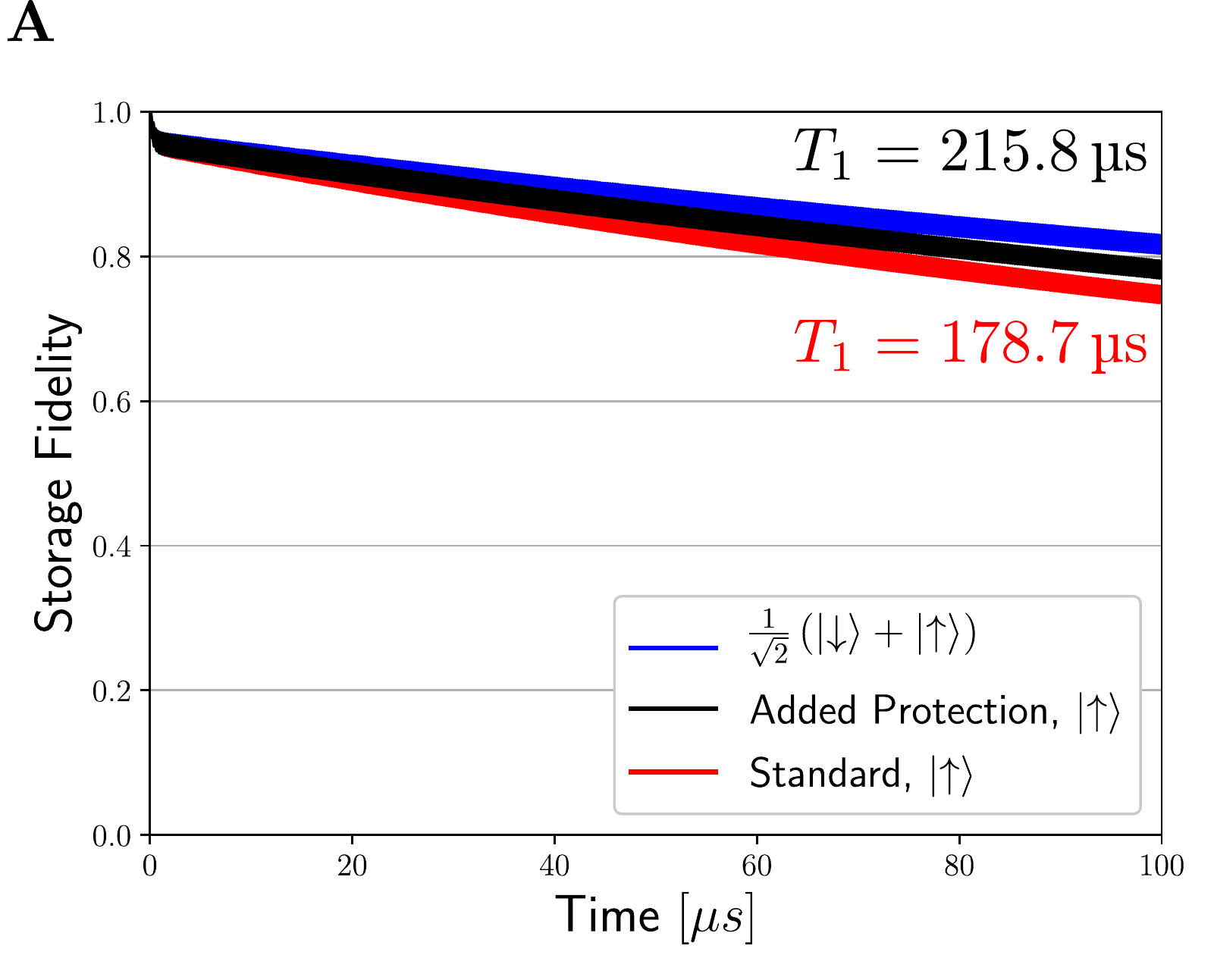}
  \captionlistentry{} \label{fig:Second_Order_Scheme_Graph}
\end{subfigure}\\ \vspace{1.4cm}
\begin{subfigure}{.49\textwidth}
   \includegraphics[scale=0.52]{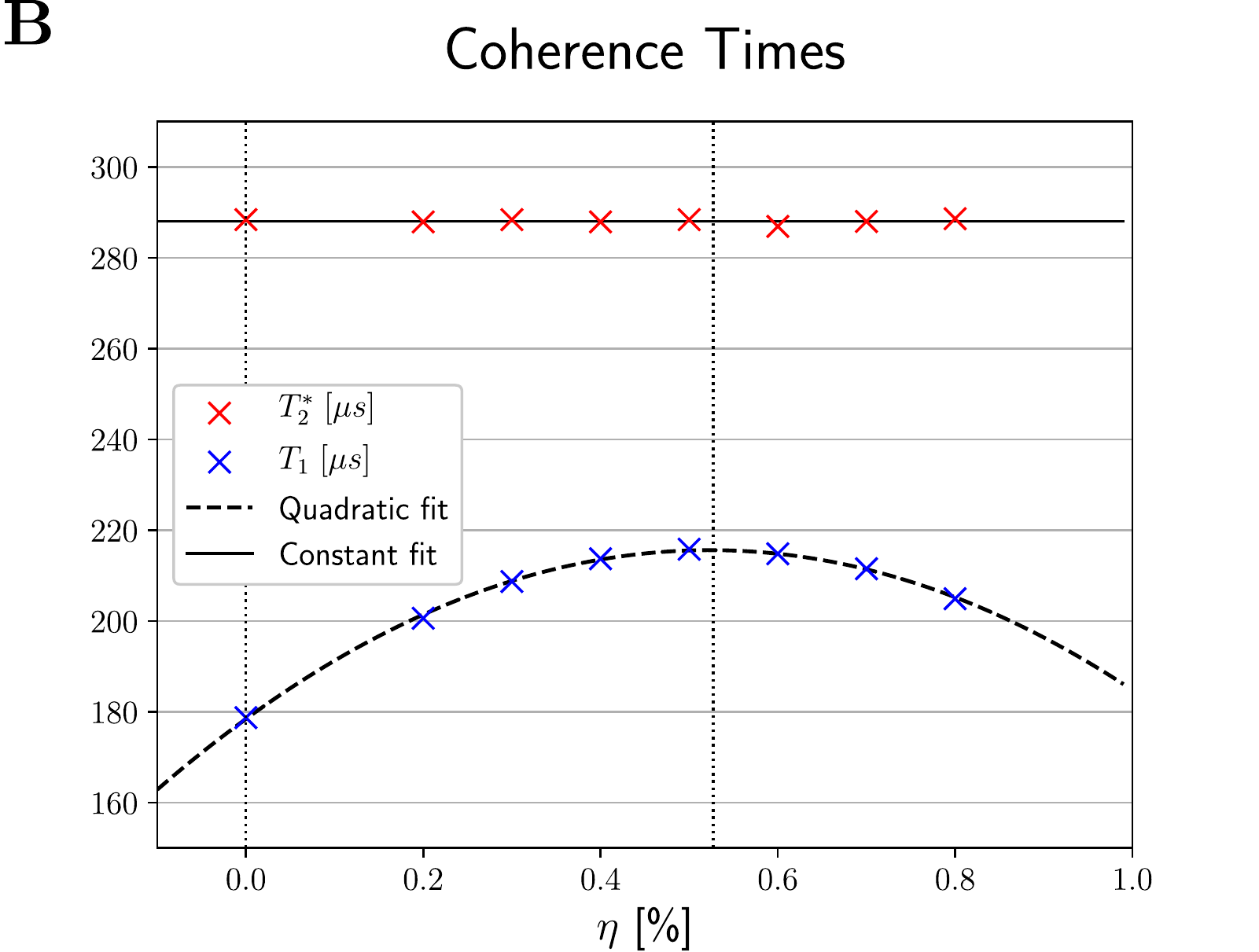}
        \captionlistentry{} \label{fig:Eta_Scaling}
\end{subfigure}%
\begin{subfigure}{.49\textwidth}
   \includegraphics[scale=0.52]{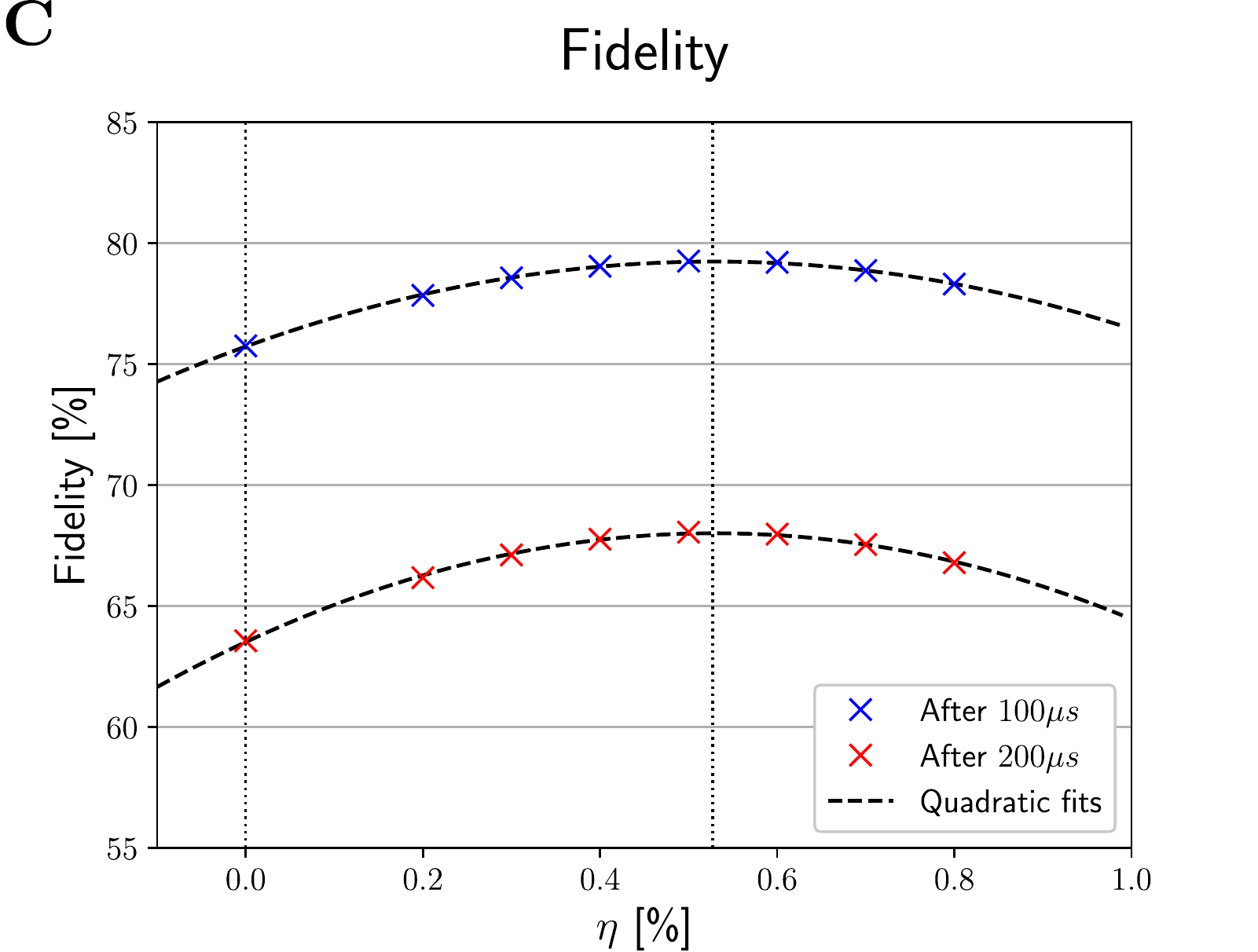}
     \captionlistentry{} \label{fig:Eta_Scaling_C}
\end{subfigure}
\caption{\textbf{Suppression of second-order transitions.} (\textbf{A}) Storage fidelity for the state $\left| \uparrow \right>$ as a function of time when suppression of second order effects detailed in App. \ref{sec:Second_Order_Suppression} is either enabled (black) or not enabled (red). For reference, the corresponding time-evolution of the storage fidelity for the state $\frac{1}{\sqrt{2}}\left( \left| \downarrow \right> + \left| \uparrow \right> \right)$ is plotted as well (blue). The parameters used in the simulations are the same as those used in the main text (See Fig. \ref{tbl:Parameters}), except in the case of the black line for the addition of the interaction \eqref{eq:ZZ_Sup_Hamiltonian} with a strength $\eta=0.005$. (\textbf{B}) Dependency of the coherence times $T_1$ and $T_2^*$ on the strength $\eta$ of the interactions suppressing second-order effects. Vertical lines mark both $\eta=0$ and the ideal value of $\eta$ as predicted by Eq. \eqref{eq:Ideal_Eta}. Additionally, the plots contain fits to the data of quadratic form. These fits are mostly to guide the eyes, though they were also used to extract a few key metrics used in the discussion at the end of App. \ref{sec:Second_Order_Suppression}. (\textbf{C}) Plot of the storage fidelity for the states  $\left| \uparrow \right>$ after $\SI{100}{\micro \second}$ or $\SI{200}{\micro \second}$ as a function of the strength of the second-order suppression scheme from App. \ref{sec:Second_Order_Suppression}. As in (B), $\eta$-values of zero and the value from Eq. \eqref{eq:Ideal_Eta} are marked, and simple fits to the data are added to guide the eyes.}
\label{fig:Second_Order_Suppression_Mega}
\end{figure*}

In order to investigate the magnitude of this problem, let us think of $H_{\text{pair}}$ as the dominant Hamiltonian of the system, and $H_{\text{chain}}$ as a small perturbation. In this way, we can use standard methods of perturbation-theory to extract estimates for transition-rates. Specifically, expanding the wavefunction in the eigenstates of $H_{\text{pair}}$:
\begin{align}
\left| \psi ( t) \right> &= \sum_a c_a(t) \left| \psi_a \right> \; ,
\label{eq:State_Expansion}
\end{align}
one can write the following perturbative expansion for the evolution of the coefficients~\cite{Bransden2003}:
\begin{align}
c_a(t) =\, c_a(0) + \frac{1}{i \hbar} \sum_b \int_0^t c_b(0) \, \left[H_{\text{chain}}\right]_{a,b} \, \exp\left(\frac{i E_{ab} t'}{\hbar} \right) dt' \nonumber\\
+ \frac{1}{(i\hbar)^2} \sum_b \sum_c \int_0^t \int_0^{t'}\left[ c_c(0) \, \left[H_{\text{chain}}\right]_{a,b} \left[H_{\text{chain}}\right]_{b,c} \vphantom{\exp\left(\frac{i E_{b,c} t''}{\hbar} \right)} \right. \nonumber\\
\left.  \times \exp\left(\frac{i E_{a,b} t'}{\hbar} \right)  \exp\left(\frac{i E_{b,c} t''}{\hbar} \right) \right] dt'' dt' \nonumber\\
+ \text{O}\left(H_{\text{chain}}^3\right)\; ,
\label{eq:Second_Order_Dynamics_Expansion}
\end{align}
where $\left[H_{\text{chain}}\right]_{a,b} = \left< \psi_a \right| H_{\text{chain}} \left| \psi_b \right>$ are matrix-elements of the perturbation in the unperturbed basis and $E_{a,b} = E_a - E_b$ the change of $H_{\text{pair}}$-energy related to the transition $\left| \psi_b \right> \rightarrow \left| \psi_a \right>$. Note that we have kept terms up to second order in our small perturbation $H_{\text{pair}}$ because the effect we are looking for is exactly a second order transition---It requires at least two applications of $H_{\text{pair}}$ to move from the left-hand side to the right-hand side in Eq. \eqref{eq:Second_Order_Full_Process}. On a related note, lets assume the indices of Eq. \eqref{eq:State_Expansion} are such that
\begin{align*}
\left| \psi_0 \right> &= \left| + \, + \, + \right>\\
\left| \psi_1 \right> &= \left| - \, - \, + \right> \, ,
\end{align*}
and that the initial state of the system is that we are fully in the $\left| \psi_0 \right>$-state, i.e. $c_a(0) = \delta_{a,0}$. The central question is then how quickly this population will tend to transfer to the error-state $\left| \psi_1 \right>$, i.e. how quickly $c_1(t)$ will tend to grow as time passes. Plugging the assumptions about the initial state into the perturbative expansion yields
\begin{align*}
c_1(t) =\,  \frac{1}{i \hbar} \int_0^t  \, \left[H_{\text{chain}}\right]_{1,0} \, \exp\left(\frac{i E_{1,0} t'}{\hbar} \right) dt'\\
+ \frac{1}{(i\hbar)^2} \sum_b  \int_0^t \int_0^{t'}\left[ \left[H_{\text{chain}}\right]_{1,b} \left[H_{\text{chain}}\right]_{b,0} \vphantom{\exp\left(\frac{i E_{b,0} t''}{\hbar} \right)} \right.\\
\left.  \times \exp\left(\frac{i E_{1,b} t'}{\hbar} \right)  \exp\left(\frac{i E_{b,0} t''}{\hbar} \right) \right] dt'' dt' \\
+ \text{O}\left(H_{\text{chain}}^3\right)\; .
\end{align*}
The two states $\left| \psi_0 \right>$ and $\left| \psi_1 \right>$ are not coupled directly, so $\left[H_{\text{chain}}\right]_{1,0} = 0$, and thus the first term vanishes. For the second term we get two nonzero contributions, namely $\left| \psi_b \right> = \left| S \, S \, + \right>$ and $\left| \psi_b \right> = \left| T \, T \, + \right>$. For both of these contributions, $\left[H_{\text{chain}}\right]_{1,b} \left[H_{\text{chain}}\right]_{b,0} = -4 \delta^2$, meaning \footnotesize
\begin{align*}
c_1(t) &\simeq -\frac{4\delta^2}{(i\hbar)^2} \sum_b  \int_0^t \int_0^{t'} \exp\left(\frac{i E_{1,b} t'}{\hbar} \right)  \exp\left(\frac{i E_{b,0} t''}{\hbar} \right)  dt'' dt' \\
&= -\frac{4\delta^2}{(i\hbar)^2} \sum_b  \int_0^t  \exp\left(\frac{i E_{1,b} t'}{\hbar} \right)  \frac{\hbar}{i E_{b,0}} \left[ \exp\left(\frac{i E_{b,0} t'}{\hbar} \right) - 1 \right]  dt' \\
&=  \frac{4\delta^2}{i\hbar} \sum_b \frac{1}{E_{b,0}}  \int_0^t  \left[ 1 - \exp\left(\frac{i E_{1,b} t'}{\hbar} \right) \right]  dt' \\
&= \frac{4\delta^2}{i\hbar} \sum_b \frac{1}{E_{b,0}}  \left[ t - \frac{\hbar}{i E_{1,b}} \left(\exp\left(\frac{i E_{1,b} t}{\hbar} \right)
- 1 \right) \right] \; ,
\end{align*} \normalsize
where we have used that $E_{1,b}+E_{b,0}=E_1-E_0=0$ because the initial and final states are degenerate with respect to $H_{\text{pair}}$. Defining 
\begin{align*}
\Delta_T &= 4 J_z - 2 J_x \\
\Delta_S &= 4 J_z + 2 J_x \; ,
\end{align*} 
we see that three terms emerge:
\begin{align*}
c_1(t) \simeq& \, -\frac{4\delta^2}{i\hbar} \left(\frac{1}{\Delta_T} + \frac{1}{\Delta_S} \right) t \\
&- \frac{8 i \delta^2}{ \Delta_T^2} \exp\left(\frac{i \Delta_T t}{2 \hbar} \right)  \sin \left(\frac{i \Delta_T t}{2 \hbar} \right)\\
&- \frac{8 i \delta^2}{ \Delta_S^2} \exp\left(\frac{i \Delta_S t}{2 \hbar} \right)  \sin \left(\frac{i \Delta_S t}{2 \hbar} \right)
\end{align*}
The last two terms represent rapid oscillations in the population of the error state. However, the prefactors are only on the order of $1\cdot 10^{-5}$ for the standard parameters from Fig. \ref{tbl:Parameters}. In comparison, the coefficient of the linear term is $\SI{33}{\kilo \hertz}$, leading to a more sizeable contribution of about $8.3 \cdot 10^{-3}$ during the time $T_A$. Of course, the probability of the syndrome measurements collapsing the system to the error state is quadratic in this amplitude, so only a population on the order of $10^{-3} \, \%$ is expected to be lost at each active correction step. However, over the course of $\SI{100}{\micro \second}$ we will have about 400 of these corrections, meaning the lost fidelity accumulates to about $2.8 \%$. Additionally, an identical analysis reveals a similar rate of error induced by the process
\begin{align*}
\left| + \, + \, + \right> \; \longrightarrow \; \left| + \, - \, - \right> \; .
\end{align*}
Thus in total we expect these second-order processes to be able to reduce the storage-fidelity over $\SI{100}{\micro \second}$ by about $5\%$. Note that a similar process corrupts $\left| - \, - \, - \right>$ in a completely symmetrical way. Indeed, as mentioned above, this rigid symmetry implies that the states $\frac{1}{\sqrt{2}} \left( \left| + \, + \, + \right> \pm \left| - \, - \, - \right> \right)$ should be unaffected by this error channel. In other words, these second order processes explain some of the reduced performance of storing $\left| \uparrow \right>_L$ compared to $\frac{1}{\sqrt{2}} \left( \left| \downarrow \right>_L + \left| \uparrow \right>_L \right)$, though App. \ref{sec:TA_Dependency} will reveal that there  are other factors at play in this discrepancy as well. Nevertheless, it would be nice if the detrimental second-order effects could be suppressed. As we will now see, one way to do this is to take advantage of the first-order term in Eq. \eqref{eq:Second_Order_Dynamics_Expansion}. In the calculations above, this term vanished due to the fact that there was no direct couplings between the initial and final states. However, adding such a direct coupling essentially allow us to cancel the second-order effects using a weak first-order effect. Specifically, we consider an additional term in the Hamiltonian of the form:
\begin{align}
H_{\text{sup}} = \eta \, \delta  \, \sum_{j=1}^2  \left( \sigma_{A,j}^z \sigma_{B,j}^z +  \sigma_{B,j}^z \sigma_{C,j}^z \right) \; ,
\label{eq:ZZ_Sup_Hamiltonian}
\end{align}
that is, we add a $\sigma^z \sigma^z$-style interaction to the chain-Hamiltonian, with relative strength $\eta$ compared to the original contributions. This induces a direct coupling between the initial and final states:
\begin{align*}
\left[ H_{\text{sup}} \right]_{0,1} = 2 \, \eta \, \delta \; .
\end{align*} 
Adding this term to our perturbation therefore yields the following extra first-order term to the analysis above
\begin{align*}
\frac{1}{i \hbar} \int_0^t  \, \left[H_{\text{sup}}\right]_{1,0} \, \exp\left(\frac{i E_{1,0} t'}{\hbar} \right) dt'= \frac{2 \eta \delta}{i \hbar} t \; .
\end{align*}
To second order no other changes occur as a result of the new term, and thus we get
\begin{align}
\label{eq:Pertubation_Result}
c_1(t) \simeq& \, \frac{2 \delta}{i \hbar} \left[  \eta - 2 \delta \left(\frac{1}{\Delta_T} + \frac{1}{\Delta_S} \right) \right] t\nonumber \\
&- \frac{4 i \delta^2}{ \Delta_T^2} \exp\left(\frac{i \Delta_T t}{2 \hbar} \right)  \sin \left(\frac{i \Delta_T t}{2 \hbar} \right)\\
&- \frac{4 i \delta^2}{ \Delta_S^2} \exp\left(\frac{i \Delta_S t}{2 \hbar} \right)  \sin \left(\frac{i \Delta_S t}{2 \hbar} \right) \; . \nonumber
\end{align}
From this we conclude that by setting
\begin{align}
\eta = 2 \delta \left(\frac{1}{\Delta_T} + \frac{1}{\Delta_S} \right) \; ,
\label{eq:Ideal_Eta}
\end{align}
we should be able to cancel the dominant second order effect. For the parameters used in the main text, this requires $\eta = 5.3 \cdot 10^{-3}$. In other words, the new term needs to be very weak compared to the original interactions in $H_{\text{chain}}$ . This is not surprising, since our new interaction is a direct and unsuppressed coupling while the interactions from $H_{\text{chain}}$ that it needs to cancel only arise as highly suppressed second-order effects. Note that for implementations such as superconducting circuits, small couplings of the $\sigma^z \sigma^z$-variety may in fact be unavoidable~\cite{Zhang2018}, meaning extra complexity is not necessarily incurred by adding $H_{\text{sup}}$ to the model, except perhaps through the complexity related to controlling the relative strength compared to $H_{\text{chain}}$. However, proposals for implementations where this relative strength is tuneable is possible already exist~\cite{Kounalakis2018,Geller2015}. Additionally, we see from Eq. \eqref{eq:Pertubation_Result} and Fig. \ref{fig:Eta_Scaling} and \ref{fig:Eta_Scaling_C} that any value of $\eta$ between $0$ and $1\%$ will lead to a reduced rate of second order transitions, meaning limited fine-tuning is required. We are therefore hopeful that the second-order suppression scheme presented here may be implementable without too much overhead.

Of course, before any such efforts are undertaken, we need to  sure that the suppression-scheme actually works and leads to the expected improvements in performance. Simulation-data showing the effect of $\eta=5.0 \cdot 10^{-3}$ when otherwise using the parameters from the main text is depicted on Fig. \ref{fig:Second_Order_Scheme_Graph}. As can be seen from this figure, our suppression scheme indeed manages to improve the storage-performance for the state $\left| \uparrow \right>_L$, bringing it closer to the performance related to the state $\frac{1}{\sqrt{2}} \left( \left| \downarrow \right> + \left| \uparrow \right> \right)$. Specifically, the $T_1$ is increased from $178.7 \pm \SI{0.2}{\micro \second}$ to $215.8 \pm \SI{0.4}{\micro \second}$.

To further illustrate the validity of the considerations above, further simulations were performed using different values of $\eta$ between $0$ and $1\%$. The coherence times and final storage-fidelities extracted from these simulations are depicted on Fig. \ref{fig:Eta_Scaling} and \ref{fig:Eta_Scaling_C}. From these figures, we see that the value of $\eta$ calculated above using perturbation-theory very closely matches the value resulting in the best storage-performance. In fact, quadratic fits to the fidelities yield maximas at $\eta = 0.529 \pm 0.002 \%$ and $\eta = 0.530 \pm 0.004 \%$, while the coherence times seems to have a maximum at $\eta=0.524 \pm 0.003 $---all values in very good agreement with the perturbation-theory value of $\eta=0.527$. Additionally, the simulations allow us to estimate improvement to storage-fidelity and coherence time achievable for the default parameters when applying the second-order suppression scheme. Specifically, we see that $T_1$ can be improved by $37.2 \pm \SI{0.8}{\micro \second}$ without altering the $T_2^*$-time at all, and that the storage fidelity of the state $\left| \uparrow \right>$ after $\SI{100}{\micro \second}$ can be improved by $3.51 \pm 0.06 \%$, which is close to the $\sim 5 \%$ improvement we were hoping to be able to achieve by correcting for second-order processes.

\subsection{AC Stark Shifts}
\label{sec:AC_Stark_Shifts}
As explained in \ref{sec:Details_on_Main_Scheme}, the effects of the driving are constrained through detuning due to the fact that the scale of the pairwise interaction Hamiltonian $H_{\text{pair}}$ is much larger than the scale of the driving Hamiltonian $H_{d}$, i.e. $A\ll J_x, J_z$. However, as illustrated in App. \ref{sec:Second_Order_Suppression} above, this suppression works best at suppressing first order transitions. Indeed, as illustrated in this section, combining two first order transitions into a second order effect allowed us to start and end in states that are not detuned from each other, and thus to escape the effects of detuning at the cost of a significantly reduced coupling strength. While App. \ref{sec:Second_Order_Suppression} focused on second order processes with differing start- and end-points, i.e. transitions, it is also possible for second order processes to start and end at the same state. In this case, the result is a shift in the energy of the corresponding state, known as the AC Stark shift. In the case of our encoding into the states $\left| \pm \right>$, such a shift could be problematic. After all, our driving Hamiltonians leave states of the form $\left| \uparrow \; \uparrow \right> \left| \downarrow \right>$ completely untouched:
\begin{align*}
H_d (t) \left| \uparrow \; \uparrow \right> \left| \downarrow \right> &= 2 A \left( \sigma_2^+ \sigma_S^z + \sigma_2^- \sigma_S^- \right) \cos \left( \omega t \right) \left| \uparrow \; \uparrow \right> \left| \downarrow \right>\\
&= 0 \; ,
\end{align*}
while it does induce couplings for the state $\left| \downarrow \; \downarrow \right> \left| \downarrow \right>$: 
\begin{align*}
H_d (t) \left| \downarrow \; \downarrow \right> \left| \downarrow \right> &= 2 A \left( \sigma_2^+ \sigma_S^z + \sigma_2^- \sigma_S^- \right) \cos \left( \omega t \right) \left| \downarrow \; \downarrow \right> \left| \downarrow \right>\\
&= \sqrt{2} A \cos \left( \omega t \right) \left( \left| T \right> - \left| S \right> \right) \left| \uparrow \right>
\end{align*}
In other words, the $\left| \downarrow \; \downarrow \right> \left| \downarrow \right>$-state is susceptible to AC-Stark shifts while the state $\left| \uparrow \; \uparrow \right> \left| \downarrow \right>$ is not, leading to differing energies of these states and thus to oscillations between the two states
\begin{align*}
\left| \pm \right>\left| \downarrow \right> &= \frac{1}{\sqrt{2}}\left( \left| \uparrow \; \uparrow \right> \pm \left| \downarrow \; \downarrow \right> \right) \left| \downarrow \right> \; .
\end{align*}
Since flipping these signs is equivalent to a phase error, the effect of the AC-Stark shift will be to introduce an effective additional contribution to the phase noise of the driven qubit pairs, i.e. to the two pairs at the ends of the 3-pair chain of the scheme presented in the main text.

In order to determine the severity of this problem, the magnitude of the Stark shift is estimated using second order pertubation theory. Specifically, consider a system that evolves under a Hamiltonian of the form
\begin{align*}
H(t) = H_0 + \tilde{H} \cos\left(\omega t \right) \; ,
\end{align*}
with $H_0$ and $\tilde{H}$ time-independent and the magnitude of the driving-term $\tilde{H}$ small compared to the magnitude of $H_0$. Additionally, consider an eigenstate $\left| \psi_a \right>$ of $H_0$ with the property that the driving-term only couples it off-resonantly to the other $H_0$-eigenstates $\left\{\left| \psi_k \right>\right\}_k$, and where $\left< \psi_a \right| \tilde{H} \left| \psi_a \right>=0$. The Stark shift of such a state as a result of the driving term can then be written on the form~\cite{Bransden2003}
\begin{align*}
\Delta E_a &= - \frac{1}{2} \sum_{k \neq a} \left| \left< \psi_k \right| \tilde{H} \left| \psi_a \right> \right|^2 \frac{E_k-E_a}{\left(E_k-E_a\right)²-\hbar^2 \omega^2} \; ,
\end{align*} 
where $E_k$ is the unperturbed energy of the state $\left| \psi_k \right>$, i.e.
\begin{align*}
H_0 \left| \psi_k \right> = E_k \left| \psi_k \right> \; .
\end{align*}
Applying this expression to a system consisting of a qubit-pair coupled to a shadow qubit through the Hamiltonians
\begin{align*}
H_0 &= J_x \left( \sigma_1^x \sigma_2^x + \sigma_1^y \sigma_2^y \right) + J_z \sigma_1^z \sigma_2^z + \Omega_S \sigma_S^z\\
\tilde{H} &= 2 A \left( \sigma_2^+ \sigma_S^z + \sigma_2^- \sigma_S^- \right)\\
\omega_{\pm} &= \frac{\Omega_S + 2 J_z \pm J_x }{\hbar} \; ,
\end{align*}
i.e. considering one of the end qubit-pairs in isolation, gives a shift of the state $\left| \downarrow \; \downarrow \right> \left| \downarrow \right>$ of
\begin{align*}
\Delta E_{\pm} &= \frac{A^2}{4} \left( \frac{\Omega_S - 2 J_z \pm J_x}{2 J_z \left( \Omega_S \pm J_x \right)} + \frac{\Omega_S - 2 J_z \mp J_x}{\Omega_S \left( 2 J_z \pm J_x \right)}\right) \; .
\end{align*}
Plugging in the parameter-values of Fig. \ref{tbl:Parameters}, this gives
\begin{align*}
\Delta E_+ &= \SI{4.4}{\kilo \hertz} \times 2 \pi \hbar &  \Delta E_- &= \SI{5.9}{\kilo \hertz} \times 2 \pi \hbar \; ,
\end{align*}
corresponding to oscillations in the $\left| \pm \right>$-subspace with periods of
\begin{align*}
T_+ &= \SI{230}{\micro \second} & T_+ &= \SI{170}{\micro \second}
\end{align*}
depending on whether the $S$- or $T$-correcting driving is applied. These predictions are borne out by simulations of the system.

To give an estimate of how severe an effect such oscillations have on our error correction scheme, lets consider the probability that a state originally in the state $\left| + \right> \left| \downarrow \right>$ is still in this state after a time $t$. Simple quantum mechanics predicts that this is given by
\begin{align}
P_\pm(t) = \left| \cos \left( \pi \frac{t}{T_{\pm}} \right) \right|^2 \; .
\label{eq:UnitaryDephasing}
\end{align}
The reasonable thing to compare this effect to is the dephasing effects due to noise. These would reduce the probability as
\begin{align}
P(t) = \frac{1}{2} \left( 1 + e^{-\frac{t}{T}} \right) \; .
\label{eq:NoiseDephasing}
\end{align}
with the dephasing from the simulations in the main text happening on the timescale $T=\SI{20}{\micro \second}$. Thus by comparing equations \eqref{eq:UnitaryDephasing} and \eqref{eq:NoiseDephasing}, we can extract an effective dephasing timescale $T_{\text{eff}}$:
\begin{align*}
\frac{1}{2} \left( 1 + e^{-\frac{t}{T_{\text{eff},\pm}}} \right) &= \, \left| \cos \left( \pi \frac{t}{T_{\pm}} \right) \right|^2\\
\Longrightarrow T_{\text{eff},\pm} &= \frac{-t}{\log\left(2 \left| \cos \left( \pi \frac{t}{T_{\pm}} \right) \right|^2 - 1 \right)} \; .
\end{align*}
The interpretation of this effective timescale is that it is the timescale that phase-noise would need to have in order to be able to corrupt the stored $\left|+\right>$-state as much as the unitary dynamics do given the same amount of time $t$. The relevant timescale in which both noise and unitary dynamics are allowed to run rampant must be the time between active correction steps, i.e. $t=\SI{0.25}{\micro \second}$. Plugging this into the expressions above yields an effective timescale of the Stark-shift induced dephasing of
\begin{align*}
T_{\text{eff},+} &= \SI{11}{\milli \second} & T_{\text{eff},-} &= \SI{5.9}{\milli \second} \; .
\end{align*}
Thus, the dephasing due to the Stark effect happens on a timescale that is almost 3 orders of magnitude longer than the dephasing resulting from the noise affecting the individual qubits. In other words, it does not add a significant contribution to the phase-errors that our code needs to correct, and can to good approximation simply be ignored.

\clearpage

\section{Scaling of Performance with Key Parameters}
\label{sec:Scalings_with_Ts}
While App. \ref{sec:Main_Results} provided information on how the performance of the scheme scales with the overall strength of the applied error correction, it did not comment on how the performance of the scheme depends on a pair of parameters of key practical importance, namely the coherence times of the underlying physical qubits and the frequency with which the active correction-steps can be applied. In this section, these dependencies are investigated, resulting in estimates for the tolerances of the scheme with respect to these parameters, as well as some further insights into the dominant residual errors affecting different logical states.

\subsection{Dependency on $T_A$}
\label{sec:TA_Dependency}
A central question in schemes that rely on repeated measurements is how often these measurements need to be performed in order for the scheme to work. After all, there is a limit to how often it is feasible to perform measurements in many architectures---Indeed, the difficulty of performing frequent measurements was a central part of the motivation for using autonomous correction-methods in the first place. A reasonable question to ask is therefore how the lifetime of encoded information scales when the time $T_A$ between the measure-and-correct steps is varied. Additionally, it tuns out that answering this question will also uncover some surprising properties of the scheme, as well as help elucidate why the operator $g_0=(Z)(ZZZZZZ)(Z)$ is preferable to the simpler operator $g_0=(I)(ZZZZZZ)(I)$.

\begin{figure*}[hbtp]
\renewcommand*{\arraystretch}{1.3}
  \centering
\begin{subfigure}{.50\textwidth}
   \includegraphics[scale=0.5]{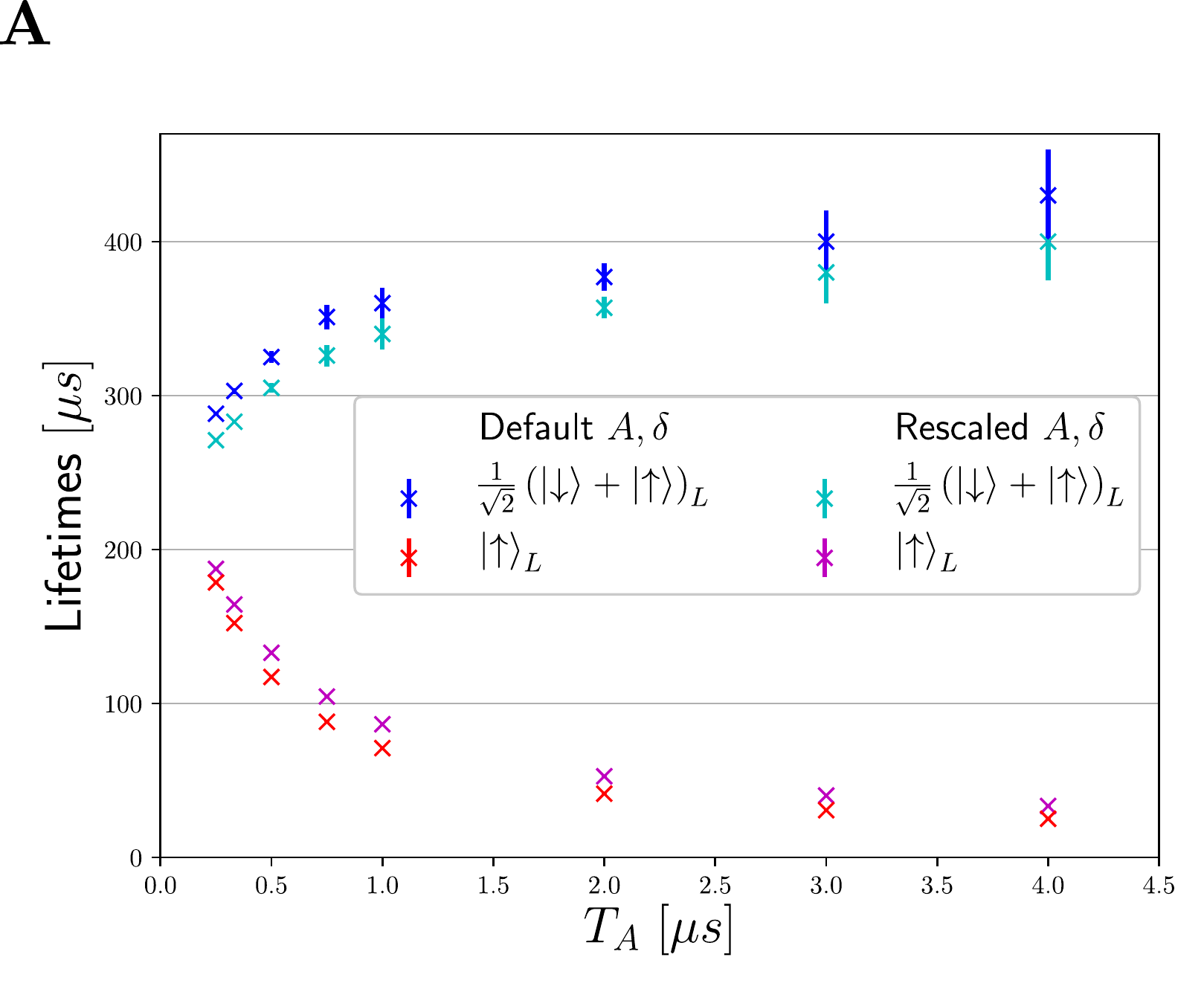}
     \captionlistentry{} \label{fig:TA_Dependency}
\end{subfigure}%
\begin{subfigure}{.50\textwidth}
   \includegraphics[scale=0.5]{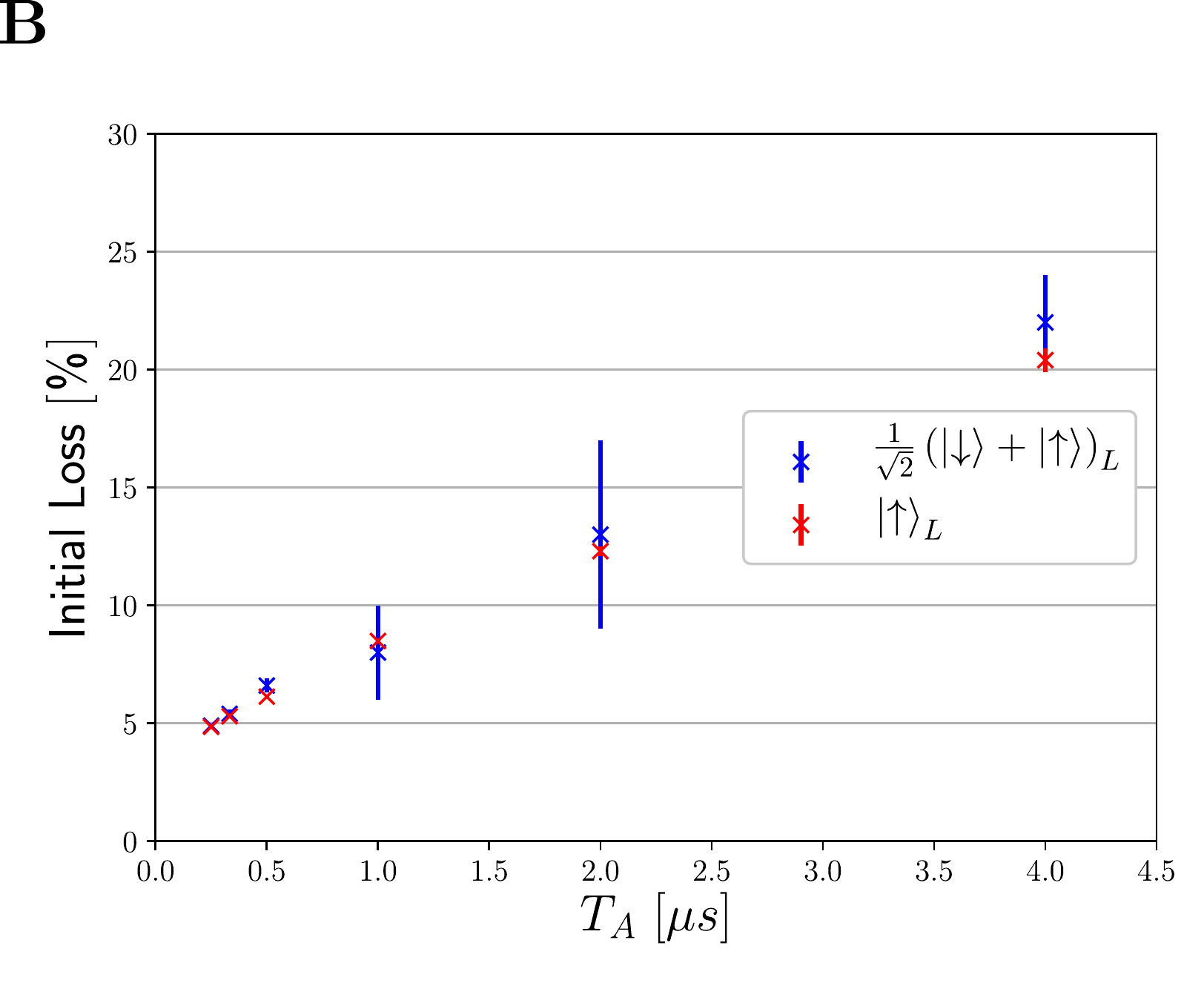}
   \captionlistentry{} \label{fig:TA_Dependency_2}
\end{subfigure}  \\ \vspace{0.4cm}
\begin{subfigure}[b]{1.0\textwidth}
   \includegraphics[scale=0.6]{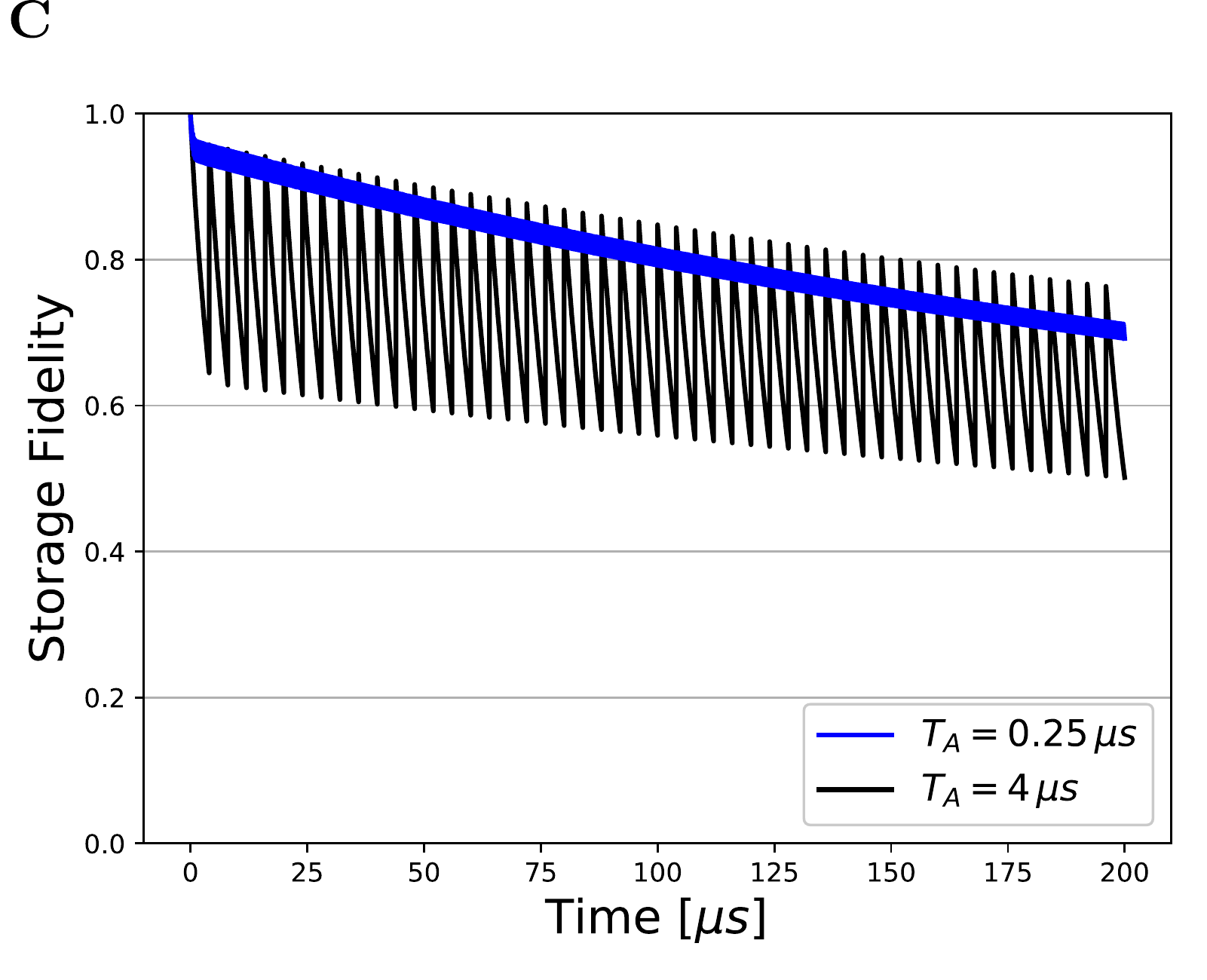}
  \captionlistentry{}  \label{fig:Long_v_Short_TA}
\end{subfigure}\\ \vspace{0.1cm}
\begin{subfigure}[b]{1.0\textwidth}
   \includegraphics[scale=0.9]{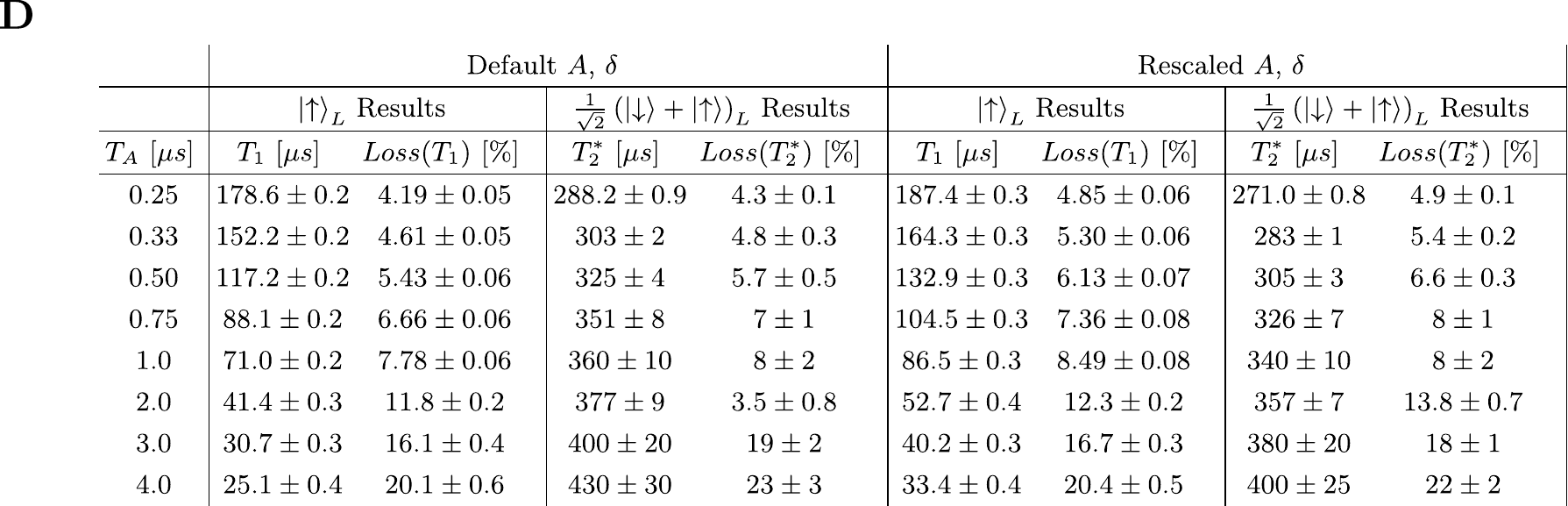}
  \captionlistentry{} \label{tbl:TA_Dependency}
\end{subfigure}\\ \vspace{0.4cm}
\caption{\textbf{Performance-scaling with frequency of active correction.} (\textbf{A}) Characteristic timescales for the storage of the states $\left| \uparrow \right>$ and $\frac{1}{\sqrt{2}} \left( \left| \uparrow \right> + \left| \downarrow \right> \right)$ as a function of the time $T_A$ between active correction steps. Data for two different parameters sets are used, corresponding to the default parameters used in the main text ("Default $A,\delta$") and parameters where the parameters $A$ and $\delta$ have been rescaled to $75\%$ of their default value in order to improve storage of the $\left| \uparrow \right>$-state. (See Fig. \ref{tbl:Parameters_large} and (C) for further details). (\textbf{B}) The initial fidelity-loss as a function of $T_A$ for the same states as parameters as those used in (A). (\textbf{C}) Comparison of the storage-fidelity as a function of time when the state $\frac{1}{\sqrt{2}} \left( \left| \downarrow \right> + \left| \uparrow \right> \right)_L$ is stored using either two different values of $T_A$. (\textbf{D}) Data used to generate (A) and (B).}
\label{fig:TA_Megafig}
\end{figure*}

To investigate the questions above, simulations similar to those presented in the main text were performed for different values of $T_A$, with all other parameters fixed to the values given in Fig. \ref{tbl:Parameters} except for a rescaling of $A$ by $75\%$ to improve storage of $\left| \uparrow \right>_L$ (see Fig. \ref{tbl:TA_Dependency}). From these simulations, the decay lifetimes and the initial fidelity-loss was then extracted using the methods explained in App. \ref{sec:Methods}. The main results from this process are given in Fig \ref{fig:TA_Megafig}. From these data we ascertain two things. Firstly, we see that the storage-performance for the state $\left| \uparrow \right>_L$ decays as you increase the time between corrections and thus allow more and more errors to accumulate. In other words, successfully storing this state requires corrections to happen as frequently as possible. Surprisingly, however, the opposite trend seems to be in effect for the state $\frac{1}{\sqrt{2}} \left( \left| \downarrow \right> + \left| \uparrow \right> \right)_L$. For this state, the coherence time seems to increase the longer you wait between active correction-steps. Of course, the escalating initial loss of fidelity during the first few microseconds probably renders this effect impractical to use (see Fig. \ref{fig:Long_v_Short_TA}), but from a fundamental standpoint it seems  odd that less frequent correction should result in the accumulation of fewer errors instead of more. The explanation for this effect seems to require two insights. Firstly, consider the state $\frac{1}{\sqrt{2}} \left( \left| \downarrow \right> + \left| \uparrow \right> \right)_L$:
\begin{align*}
\frac{1}{\sqrt{2}} \left( \left| \downarrow \right> + \left| \uparrow \right> \right)_L = \frac{1}{\sqrt{2}} \left( \left| + \, + \, + \right> + \left| - \, - \, - \right> \right) \; .
\end{align*}
In the event of photon-losses, the autonomous dynamics would partially remove this error and partially convert it into phase-errors, i.e. it would introduce components of the form:
\begin{align}
\frac{1}{\sqrt{2}} \left( \left| - \, + \, + \right> + \left| + \, - \, - \right> \right) \nonumber\\
\frac{1}{\sqrt{2}} \left( \left| + \, + \, - \right> + \left| - \, - \, + \right> \right) \; .
\label{eq:Plus_Error_States_1}
\end{align}
If we assume that active corrections are rare, all three of these components would be allowed to survive for a significant amount of time, and thus further errors could occur. Another photon loss followed by a correction would then introduce a new component
\begin{align}
\frac{1}{\sqrt{2}} \left( \left| - \, + \, - \right> + \left| + \, - \, + \right> \right)
\label{eq:Plus_Error_States_2}
\end{align}
as well as shuffle the population around among the three other components. Similarly, phase errors would also shuffle between these four states. In fact, the only way to exit the subspace spanned by the four states would be to experience multiple decays within the $\sim \SI{1.5}{\micro \second}$ time-frame that it takes for the autonomous correction to operate. As long as this does not happen, the large degree of symmetry in how the autonomous correction treats $\left|+ \right>$ and $\left|- \right>$ guarantees that we stay within the subspace. But looking at each of the four contributing states, we see that they are all corrected back to $\frac{1}{\sqrt{2}} \left( \left| \downarrow \right> + \left| \uparrow \right> \right)_L$ by the measure-correct step detailed in the main text. In other words, allowing a long time to pass between active corrections will result in more population being picked up by the states in \eqref{eq:Plus_Error_States_1} and \eqref{eq:Plus_Error_States_2}, but this will not matter---As soon as we actually do the active correction step, this population will all be corrected back into the original, errorless state anyway. This explains how both $\frac{1}{\sqrt{2}} \left( \left| \downarrow \right> + \left| \uparrow \right> \right)_L$ and $\frac{1}{\sqrt{2}} \left( \left| \downarrow \right> - \left| \uparrow \right> \right)_L$ tend to be immune to detrimental effects of long $T_A$-times. Contrast this with the state
\begin{align*}
\left| \uparrow \right>_L = \left| - \, - \, - \right>
\end{align*} 
which will tend to pick up components of the form 
\begin{align}
\left| + \, - \, - \right> & & \left| - \, + \, - \right> & & \left| - \, - \, + \right>
\label{eq:One_Error_States_1}
\end{align}
from a single error, and components  of the form
\begin{align}
\left| + \, + \, - \right> & & \left| - \, + \, + \right> \nonumber \\
\left| + \, - \, + \right> & & \left| + \, + \, + \right> 
\label{eq:One_Error_States_2}
\end{align}
from multiple errors. Of these components, only the ones in \eqref{eq:One_Error_States_1} are corrected back to the initial state by an active correction step, while the others would instead be mapped to the state $\left| \downarrow \right>_L = \left| + \, + \, + \right>$. In other words, the active correction needs to be applied before a second error can occur, otherwise the active correction inadvertently introduce a $\sigma_L^x$-operation on the state $\left| \uparrow \right>_L$. As a result, quick correction, and thus low $T_A$, is highly beneficial to the storage of this state.

A convenient way to summarize the results above is to think of the residual errors of the logical qubit as consisting mainly of $\sigma^x_L$-like noise which can be suppressed by short $T_A$'s, or which could alternatively be dealt with by concatenating the hybrid-code with a bit-flip code like the 3-qubit repetition code~\cite{Shor1995,Nielsen2002}. From this and the more detailed considerations above, it is no longer surprising that the coherence time of $\frac{1}{\sqrt{2}} \left( \left| \downarrow \right> + \left| \uparrow \right> \right)_L$ is not adversely affected by long $T_A$-times while the coherence time of $\left| 1 \right>_L$ is. However, what we observe from Fig. \ref{fig:TA_Dependency} is that storage of $\frac{1}{\sqrt{2}} \left( \left| \downarrow \right> + \left| \uparrow \right> \right)_L$ is not just unaffected by large values of $T_A$, but actually seems to benefit from them. To see why this is the case, we first note that the problem persists even if the correction-step is switched off, indicating that the detrimental effect to the performance stems from the measurement-part of the measure-correct steps. This indicates that the decreased performance for low values of $T_A$ may be the result of quantum-Zeno effects interfering with the autonomous correction~\cite{Itano1990}. To investigate this possibility further, a set of simulation of how the system behaves when starting in the error-state $1/\sqrt{2} \left( \left| T \, + \, + \right> + \left| T \, - \, - \right> \right)$ was run. To simplify the interpretation of the dynamics, all other noise-channels than the ones related to the shadow qubits were switched off. Since the property that we are interested in investigating is how the autonomous correction is influenced by measurements related to the active error correction, so a reasonable quantity to look at is the population of the error states, i.e. the subspace spanned by the states
\begin{align}
\label{eq:Zeno_Errors}
\frac{1}{\sqrt{2}} \left| 0 \right> \left( \left| T \, + \, + \right> + \left| T \, - \, - \right> \right)\left| 0 \right> \nonumber\\
\frac{1}{\sqrt{2}} \left| 0 \right>\left( \left| + \, T \, + \right> + \left| - \, T \, - \right> \right)\left| 0 \right>  & & \text{Loss errors} \\
\frac{1}{\sqrt{2}} \left| 0 \right>\left( \left| + \, + \, T \right> + \left| - \, - \, T \right> \right)\left| 0 \right> \nonumber \\
\frac{1}{\sqrt{2}} \left| 0 \right> \left( \left| + \, + \, 11 \right> + \left| - \, - \, 11 \right> \right)\left| 1 \right> \; , \nonumber
\end{align}
as well as the populations of the corrected state and the state where the $T$-error has been converted to a phase-error:
\begin{align}
\label{eq:Zeno_Phase_Errors}
\frac{1}{\sqrt{2}} \left| 0 \right>\left( \left| + \, + \, + \right> + \left| - \, - \, - \right> \right)\left| 0 \right> & & \text{Corrected} \\
\frac{1}{\sqrt{2}} \left| 0 \right>\left( \left| + \, + \, - \right> + \left| - \, - \, + \right> \right)\left| 0 \right> & & \text{Phase error} \label{eq:Zeno_Corrected}
\end{align}

\begin{figure*}[hbtp]
  \centering
\begin{subfigure}{.5\textwidth}
   \includegraphics[scale=0.5]{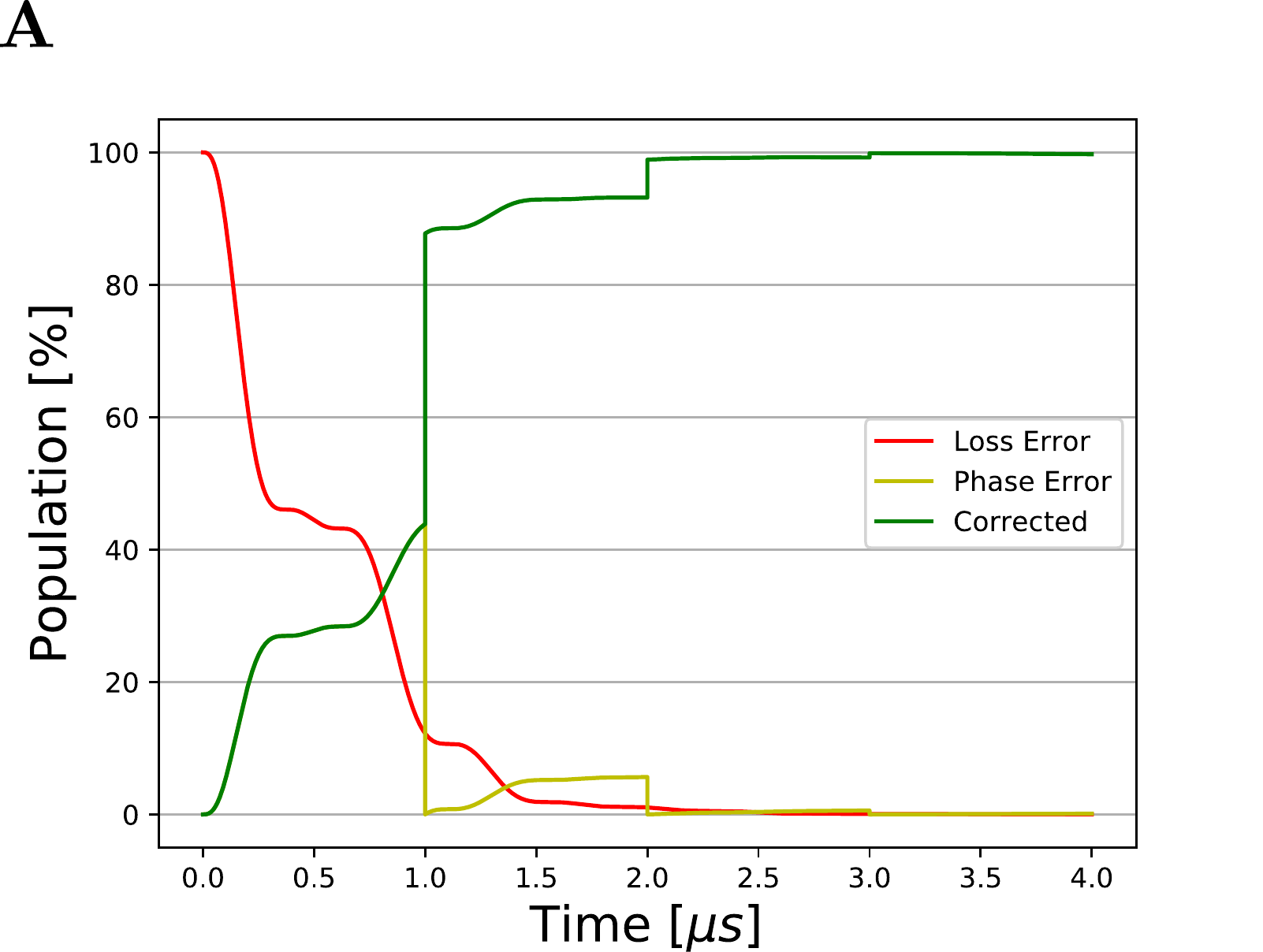}
     \captionlistentry{} \label{fig:Non_Zeno_Dynamics_Example}
\end{subfigure}%
\begin{subfigure}{.5\textwidth}
   \includegraphics[scale=0.5]{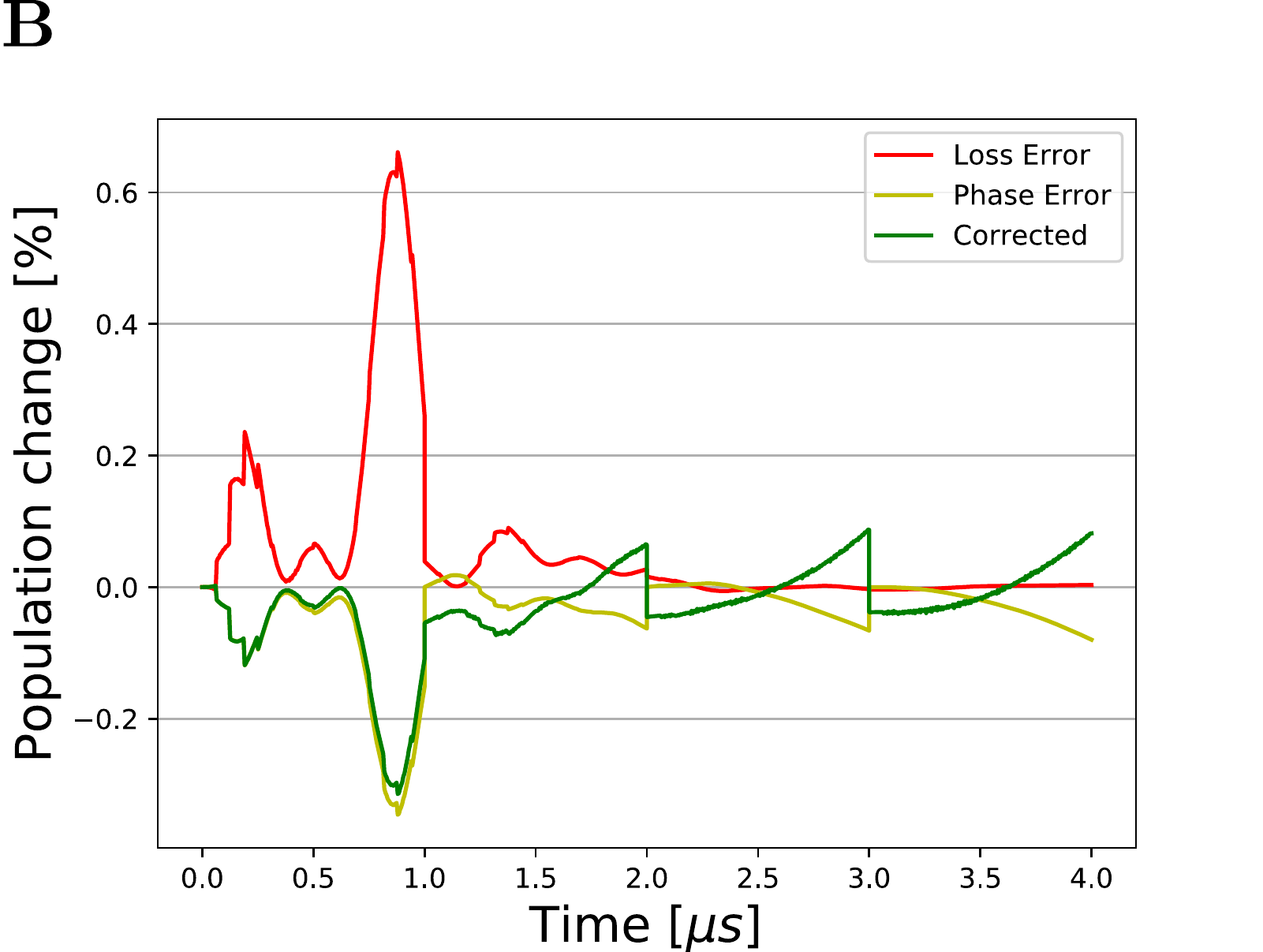}
   \captionlistentry{} \label{fig:Zeno_Changes}
\end{subfigure} \\ \vspace{0.8cm}
\begin{subfigure}{.5\textwidth}
   \includegraphics[scale=0.5]{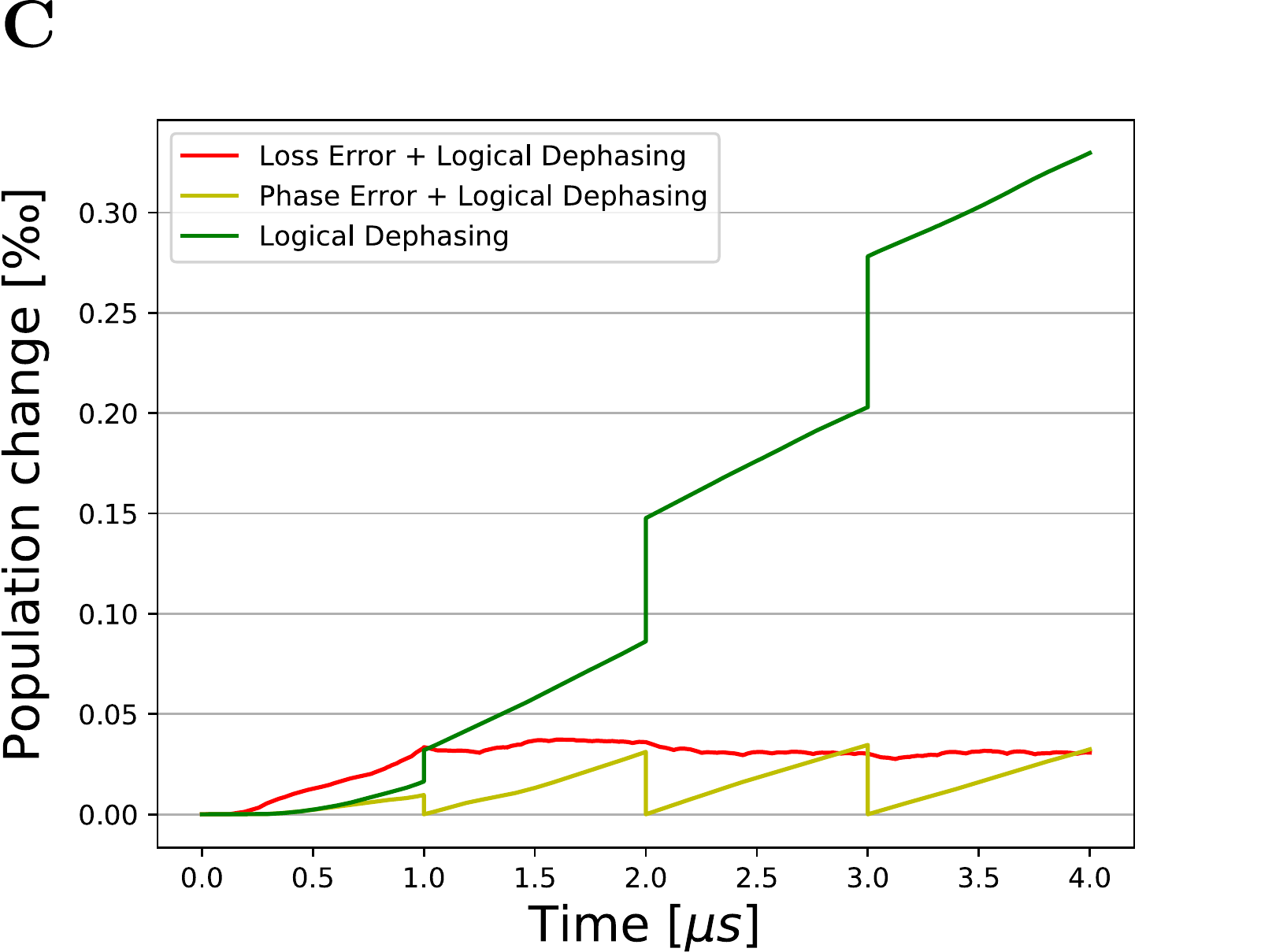}
     \captionlistentry{} \label{fig:Zeno_Dephasing}
\end{subfigure}%
\begin{subfigure}{.5\textwidth}
   \includegraphics[scale=0.5]{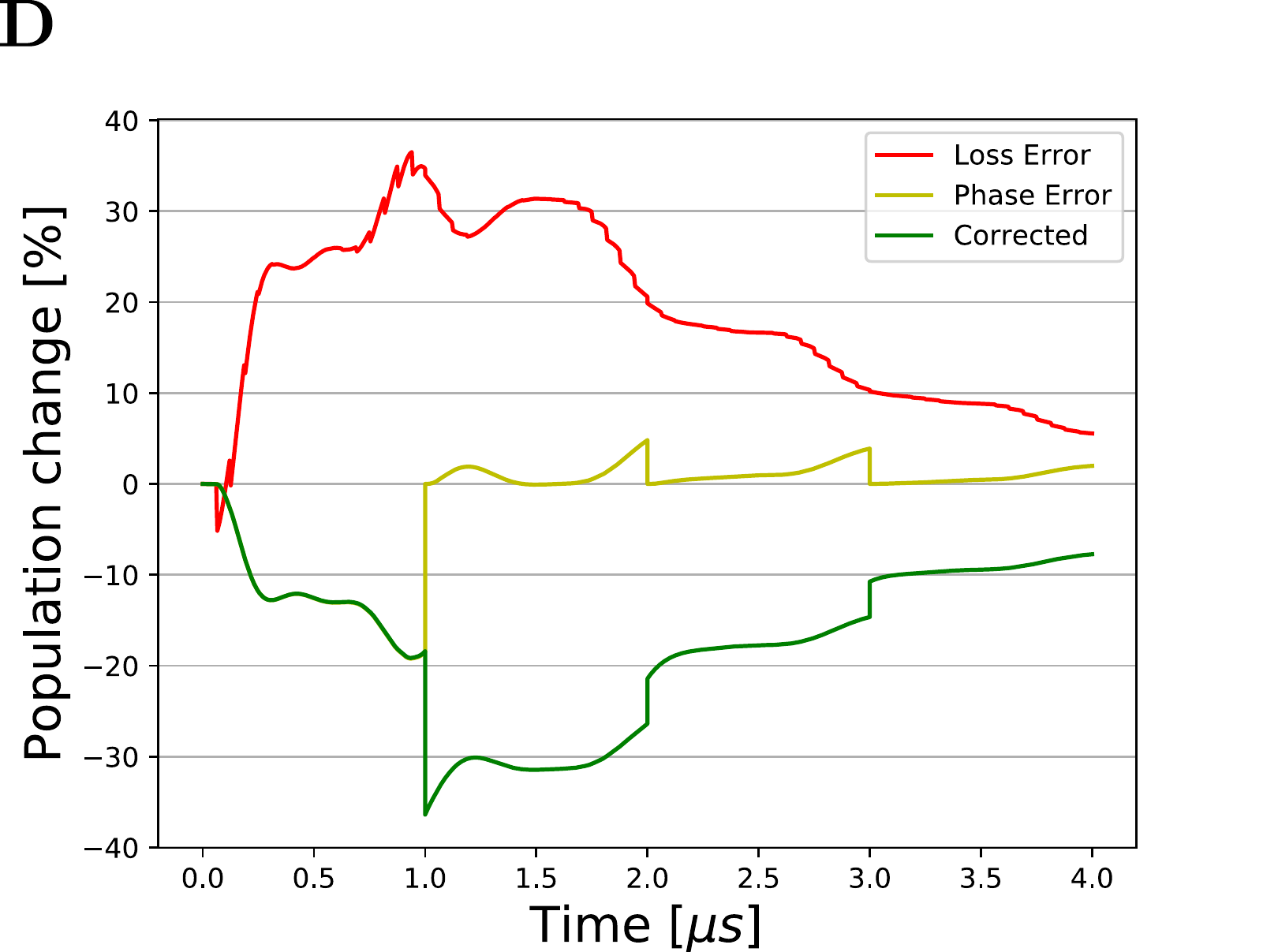}
   \captionlistentry{} \label{fig:Bad_g0_Zeno}
\end{subfigure} \vspace{0.6cm}
\caption{\textbf{Quantum-Zeno effects from repeated measurements.} (\textbf{A}) Population in the loss-error states of \eqref{eq:Zeno_Errors} (red), the phase-error states of \eqref{eq:Zeno_Phase_Errors} (yellow), and the corrected state \eqref{eq:Zeno_Corrected} (green) as a function of time when the system is initialized with an error and allowed to evolve without noise and decays in the data qubits. The parameters used are the same as in the main text except the time between corrections ($T_A$) is $\SI{1}{\micro \second}$. (\textbf{B}) Changes in the populations of the three classes of states introduced in (A) as a result of 15 additional equally spaced measurements being inserted between each correction-step. (\textbf{C}) Changes in populations of the logically dephased versions of the subspaces introduced in (A) due to the same increase in measurement-frequency as in (B). Note the change in scaling of the population-axis compared to (B) and (D). (\textbf{D}) A plot of the same quantities as in (B), but for a scheme where the reduced operator $g_0 =(I)(ZZZZZZ)(I)$ is used instead of the operator $g_0 =(Z)(ZZZZZZ)(Z)$ used in the rest of the paper.}
\label{fig:Zeno_Mega}
\end{figure*}

A sample simulation showing the evolution of these populations is shown on Fig. \ref{fig:Non_Zeno_Dynamics_Example}. During the evolution depicted on this figure, combined measurement- and correction-steps are applied after every microsecond. Comparing this reference evolution to similar evolutions where additional measurements are inserted (see Fig. \ref{fig:Zeno_Changes}), we see that the measurements indeed seem to disturb the dynamics slightly, leading to an increased population of the error states and a decreased population of the corrected states. Since the error states are the ones that can potentially lead to corruption of the encoded information when additional errors take place, it makes sense that an increased population in these states would result in a decreased coherence time. Note that from a physical standpoint, these effects do make sense---by observing $g_0$, we are repeatedly detecting if the system is in the subspace of loss-error states (i.e. \eqref{eq:Zeno_Errors}) or in the subspace of the dephased and corrected states (i.e. \eqref{eq:Zeno_Phase_Errors} and \eqref{eq:Zeno_Corrected}). From quantum-Zeno considerations~\cite{Itano1990}, we would expect such observations to freeze dynamics in place and inhibit transitions from one of these subspace to the other. In other words, the repeated measurements inhibit the decays responsible for correcting our error, resulting in the increased population in the error states and a decreased population in the corrected states that we observe. Looking at Fig. \ref{fig:Zeno_Changes}, we see that this is a relatively small effect, even for the rapid $\SI{16}{\mega \hertz}$ measurement-rate used in this figure. Nevertheless, a quick order-of-magnitude estimate shows that the effects observed on Fig. \ref{fig:Zeno_Changes} should indeed be able to result in decreases in fidelity of the magnitude observed on Fig. \ref{fig:Long_v_Short_TA}. Specifically, since each qubit decays about once every $\SI{40}{\micro \second}$, we would expect photon-loss errors in the six qubits to occur at a rate of about $R=(6/40) \mu s^{-1}$. After each decay, there is a period of about half a microsecond where there is an increase of population in susceptible states of about $0.3\%$. The risk that a decay occurs for this population during that $\SI{0.5}{\micro \second}$ window must be about $P\simeq R \cdot \SI{0.5}{\micro \second}$, leading to a population-loss of $0.4\% \cdot P$. Thus we can estimate the rate at which additional corruptions occur as
\begin{align*}
R \cdot 0.3\% \cdot P = R^2 \cdot \SI{0.5}{\micro \second} \cdot 0.3\% \; ,
\end{align*}
and thus the total amount of lost fidelity over $\SI{200}{\micro \second}$ will be 
\begin{align*}
R^2 \cdot \SI{0.5}{\micro \second} \cdot \SI{200}{\micro \second} \cdot 0.3\% \simeq 1\% \; .
\end{align*}
The actual observed decrease in fidelity at the top of the active-correction spikes for these parameters is approximately $5\%$. In other words, the effect of the additional population in the error states has the right order of magnitude. However, the effect does not seem quite severe enough to be the sole explanation for the decrease in fidelity---at least if our crude estimates are to be believed. Thus it is likely that another mechanism exist which contributes to the corruption of encoded information in a similar manner. A promising candidate for this would be logical dephasing, i.e. the accumulation of population in states similar to those in \eqref{eq:Zeno_Errors}, \eqref{eq:Zeno_Phase_Errors} and \eqref{eq:Zeno_Corrected}, but with the relative sign of the two components in these expressions reversed. Indeed, simulations show that the repeated measurements results in an ever-increasing additional population in these logically dephased states (see Fig. \ref{fig:Zeno_Dephasing}). While the rate of accumulation is relatively slow (about $0.01\frac{\%}{ \mu s}$), the long $\SI{200}{\micro \second}$ timescales involved in the storage of information in our scheme could result in this effect giving rise to a $1-2\%$ decrease in fidelity, and thus to decoherence on the same scale as the effect investigated above.

One thing to note is that the degree to which the measurements impact the dynamics depends strongly on the type of measurements performed. As explained above, the process inhibited by the repeated measurements of $g_0=(Z)(ZZZZZZ)(Z)$ is the decay of the shadow qubit. However, it is in principle possible to operate the error correction code with a reduced form of the operator, namely $g_0=(I)(ZZZZZZ)(I)$. In this case, the transition from $g_0=-1$ to $g_0=+1$ would happen already at the oscillations
\begin{align*}
\left| \downarrow \right> \left| \pm \pm T \right>\left| \downarrow  \right> \leftrightarrow \left| \downarrow  \right>  \left( \left| \pm \pm + \right> + \left| \pm \pm - \right>  \right) \left| \uparrow \right> 
\end{align*}
rather than at the shadow-qubit decay. This oscillation occurs on the timescale $\frac{\pi \hbar}{\frac{A}{\sqrt{2}}} = \SI{0.35}{\micro \second}$, which is relatively long compared to the timescale $T_S= \SI{0.08}{\micro \second}$ of shadow-qubit decay. Since slower dynamics are expected to suffer more severely from Zeno-effects, we therefore expect this version of the scheme to suffer more heavily when frequent measurements are introduced---a prediction which is confirmed by the plot on Fig. \ref{fig:Bad_g0_Zeno}.

To sum up, it seems reasonable to expect the storage-performance of the state $\frac{1}{\sqrt{2}} \left( \left| \downarrow \right> + \left| \uparrow \right> \right)_L$ to be relatively independent of $T_A$, because as long as the autonomous correction works as it should, this state will be immune to the errors that result from large values of $T_A$. However, as we have seen above, performing frequent measurements disturbs the dynamics responsible for the autonomous correction, leading to both a slower correction of errors and an increased rate of logical dephasing. Combining these two facts explains the trend for $\frac{1}{\sqrt{2}} \left( \left| \downarrow \right> + \left| \uparrow \right> \right)_L$-storage depicted in Fig. \ref{fig:TA_Dependency}. In contrast, $\left| \uparrow \right>_L$ is highly sensitive to additional errors occurring before an active correction-step, meaning  low $T_A$-values are a required to store this state well. Additionally, it is not sensitive to logical dephasing, and is thus immune to part of the detrimental effects of frequent measurements. Combining these facts fully explain the data depicted on Fig. \ref{fig:TA_Dependency}. Finally, we have concluded that the choice of operator for $g_0$ has a large influence on the severity of any quantum Zeno effects, and thus motivated the choice of $g_0$ made in the main text.\\

\subsection{Scaling with Coherence Times of Physical Qubits}
\label{sec:Scaling_with_Phys_Coh}

\begin{figure*}[hbtp]
  \centering
  \begin{subfigure}{.48\textwidth}
   \includegraphics[scale=0.5]{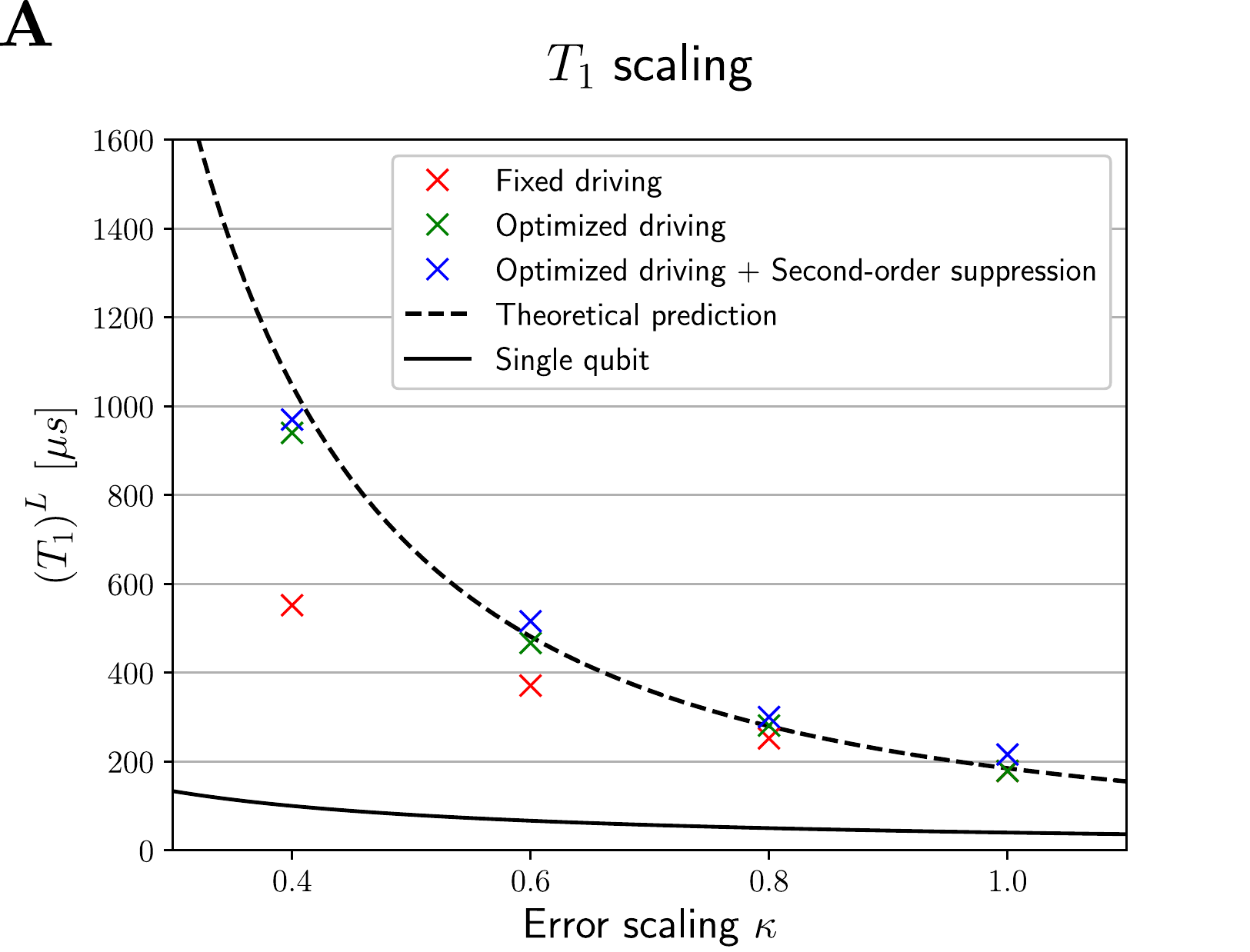}
   \captionlistentry{} \label{fig:Error_Scaling_Absolute}   
\end{subfigure}%
\begin{subfigure}{.48\textwidth}
   \includegraphics[scale=0.5]{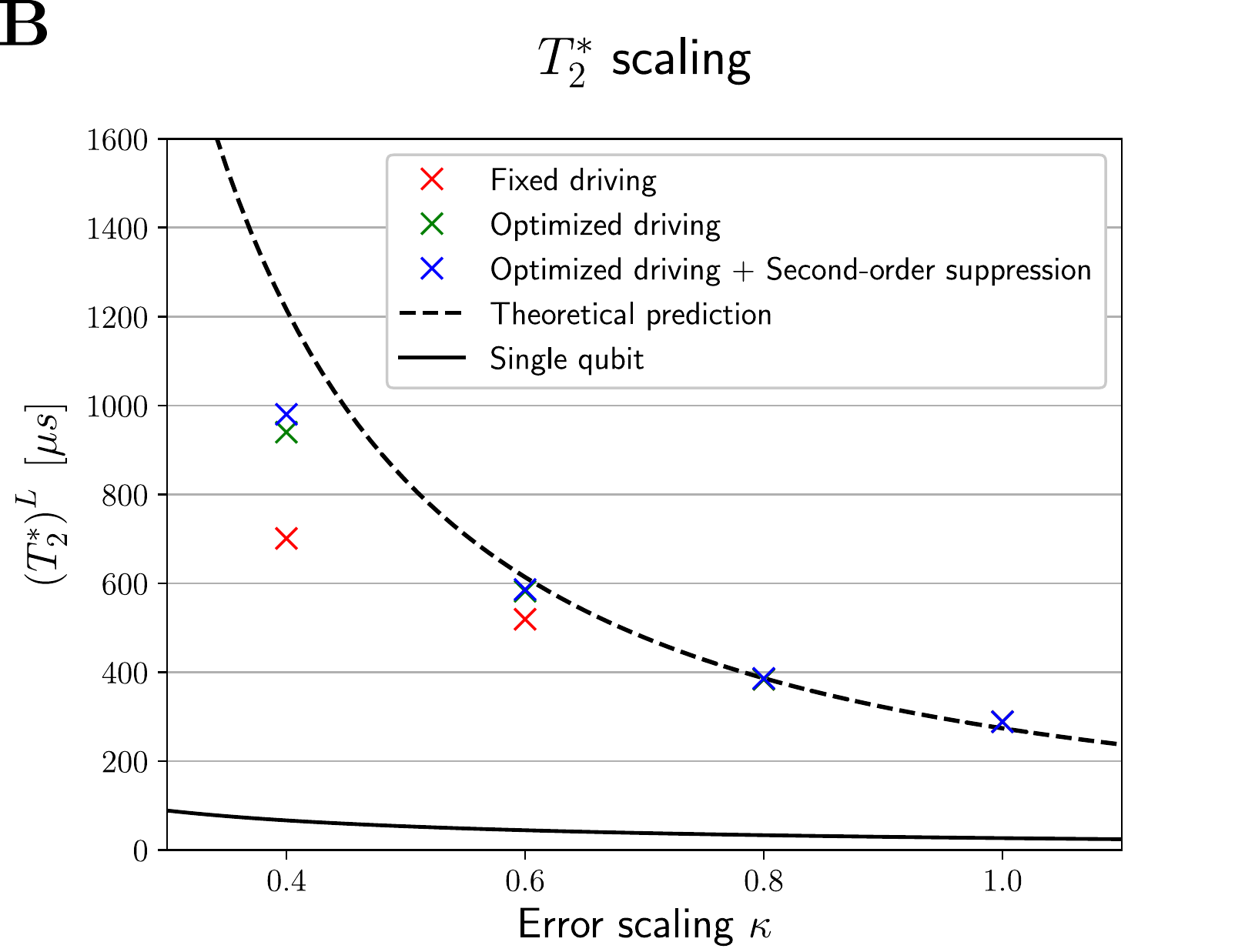}
   \captionlistentry{} \label{fig:Error_Scaling_Absolute_B}
\end{subfigure}\\ \vspace{0.9cm}
\begin{subfigure}{.48\textwidth}
   \includegraphics[scale=0.5]{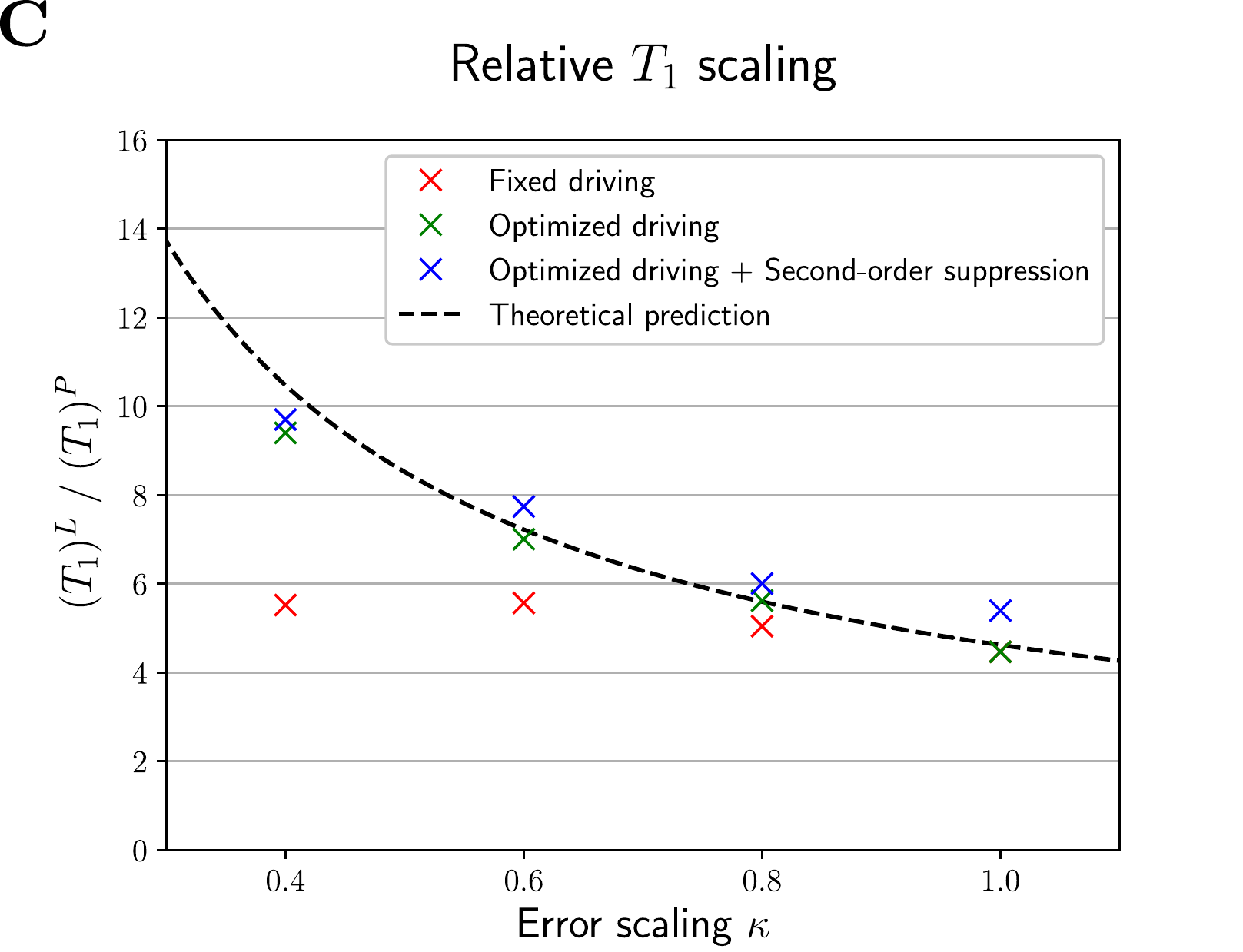}
      \captionlistentry{} \label{fig:Error_Scaling_Relative}
\end{subfigure}%
\begin{subfigure}{.48\textwidth}
   \includegraphics[scale=0.5]{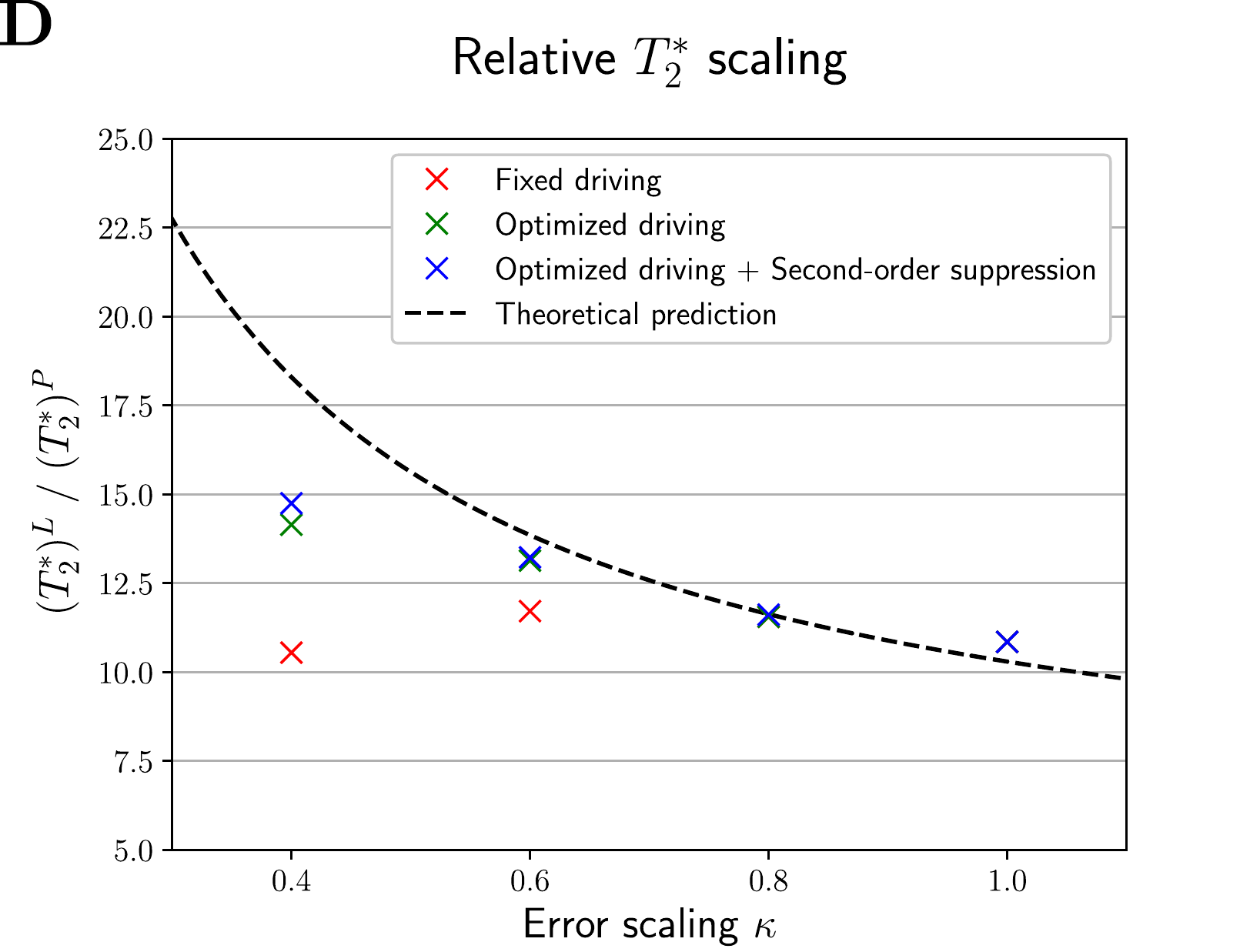}
   \captionlistentry{} \label{fig:Error_Scaling_Relative_D}
\end{subfigure} \vspace{0.6cm}
\caption{\textbf{Scaling of coherence times with the quality of the underlying physical qubits.} (\textbf{A} and \textbf{B}) Coherence times for a logical qubit encoded using the scheme from the main text when the error-rates of the physical qubits are rescaled by a factor $\kappa$. The plots contain data from both a scheme with parameters fixed to those of Fig. \ref{tbl:Parameters} (red), a similar scheme where the driving-strength $A$ is adjusted for better performance (green), and a scheme where this adjustment is combined with the protection from second-order processes detailed in App. \ref{sec:Second_Order_Suppression} (blue). For reference, the lifetime of the physical qubits is added, as well as a theoretical prediction of the performance of the scheme (see App. \ref{sec:TA_Dependency} for details). (\textbf{C} and \textbf{D}) A plot of the same information as depicted on (A) and (B), except all coherence times are measured relative to the corresponding coherence time of a single physical qubit.}
\label{fig:Error_Scaling_Mega}
\end{figure*}

As we have seen in previous sections, the dynamics related to the autonomous correction is not infallible, but tends to introduce small errors on its own through small second-order effects, as well as a small amount of logical dephasing when combined with the measurements of the measure-correct part of the scheme. A central question is the combined severity of these effects. Since the resulting decoherence will persist even as the coherence times of the physical qubits are scaled up, a way to quantify this severity is to look at how the storage-performance of the scheme depends on the coherence times of the underlying physical qubits. Simulations investigating this question are depicted on Fig. \ref{fig:Error_Scaling_Mega}. On this figure, Fig. \ref{fig:Error_Scaling_Absolute} and \ref{fig:Error_Scaling_Absolute_B} depict how the coherence times of the encoded logical qubit change as the rate of physical errors are scaled down. Specifically, each data-point represents the coherence times $(T_1)^L$ and $(T_2^*)^L$ of the logical qubit when the rate of photon-loss errors $\gamma_1=\frac{1}{T_1}$ and the rate of dephasing errors $\gamma_\phi = \frac{1}{T_\phi}$ of the physical qubits are rescaled by a factor $\kappa \in \{0.4,0.6,0.8,1.0\}$. For the red datapoints, no adjustments were made to the scheme except this rescaling of the errors. However, as can be seen from Fig. \ref{fig:Error_Scaling_Absolute} and \ref{fig:Error_Scaling_Absolute_B}, this leads to the coherence times of the scheme saturating at about $\SI{600}{\micro \second}$ even in the limit where the coherence times of the physical qubits reach $\SI{100}{\micro \second}$. To improve upon this, we used the fact that progressively lower physical error rates require progressively weaker active correction. Thus by slowly decreasing the strength of the driving-parameter $A$ (and hence also $\delta$) in tandem with the reduction of physical error rates, the rate of problematic transitions induced by the autonomous driving can be kept in check while still providing sufficiently quick autonomous correction to combat the reduced error rates. The resulting performance is depicted in green on Fig. \ref{fig:Error_Scaling_Mega}. Finally, a small additional improvement could be achieved by adding the protection from second-order processes detailed in App. \ref{sec:Second_Order_Suppression}, leading to the blue datapoints on Fig. \ref{fig:Error_Scaling_Mega}. For reference, the plots also contain the physical-qubit lifetimes:
\begin{align*}
(T_1)^P(\kappa) &= \frac{\SI{40}{\micro \second}}{\kappa}\\
(T_2^*)^P(\kappa) &= \frac{\SI{26.6}{\micro \second}}{\kappa}
\end{align*}
and a theoretical estimate of how the coherence times are expected to increased based on how they depend on the strength of the parameters of the autonomous scheme. This prediction is based on the realization that rescaling the unit of time $\left[ t \right]$ as follows:
\begin{align*}
\left[ t' \right] = \frac{1}{\kappa} \left[ t \right]
\end{align*}
results in the unit-less numerical values representing the coupling-strengths of the scheme having to be rescaled by a similar factor, since they essentially represents oscillation frequencies:
\begin{align*}
\tilde{J} \sim \frac{\text{oscillations}}{\left[ t \right] } = \frac{1}{\kappa} \frac{\text{oscillations}}{\left[ t' \right] } \sim \frac{1}{\kappa} \tilde{J}' \; .
\end{align*}
Here, we have used a tilde to signify the fact that these are the unitless quantities used in the simulation. Thus a rescaling of the unit of time results in a rescaling of the entire Hamiltonian related to the autonomous part of the correction:
\begin{align*}
\tilde{H}' = \frac{1}{\kappa} \tilde{H} \; ,
\end{align*}
In a similar vein, the change of units results in a rescaling of the decay-rates:
\begin{align*}
\tilde{\gamma}_1' &= \frac{1}{\kappa} \tilde{\gamma}_1 \\
\tilde{\gamma}_\phi' &= \frac{1}{\kappa} \tilde{\gamma}_\phi \\
\tilde{\gamma}_S' &= \frac{1}{\kappa} \tilde{\gamma}_S \; .
\end{align*}
In contrast, $T_A$ is rescaled as:
\begin{align*}
T_A = \tilde{T}_A \left[ t \right] &= \tilde{T}_A' \left[ t' \right]  = \frac{1}{\kappa} \tilde{T}_A' \left[ t \right] \\
& \Rightarrow \tilde{T}_A' = \kappa \tilde{T}_A \; . 
\end{align*}
Note that all of these rescalings cannot have any influence on the actual physical results from the simulation, since they simply represent a different choice of units. As a result, any physical time-scale $T$ that we extract from running simulations with a given set of parameters $\{ \tilde{H}, \tilde{\gamma}_1, \tilde{\gamma}_\phi, \tilde{\gamma}_S, \tilde{T}_A \}$ must fulfil
\begin{align*}
T \left( H, \gamma_1 , \gamma_\phi , \gamma_S , T_A \right) &= \tilde{T} \left( \tilde{H}, \tilde{\gamma}_1, \tilde{\gamma}_\phi, \tilde{\gamma}_S, \tilde{T}_A \right) \left[ t \right] \\
&= \tilde{T} \left( \tilde{H}', \tilde{\gamma}'_1, \tilde{\gamma}'_\phi, \tilde{\gamma}'_S, \tilde{T}'_A \right) \left[ t' \right] \; ,
\end{align*}
which results in the relation
\begin{align*}
\tilde{T} \left( \tilde{H}, \kappa \tilde{\gamma}_1, \kappa \tilde{\gamma}_\phi, \tilde{\gamma}_S, \tilde{T}_A \right) = \frac{1}{\kappa} \tilde{T} \left( \frac{1}{\kappa} \tilde{H}, \tilde{\gamma}_1, \tilde{\gamma}_\phi, \frac{1}{\kappa} \tilde{\gamma}_S, \kappa \tilde{T}_A \right) \; .
\end{align*}
Now, we arrived at this result by imagining a rescaling of units, but the above property must hold no matter the reason that the input-parameters are rescaled. In other words, the above also holds if the unit of time is fixed but the physical parameters of the model are changed, meaning we are allowed to conclude:
\begin{align*}
T \left( H, \kappa \gamma_1, \kappa \gamma_\phi, \gamma_S, T_A \right) = \frac{1}{\kappa} T \left( \frac{1}{\kappa} H, \gamma_1, \gamma_\phi, \frac{1}{\kappa} \gamma_S, \kappa T_A \right)
\end{align*}
What we see from this is that a rescaling of the error-rates of the physical qubits by $\kappa$ can be related to a rescaling of the strength of the autonomous correction by $\frac{1}{\kappa}$ through a rescaling of the hQEC-scheme by $\frac{1}{\kappa}$:
\begin{align*}
H \; &\rightarrow \; \frac{1}{\kappa}\, H\\
\gamma_S \; &\rightarrow \; \frac{1}{\kappa} \, \gamma_S\\
T_A \; &\rightarrow \; \kappa \, T_A \; .
\end{align*}
In other words, we have found a precise relation between the effects of scaling the strength of error correction and the effects of scaling the strength of single-qubit errors. since $(T_1)^L$ and $(T_2^*)^L$ are both physical timescales extracted from our simulations, the scaling-result applies to both of these quantities. Given a lot of data has already been gathered about the scaling of coherence times when the strength of correction is changed (see App. \ref{sec:Params}), this allows us to give a theoretical estimate for how the coherence times are expected to scale as a function of $\kappa$. Specifically, fits yield
\begin{align*}
(T_1)^L \left( \frac{1}{\kappa} H, \gamma_1, \gamma_\phi, \frac{1}{\kappa} \gamma_S, \kappa T_A \right) &\simeq \left( \SI{78}{\micro \second} \right) \frac{1}{\kappa} + \SI{29}{\micro \second} \\
(T_2^*)^L \left( \frac{1}{\kappa} H, \gamma_1, \gamma_\phi, \frac{1}{\kappa} \gamma_S, \kappa T_A \right) &\simeq \left( \SI{144}{\micro \second} \right) \frac{1}{\kappa} + \SI{134}{\micro \second}  \; ,
\end{align*}
and thus
\begin{align*}
(T_1)^L \left( H, \kappa \gamma_1, \kappa \gamma_\phi,  \gamma_S,  T_A \right) &\simeq \left( \SI{78}{\micro \second} \right) \frac{1}{\kappa^2} + \left( \SI{29}{\micro \second}\right) \frac{1}{\kappa}  \\
(T_2^*)^L \left( H, \kappa \gamma_1, \kappa \gamma_\phi,  \gamma_S,  T_A \right) &\simeq \left( \SI{144}{\micro \second} \right) \frac{1}{\kappa^2} + \left( \SI{134}{\micro \second}\right) \frac{1}{\kappa}  \; . \\
\end{align*}
It is these trendlines that are depicted on Fig. \ref{fig:Error_Scaling_Absolute} and \ref{fig:Error_Scaling_Absolute_B}. Note that while the data fitted to yield these estimates were all in the range $\kappa \in \left[ 0.8, 2.0 \right]$, the model seems to fit well with blue and green data points even for much lower values of $\kappa$.

With the content of Fig. \ref{fig:Error_Scaling_Absolute} and \ref{fig:Error_Scaling_Absolute_B} now explained in detail, what can we actually conclude from this figure? With respect to the $T_1$-time of the encoded qubit, we see that this follows the theoretical prediction very well once the driving-strength is adjusted, which indicates that the scheme still works in essentially the same way for the low physical error-rates as for the larger error rates from which the theoretical prediction was extracted. In other words, it does not seem that the errors induced by the scheme itself dominates until at values of $\kappa$ lower than the $0.4$ reached in our simulations. With respect to the actual coherence times, we see a steady increase as the lifetimes of the physical qubits are increased, culminating in a $T_1$-time of $\SI{970}{\micro \second}$ in the situation where $(T_1)^P = \SI{100}{\micro \second}$ and $\left(T_2^*\right)^P = \SI{66}{\micro \second}$. Similarly, we see that milli-second coherence times can also be reached with respect to $T_2^*$, but that these data display a larger deviation from the predicted behaviour, indicating that performance may be saturating for $\kappa$ less that $0.4$ due to the  second order transitions and quantum-Zeno effects mentioned at the beginning of this section. Nevertheless, coherence times on the order of a millisecond should be achievable for our scheme given physical qubits of sufficiently high quality.

While the above considerations gives a good idea of the absolute level of performance one can hope to extract from the code, another important question is how large of an improvement the scheme provides compared to a single physical qubit with a similarly increased coherence time. After all, it is the improvement compared to an uncorrected qubit that is the central figure of merit for any error correction scheme. In Fig. \ref{fig:Error_Scaling_Relative} and \ref{fig:Error_Scaling_Relative_D}, we therefore plot the ratio between the coherence time of the scheme and that of a single physical qubit. As can be seen from this plot, this ratio actually increases as the error rates of the physical qubits are decreased, indicating that the coherence times of the error-correction scheme increase more rapidly than those of the physical qubits. Of course, in the limit $\kappa=0$ the single qubit will have infinite coherence time while second order processes will result in our scheme having a finite coherence time, meaning the ratio of coherence times should tend to zero in this limit. As a result, the increasing ratio with decreasing $\kappa$ cannot hold for very small values of $\kappa$, a fact already indicated by the $T_2^*$-data. Nevertheless, we see that the scheme can provide improvements to coherence times of more than an order of magnitude if given access to sufficiently good physical qubits

Note that the theoretical predictions for the ratios actually says a lot about the capabilities of the correction schemes. Specifically, since
\begin{align*}
\left(\frac{(T_1)^L}{(T_1)^P}\right) \left( \kappa \right) &\simeq 4.9 \, \frac{1}{\kappa} + 0.72\\
\left(\frac{(T_2^*)^L}{(T_2^*)^P} \right) \left( \kappa \right) &\simeq 5.3 \, \frac{1}{\kappa} + 4.9\; ,
\end{align*}
we see that our scheme will tend to provide an improvement to $T_2^*$-times of about a factor 5 even in the presence of strong noise, while improvements to $T_1$-times are more fragile towards noisy physical qubits, and for sufficiently strong noise may degrade to the point of having worse relaxation-time than a single physical qubit.

\clearpage
\bibliography{bach}

\end{document}